\pdfoutput=1

\documentclass[letterpaper]{JHEP3}

\newcommand{\beq}{\begin{equation}}
\newcommand{\eeq}{\end{equation}}
\newcommand{\phib}{\ensuremath{\overline{\phi}}}
\newcommand{\psib}{\ensuremath{\overline{\psi}}}
\newcommand{\one}{\hat{1}}
\usepackage{graphicx}
\usepackage{amsmath}
\usepackage{amsfonts}
\usepackage{epsfig}

\title{Towards lattice simulation of the gauge theory duals to black holes and hot strings}

\author{Simon Catterall \\
Department of Physics, Syracuse University, Syracuse, NY13244, USA \\
E-mail: \email{smc@phy.syr.edu}
}

\author{Toby Wiseman \\
Theoretical Physics Group, Blackett Laboratory, Imperial College, London SW7 2AZ, UK \\
E-mail: \email{t.wiseman@imperial.ac.uk}
}

\preprint{hep-th/yymmddd\\ Imperial/TP/07/TW/03}

\date{June 2007}

\abstract{

A generalization of the AdS/CFT conjecture postulates a duality between IIA string theory and $16$ supercharge Yang-Mills quantum mechanics in the large $N$ 't Hooft limit. At low temperatures string theory describes black holes, whose thermodynamics may hence be studied
using the dual quantum mechanics.
This quantum mechanics is strongly coupled
which motivates the use of lattice techniques.
We argue that, contrary to expectation, the theory when  discretized naively
will nevertheless recover continuum supersymmetry as the lattice spacing is
sent to zero.
We test these ideas by 
studying the $4$ supercharge version of this Yang-Mills quantum mechanics in the 't Hooft limit. 
We use both a naive lattice action and a
manifestly supersymmetric action. Using Monte Carlo methods we simulate the Euclidean theories, and study the lattice continuum limit, for both thermal
and non-thermal periodic boundary conditions, confirming continuum
supersymmetry is recovered for the naive action when appropriate. We obtain results for the thermal
system with $N$ up to 12. These favor the existence of
a single deconfined phase for all non-zero temperatures.
These results are an encouraging indication that the $16$ supercharge
theory is within reach using similar methods and resources.

}

\begin{document}

%
\section{Introduction}
%

Despite progress in understanding quantum gravity there has been relatively little progress in theoretically realizing a quantum black hole, the key object in quantum gravity. Substantial progress was afforded by the work of Strominger and Vafa \cite{StromVafa} who showed that for certain black holes in string theory the constituent microstates can be explicitly
counted and the result is in agreement with the usual
Bekenstein-Hawking entropy \cite{Bekenstein}. The calculation relies on a continuation to a weakly coupled theory where states can be explicitly counted, and this number is
then argued to be independent of the continuation. This
gives an indirect understanding of the microstates of the black hole. Another beautiful result of Strominger is that the entropy of black holes with an $AdS_3$ near horizon geometry can be computed on very general grounds as the gravitational theory describing the black hole can be argued to be a 2-d conformal field theory \cite{Strom}. One does not know the theory, but does know the central charge, which is sufficient to compute the entropy using Cardy's result. Thus again we gain insight into the microstates, although we are not able to solve the theory governing the quantum black hole. For a review of these approaches we refer the reader to \cite{BHreview} and for an interesting new proposal concerning understanding the microscopic description without resorting to a weak coupling continuation see the work of Mathur \cite{Mathur}.

A very promising new direction for the study of quantum gravity is the AdS/CFT correspondence and its generalizations \cite{Mald, Itzhaki}. Here one can describe certain black holes in terms of the worldvolume theory of the D-branes that compose them. These are 16 supercharge gauge theories in various dimensions, taken in the large $N$ 't Hooft limit and at finite temperature.
The regime in which they describe string theory black holes is one in which they are strongly coupled, and solving this theory would allow one to directly study the quantum properties of the dual black hole, including its thermodynamic properties. The problem of computing directly in this strongly coupled theory is a technical one, not one of principle. While analytic methods have made much progress in describing the planar limit of the 4 dimensional version of this correspondence using integrability techniques \cite{Minahan} it is not generally understood how to go beyond this limit. It particular black holes are precisely objects that live beyond this limit, having dual energy densities of order $O(N^2)$. Even protected 
quantities that can give information about the strongly coupled theories through weak coupling calculations have not yet revealed information about black holes \cite{MaldMinw} and if they do in the future, they will remain indirect probes of quantum black hole physics. Another direction for direct calculation is the use of the Gaussian approximation, pioneered in this context for
Yang-Mills quantum mechanics in \cite{Kabat1}. Certain agreement with black hole thermodynamics was extracted \cite{Kabat2}, although the approximation scheme is essentially uncontrolled, having no small parameter.

The aim of this paper is to investigate the feasibility of using lattice methods to directly simulate the strongly coupled worldvolume theories. Hiller et al \cite{Hiller} have used lightcone methods to give evidence that the graviton propagator is correctly reproduced in the 2 dimensional version of the correspondence. Here, we test the use 
of direct Monte Carlo simulation of the lattice regulated path
integral to investigate these systems.
Since we are interested in studying the theory at finite temperature, Monte Carlo techniques seem the most natural as they allow us to study observables directly in the thermal ensemble.
In addition, Monte Carlo
methods are often the only feasible approach in higher dimensions.

These theories are necessarily supersymmetric, and this presents technical challenges for the lattice. 
The first problem encountered in such a study is how best to discretize
such a supersymmetric theory.
There has been much interest in the lattice study of supersymmetric theories. Aside from our current application, supersymmetry plays a prominent role in efforts to construct theories which go beyond the standard model of particle physics and the lattice provides the possibility of studying nonperturbative effects such
as dynamical supersymmetry breaking, of relevance for phenomenology.
Unfortunately, in many cases, supersymmetry is broken at the 
classical level in such discretizations and can only be regained in the quantum continuum limit after
a great deal of fine tuning in the lattice theory. 
Recently significant effort has gone into attempts to construct lattice theories which retain an exact
supersymmetry at non-zero lattice spacing. Two approaches have
been used; lattice models arising from orbifolding a supersymmetric
matrix model \cite{Kaplan1,Kaplan2,Kaplan3,Damgaard,Giedt_rev2} and
constructions based on
discretizing a topological or twisted form of the continuum supersymmetric
theory \cite{Sugino1,Sugino2,Kawamoto,Catterall_2d,Catterall_4d}.
In the case of Yang-Mills theories it appears that these approaches are
intimately connected \cite{Unsal,Takimi}. The approach based on twisting
has also been studied in the case of Wess-Zumino and sigma models
\cite{Catterall_wz,Catterall_top,Giedt_wz,Catterall_sig,Bergner}. 
The philosophy
behind these approaches is that this residual lattice supersymmetry will help to protect the theory
from the dangerous radiative corrections which lead to fine tuning.

However, in a sufficiently low number of dimensions the super renormalizable nature of the models leads
to much reduced fine tuning problems -- typically there are only a finite number of divergences in such theories
which may be computed in perturbation theory \cite{Giedt}. 
In the case considered here, supersymmetric Yang-Mills
quantum mechanics,
we argue that {\it no} relevant supersymmetry breaking counterterms
can be written down. Naive discretizations of such a theory, while breaking supersymmetry classically
by terms of the order of the lattice spacing, flow automatically to the supersymmetric theory in the
naive continuum limit corresponding to a vanishing lattice spacing.

In this paper we have examined the effectiveness of using Monte Carlo simulation
to study super Yang-Mills quantum mechanics in the large $N$ 't Hooft limit in the case of the $4$ supercharge model. This theory, while
exhibiting many features which are qualitatively similar to its
$16$ supercharge cousin, is computationally 
easier and a useful `warm-up' exercise before tackling the 16 supercharge case. 
It has been studied using Hamiltonian methods in \cite{Campostrini1,Campostrini2} for $SU(2)$.
We have studied two different discretizations of this 
theory. One of these possesses 2 exact lattice supersymmetries corresponding to the dimensional reduction of the
twisted model derived in \cite{Catterall_2d}. The other corresponds to a naive discretization of the
continuum theory constrained only by the necessity of employing a lattice derivative which forbids fermion
doubling. Both are invariant under lattice gauge transformations.

We give analytical arguments that the naive action should not require fine tuning to attain the
correct continuum limit. We reinforce this conclusion by
conducting numerical simulations of the model on finite lattices at zero temperature.
The results of these simulations show that the vacuum energy approaches zero
as the lattice spacing is reduced and that expectation values computed in this
naive lattice theory agree with those computed with the supersymmetric lattice action.
Since the simulations of the supersymmetric system are much more computationally demanding we
have primarily used the naive action in subsequent simulations of the thermal
system. 

To set the stage for the interpretation of our thermal results we have also simulated the quenched theory. 
These confirm the previous expectation of a rapid crossover in the
thermodynamic behavior of the system at some temperature which 
becomes a sharp phase transition as $N\to\infty$ \cite{Aharony1,Aharony2}.
For high temperatures the theory appears deconfined but below
the critical temperature the system enters 
a confining phase with non-zero vacuum energy. Our new results from
simulations of the full 4 supercharge theory however favor a simple phase structure with
a single deconfined phase for all non-zero temperatures - as we would expect for the behaviour of the 16 supercharge theory from holographic considerations. We are able to simulate this 4 supercharge thermal theory with $N$ up to 12, already sufficient to see the asymptotic 't Hooft scaling.

The plan of the paper is as follows. In section 2 we introduce the Yang-Mills model in the continuum including a discussion
of its renormalizability and vacuum structure. Section 3 describes the two lattice actions we have utilized while
we present our results for the quenched, zero temperature and thermal theories in 
section 4. Our final section discusses the implication of our results both in the context of
the $4$ supercharge model and its string theory cousin with 16 supercharges.  

Finally, the paper ends with several appendices containing more technical details on
holographic duality, the quantum corrections to the classical moduli space of the theory,
and further numerical results concerning the extrapolation of our
finite lattice data to the continuum limit. It also contains a detailed description of the RHMC
algorithm used in our simulations.

While this paper was in preparation we received a paper
which utilizes momentum space methods to study the same
system for $SU(4)$ with results that are in at least qualitative
agreement with ours \cite{Nishimura}.

%
\section{4 supercharge Yang-Mills quantum mechanics}
%

The Euclidean supersymmetric $SU(N)$ Yang-Mills quantum mechanics we are interested in can be thought of as arising from the classical dimensional reduction of $\mathcal{N} = 1$ super Yang-Mills in 4 dimensions. The matter content in the quantum mechanics arises from reduction of the gauge field in 4-d, giving 3 adjoint bosonic scalar fields $X_i$, where $i = 1,\ldots,3$  and the adjoint gauge field, $A$. The 4-d adjoint fermions can be equivalently written in either a Weyl or Majorana representation. Here we will use the Weyl representation, giving a complex 2-component fermion field, ${\Psi}_{\alpha}$ with a Euclidean spinor index $\alpha$, transforming in the adjoint of the gauge group. Dimensionally reducing this, the spinor index becomes an internal symmetry. The quantum mechanics Euclidean action is given as,
\begin{equation}\label{eq:action}
    S= \frac{1}{\lambda} N \mathrm{Tr} \oint^{R} d\tau
      \left\{ \frac{1}{2} (D_{\tau} X_i)^2 - \frac{1}{4} \left[ X_i, X_j \right]^2 +  i \bar{\Psi} \bar{\sigma}^\tau D_{\tau} \Psi - \bar{\Psi} \bar{\sigma}^i \left[ X_i, \Psi \right] \right\} ,
\end{equation}
with $\tau$ being compact Euclidean time with radius $R$. The covariant derivative is defined as $D_{\tau} = \partial_{\tau} + i \left[ A(\tau),\cdot \right]$ on an adjoint field, and we use the usual (Euclidean) chiral conventions,
\begin{eqnarray}
\bar{\sigma}^{\tau\dot{\alpha}\alpha} & = & - i \mathbf{1} ,\; \bar{\sigma}^{i\dot{\alpha}\alpha} = - \underline{\sigma}^i 
\end{eqnarray}
with $\underline{\sigma}^i$ being the Pauli matrices. We have written the $SU(N)$ gauge theory using the 't Hooft coupling $\lambda = N g_{YM}^2$, with $g_{YM}$ the Yang-Mills gauge coupling. The supersymmetry transformation
derives from that of the reduction of the parent 4-d $\mathcal{N} = 1$ super Yang-Mills, giving,
\begin{eqnarray}
\delta A & = & - i \bar{\Psi} \bar{\sigma}^\tau \xi   \nonumber \\ 
\delta X_i & = & - i \bar{\Psi} \bar{\sigma}^i \xi   \nonumber \\
\delta \Psi & = & 2 \left( \sigma^{\tau i} \xi \right) \left( D_{\tau} X_i \right)  + i \left( \sigma^{ij} \xi \right)  \left[X_i, X_j\right] 
\end{eqnarray}
where we use the notation $\sigma^{\mu\nu} = \frac{1}{4} \left( \sigma^\mu \bar{\sigma}^\nu - \sigma^\nu \bar{\sigma}^\mu \right)$. The theory has a global $SO(3)$ symmetry which can be interpreted as a rotational invariance in the target space of the theory, with action,
\begin{equation}
X'_{i}  =  \Lambda_{(1)i}^j X_j \qquad \Psi'_\alpha  =  \Lambda_{(\frac{1}{2})\alpha}^\beta \Psi_\beta
\end{equation}
where $\Lambda_{(1)}$ gives the representation of $SO(3)$ acting on vectors, and $\Lambda_{(1/2)}$ the representation on spinors.

Since the Euclidean time is topologically $S^1$ we have two spin structures which give either  periodic or anti-periodic boundary conditions for the fermion $\Psi$ as we traverse the time circle. In the latter case, the path integral $Z$ is the usual thermal partition function with temperature $T$,
\begin{equation}
Z_a(R)  = {\mathrm Tr} e^{-R \hat{H}}
\end{equation}
where $R$ is the inverse temperature $1/T$, and $\hat{H}$ is the Hamiltonian operator. Taking periodic boundary conditions,
\begin{equation}
Z_p(R)  = {\mathrm Tr} (-1)^{\hat{F}} e^{-R \hat{H}}
\end{equation}
where $\hat{F}$ is the fermion operator, and hence $Z_p$ gives the Witten index, whose value counts the difference in the number of
fermionic and bosonic ground states $\hat{H} |\psi> = 0$, and hence should not depend continuously on $R$.

\subsection{The $16$ supercharge theory and its IIA closed string theory dual}\label{sec:holography}
The 16 supercharge theory, which arises as the 1 dimensional reduction of $\mathcal{N}=1$ super Yang-Mills in 10 dimensions, takes a very similar form to the above. The only difference is that the reduction of the 10-d gauge field gives rise to 9 adjoint scalars $X_{\bar{i}}$, with  $\bar{i} = 1,\ldots,9$, and the 10-d Majorana fermions yield an adjoint fermion $\Psi_{\bar{\mu}}$ now with spinor index $\bar{\mu}= 1,\ldots,16$. Note that unlike the 4 supercharge theory, there is no Weyl representation available for the fermions, and hence one obtains a Pfaffian from integrating out these fields rather than a determinant. 

Since the 16 supercharge theory is so similar to the 4 supercharge theory, one might ask why we have decided to simulate the 4 supercharge case here. Clearly
the 16 supercharge case will generate a fermion operator which is
four times larger than its 4 supercharge cousin. With current
simulation algorithms for near massless dynamical fermions this will
lead to a factor of 16 slowdown in our simulations. In addition even in the continuum Euclidean theory
the Pfaffian that results after integration over the fermions is in general complex as seen from considering the zero momentum sector of the theory \cite{Nicolai}. In contrast to this the continuum 4 supercharge theory has a  Pfaffian which can be recast as a positive definite real determinant, and we later show that our naive lattice discretization also has this property. The feasibility of
using Monte Carlo methods to simulate such a system with complex Pfaffian depends upon whether
the resulting phase fluctuations are relatively small and infrequent as we
approach the continuum limit. In the latter case standard reweighting 
techniques
can be used to evaluate expectation values generated in the phase
quenched ensemble (see eg. \cite{Catterall_sims}. Because of these technical subtleties
we have decided to initially study the 4 supercharge theory 
before moving to the 16 supercharge case.

Following the AdS-CFT conjecture Itzhaki 
et al \cite{Itzhaki} have argued that
$SU(N)$  16 supercharge Yang-Mills theories taken in the large $N$ 't Hooft limit are dual to certain closed superstring theories in the near horizon region of $N$ coincident D-branes.
In particular $SU(N)$ super Yang-Mills quantum
mechanics is supposedly dual to the IIA string theory describing $N$ $D0$-branes. We review  this correspondence  in more detail in the appendix \ref{app:dual}. Here we simply state the results that the analysis gives.

At a finite temperature $T = 1/R$ the quantum mechanics theory then
describes a gas of $N$ D0-branes in the dual IIA theory. 
In one dimension we can define a parameter,
\begin{equation}
\beta = \frac{\lambda^{\frac{1}{3}}}{T}
\end{equation}
which we can think of as a dimensionless inverse temperature, characterizing
the behavior of the theory. For large $\beta >> 1$ (but still finite as compared with $N$) the system of D0-branes should be well described by a supergravity black hole which is much larger in radius than the string length $\alpha'^{1/2}$. It is remarkable that since we know how to
compute the Bekenstein-Hawking entropy of the supergravity black hole, we can predict - assuming the holographic correspondence is correct - that in the large $\beta$ limit the precise form of the Yang-Mills entropy and free energy will be,
\begin{equation}
S = 11.5 N^2 \beta^{-9/5} , \quad f = -4.11 N^2 \beta^{-14/5}
\label{eq:stringprediction}
\end{equation}
where $f$ is the dimensionless free energy, given by $F = \lambda^{1/3} f$ where $F$ is the usual free energy \cite{Itzhaki}. For decreasing $\beta$ the curvature at the horizon radius becomes larger and the supergravity description receives string oscillator $\alpha'$ corrections, about which little is known. 

For small $\beta << 1$ the system can best be thought of as a highly excited hot ball of strings and branes. Polchinski and Horowitz have argued that the hot ball of strings for $\beta << 1$ and the black hole at $\beta >> 1$ are the same object, and the physics at the transition $\beta \sim 1$ -  the `correspondence point'  - is therefore smooth \cite{HorowitzP}.
Witten has argued that the presence of a black hole in the dual string theory (geometrically implying a contractible Euclidean time circle) indicates the Yang-Mills theory is in a deconfined phase, with thermodynamic quantities scaling as $O(N^2)$ and finite expectation value for the amplitude of the Polyakov loop $<|\frac{1}{N}\mathrm{Tr} e^{i \oint A d\tau}|>$ \cite{Witten}. Conversely the
appearance of a confined phase would correspond to the 
absence of a black hole in the dual geometry, or more precisely a non-contractible time circle, and results in thermodynamic quantities of order $O(1)$ and a vanishing expectation value for the
Polyakov loop. Hence at large $\beta >>1$ we expect the 16 supercharge theory to be deconfined as it is indeed dual to a black hole. At small $\beta<<1$ one can dimensionally reduce the theory to a bosonic matrix model, which Monte Carlo simulation shows has energies scaling as $O(N^2)$, and hence one expects that the 16 supercharge theory is likely to be deconfined for all $\beta$ \cite{Aharony1,Aharony2}, tallying with smoothness at the `correspondence point'.

Hence the key questions that would concern a study of the thermal 16 supercharge Yang-Mills theory in the 't Hooft limit would be to confirm the above thermodynamic expectations at small and large $\beta$, and to study the transition region to see if there is a phase transition in the correspondence region.

Whilst in this paper we shall not compute with this 16 supercharge theory, we will simulate its relative, the 4 supercharge Yang-Mills model
in the 't Hooft limit. These two theories have similar classical and quantum low energy dynamics  and one might expect their qualitative thermodynamic properties to be similar. However, strictly one should view the calculation in this paper as a warm-up exercise for the 16 supercharge case, and a demonstration that the 
quantities of interest in the latter theory are likely to be computable using similar methods.

\subsection{Infrared behaviour of the 4 supercharge Euclidean theory}

The classical bosonic moduli space of the theory is simply given by setting the scalar and gauge field adjoint matrices to be mutually commuting, and constant in Euclidean time. This implies they are all diagonal up to gauge transformation. Such a classical moduli space naively leads to the concern that the path integral is not well defined due to infra-red divergences. However, in both the periodic and thermal cases quantum corrections lift this classical bosonic moduli space and hence render the path integral convergent. There are two sources of quantum corrections to be considered. These are the off-diagonal elements of the constant modes about the time circle, and secondly, the non-constant modes about the Euclidean time circle. For a given $\beta$, the loop expansion parameter is determined by the separation of the diagonal elements of the constant modes. The infra-red behaviour of the theory is precisely in the regime where we take the diagonal elements of the constant modes to be well separated, and hence we can approximate the theory by a 1-loop calculation. In appendix \ref{app:moduli} we explicitly perform the computation of the effective action for the bosonic classical zero modes in both periodic and thermal cases. We now summarize the results in this appendix, beginning with the thermal theory.

\subsubsection{The finite temperature theory}

The thermal case is simpler than the periodic case, and the result we give was previously given as a special case in Aharony et al \cite{Aharony2}. Taking temperature $T$, and antiperiodic Euclidean time with radius $R = 1/T$, the thermal boundary conditions imply there are no fermion modes that are constant in time, and hence no fermion zero modes. Thinking of the adjoint fields as $N \times N$ matrices we may expand them as,
\begin{eqnarray}
A_{ab}(\tau) & = & A^a {\delta}_{ab}  + \frac{1}{\sqrt{2 \pi}} \sum_{m=-\infty}^{\infty}  \delta A^{(m)}_{ab} e^{ \frac{2 \pi}{R} i m \tau} \nonumber \\
X_{i,ab}(\tau) & = & x_i^a {\delta}_{ab}  + \frac{1}{\sqrt{2 \pi}} \sum_{m=-\infty}^{\infty}  \delta X^{(m)}_{i,ab}e^{\frac{2 \pi}{R} i m \tau} \nonumber \\
\Psi_{\alpha,ab}(\tau) & = & \frac{1}{\sqrt{2 \pi}} \sum_{m=-\infty}^{\infty}  \delta \Psi^{(m)}_{\alpha,ab} e^{\frac{2 \pi}{R} i (m +\frac{1}{2}) \tau} \end{eqnarray}
where the matrix indices $a,b = 1,\ldots,N$. The classical bosonic moduli are $A^a,x^a_i$. We take the perturbations $\delta A^{(0)}, \delta X^{(0)}_i$ to have no diagonal terms and since the gauge group is $SU(N)$, the field matrices are traceless so the sums $\sum_a A^a, \sum_a x_i^a$ vanish. We define $\Delta A^{ab} = R (A^a - A^b)$ and $\Delta x_i^{ab} = R (x_i^a - x_i^b)$, where we note that these are now dimensionless. As we show in the appendix, the effective dimensionless coupling constant governing the
fluctuations is
\begin{equation} \label{eq:effcoupling}
g_{\rm eff} \sim \frac{ \beta^3 }{  |\Delta x^{ab}|^4 }
\end{equation}
where $|\Delta x^{ab}|^2 = \sum_i (\Delta x_i^{ab})^2$. Hence we may integrate out all the fluctuations with any $m$ provided the diagonal component moduli are sufficiently well separated compared to dimensionless temperature, so,
\begin{equation}
|\Delta x^{ab}|  >> \beta^{3/4} .
\end{equation}
As noted in the appendix, this condition arises from the zero modes with $m=0$. For small circle size, $\beta << 1$ we can integrate out all `Kaluza-Klein' modes, ie. bosonic fluctuations with $m \ne 0$ and all the fermionic fluctuations, and are then left with a zero dimensional bosonic matrix theory. However, unless the moduli are well separated in the sense above, one cannot also integrate out the off-diagonal components of this matrix theory, ie. the $m=0$ fluctuations.

As shown in the appendix, when we gauge fix the action and integrate out the fluctuations at 1-loop we find an action for the bosonic moduli. Supersymmetry is broken by the thermal boundary conditions and hence
the 1-loop determinants from the bosonic and fermionic fluctuations do not cancel each other. We obtain
an effective action for the moduli,
\begin{equation}
S_{\rm 1-loop}[A^a,x^a_i]  =  \sum_{a < b} \log \left( \frac{ \cosh |\Delta x^{ab}| - \cos \Delta A^{ab} }{ \cosh |\Delta x^{ab}| + \cos \Delta A^{ab} } \right) .
\end{equation}
This potential takes a simple form, arising from a pair-wise interaction of the moduli and hence scales as $O(N^2)$. Gauge invariance implies the potential is periodic in $\Delta A^{ab}$, the moduli $A^a$ multiplied by $R$ being angular variables giving the value of the Wilson loop $P e^{i \oint d\tau A} = e^{i \beta A^a \delta_{ab}} = \mathrm{diag}\left( e^{i R A^1} , e^{i R A^1}, \ldots , e^{i R A^N}\right)$. The minimum for the potential is when the moduli are coincident. Asymptotically, for a large separation $|\Delta x^{ab}| >> 1$ the potential goes to zero as $S_{\rm 1-loop} \sim - e^{-|\Delta x^{ab}|} \cos{\Delta A^{ab}}$, and so is attractive towards the coincident point for $-\frac{\pi}{2} < \Delta A^{ab} < \frac{\pi}{2}$.
One might worry that all moduli will simply coalesce under these attractive pairwise forces. However, as discussed above, for a separation $|\Delta x^{ab}| \sim \beta^{3/4}$ the loop approximation breaks down and the theory becomes strongly coupled. Hence we see that the classical moduli space is lifted by an attractive potential that drives the infra-red dynamics to strong coupling.

Since the path integral measure over the non-compact zero modes is $\prod_{a,i} d x^{a}_i$ we see that the integral, giving the partition function, should be convergent in the infra-red due to the exponential decay of the potential to zero for large separations. Naively this fast fall off would imply that the tails of the eigenvalue distributions of the scalars should die off faster than a power law.

\subsubsection{The periodic theory}

The analysis of the periodic case is more subtle. The results we summarize here and give fully in the appendix are new, and draw on previous results of Aoki et al and Aharony et al \cite{Aoki,Aharony2}. The interesting feature of the periodic case is that in addition to the bosonic classical zero modes, there are also fermionic zero modes, where the fermion field matrices are diagonal and constant in time. We may then compute a 1-loop effective action for both the boson and fermion zero modes by integrating out fluctuations about these. Then integrating over the fermion zero modes yields an effective theory for the bosonic zero modes, which has an attractive potential that drives the theory to strong coupling.  We begin, as above, by writing our field as matrices and expanding as,
\begin{eqnarray}
A_{ab}(\tau) & = & A^a {\delta}_{ab}  + \frac{1}{\sqrt{2 \pi}}  \sum_{m=-\infty}^\infty  \delta A^{(m)}_{ab} e^{ \frac{2 \pi}{R} i m \tau} \nonumber \\
X_{i,ab}(\tau) & = & x_i^a {\delta}_{ab}  + \frac{1}{\sqrt{2 \pi}} \sum_{m=-\infty}^\infty  \delta X^{(m)}_{i,ab}e^{\frac{2 \pi}{R} i m \tau} \nonumber \\
\Psi_{\alpha,ab}(\tau) & = & \xi_{\alpha}^a {\delta}_{ab}  + \frac{1}{\sqrt{2 \pi}} \sum_{m=-\infty}^\infty  \delta \Psi^{(m)}_{\alpha,ab} e^{\frac{2 \pi}{R} i m \tau} 
\end{eqnarray}
where we have included classical fermionic moduli, $\xi^a_{\alpha}$. We take the fluctuations $\delta \Psi^{(0)}_{\alpha}$ to have no diagonal terms and since the gauge group is $SU(N)$, the matrix $\Psi_{\alpha}$ is traceless so the sum $\sum_a \xi_\alpha^a$ vanishes. The dimensionless coupling controlling the integration over the fluctuations is again given by $g_{\rm eff}$ as defined above in equation \eqref{eq:effcoupling}.

As shown in the appendix, when we gauge fix the action and integrate out the fluctuations at 1-loop we find an action for the bosonic and fermionic moduli. Supersymmetry leads to large cancellations between contributions of
the bosonic and fermionic fluctuations, and we find that this action is only non-trivial due to the presence of the fermionic moduli, $\xi^a_{\alpha}$. When we integrate over these, we then obtain an effective action for the bosonic moduli,
\begin{equation}
S_{\rm 1-loop}[A^a,x^a_i] = - \log \sum_{(a_1,a_2,\ldots,a_N) \in P} M^{a_1a_2}M^{a_2a_3} \ldots M^{a_{N-2}a_{N-1}} M^{a_{N-1}a_{N}}
\end{equation}
where $P$ is the set of permutations of $(1,2,\ldots,N)$, and
\begin{equation}
M^{ab} = \sum_{m=-\infty}^{\infty} \frac{1}{\left( (m + \Delta A^{ab})^2+(\Delta x_i^{ab})^2 \right)^3} .
\end{equation}
This potential is again periodic in $\Delta A^{ab}$. It is energetically unfavourable for large separations in $|\Delta x_i^{ab}|$, and hence is attractive. As above the classical bosonic moduli space is lifted, and the attractive potential drives the theory to strong coupling, with a moduli separation $|\Delta x^{ab}| \sim \beta^{3/4}$. Since the number of terms in the sum is $N!$, we expect the 1-loop effective action for the bosonic zero modes to have energy $O(N \log N)$.
Since the path integral measure over these remaining zero modes is $\prod_{a} \left( d A^{a} \prod_i d x^{a}_i \right)$ we see that the integral, giving the partition function, should be convergent in the infra-red. Analogous reasoning to that of Krauth and Staudacher \cite{Krauth} in the context of matrix integrals suggests, using naive power counting, that the distribution of the bosonic eigenvalues should have tails decaying as a power law $1/x^3$. In particular this means that the expectation values of the moments $\mathrm{Tr} X^{2p}$ for any positive integer $p$ do not exist, even though the partition function itself does. 

As mentioned above, for small enough circle size we may effectively ignore all the non-constant modes on the circle and the quantum mechanics should reduce to the 4 supercharge matrix integral. This has been studied beyond the 1-loop approximation both numerically and analytically. In particular Austing \cite{Austing} has proven analytically existence of a twisted version of the theory and full Monte Carlo simulation was performed by Ambjorn et al \cite{Ambjorn} for large $N$. 

At this point we should
comment that analogous calculations for the
16 supercharge theory would yield
similar results for the effective
potential governing the fluctuations of the
bosonic zero modes in both the periodic and thermal cases.

\subsection{Phase structure}

In our supersymmetric quantum mechanics with finite $N$, and hence a finite number of degrees of freedom, one cannot have a sharp phase transition, but only smooth cross-over behaviour.
However, we are interested here in the large $N$ limit of this quantum mechanics. In the infinite $N$ limit it is possible to have sharp phase transitions - ie. the cross-over becomes sharper as $N$ is increased, leading to non-analytic behaviour for $N\rightarrow\infty$. The simplest example of this is the Gross-Witten matrix integral which in its large $N$ limit exhibits a 3rd order quantum phase transition \cite{gross}. Indeed the quenched version of our quantum mechanics theory exhibits a discontinuous confinement-deconfinement transition at large $N$, as discussed in \cite{Aharony1,Aharony2}. Hence it is then an interesting question as
to whether the supersymmetric theory exhibits smooth thermal behaviour or not at large $N$.
As discussed above in section \ref{sec:holography}, holographic arguments suggest that the 16 supercharge quantum mechanics is always in a deconfined phase, ie. the free energy will scale as $O(N^2)$. 
Since the infra-red properties of the 16 and 4 supercharge theories are qualitatively similar one might expect the 4 supercharge theory to always be deconfined. Indeed we will see this is borne out in our results, and we see no evidence of any sharp phase transition.

\subsection{Ultraviolet behaviour of the theory}

We now consider the UV behaviour of theory, and will show that it is finite. This is extremely important for what follows as it allows a naive lattice discretization of the action to recover the full supersymmetry of the theory without fine tuning. Let us consider our action \eqref{eq:action}. Let us firstly gauge fix the theory so that the gauge field $A$ is constant in Euclidean time, $\partial_\tau A = 0$. This yields a trivial Jacobian, $\det \partial_\tau^2$, and allows us then to consider the quantum mechanics partition function $Z[A]$ defined as,
\begin{equation}
Z = \int DA \; Z[A] , \qquad Z[A] = \int DX_i(\tau) D\Psi(\tau) D\bar{\Psi}(\tau) e^{-S_{quad} - S_{int}}
\end{equation}
which for fixed $A$ has no gauge dynamics. 
The unitary matrix $A$ 
determines the Polyakov loop as $e^{i R A}$, and one performs the matrix integral of $Z[A]$ over $A$ to compute the full partition function $Z$. The ultraviolet behaviour of the gauged quantum mechanics is determined by the gauged fixed quantum mechanics derived from $Z[A]$. The final matrix integral over $A$ does not introduce any new high energy behaviour.

 The quadratic and interaction parts of the quantum mechanics derived from $Z[A]$ for fixed $A$ are given by,
\begin{equation}
    S_{quad} =  \mathrm{Tr} \oint^{R} d\tau
      \left\{ \frac{1}{2} (D_{\tau} \tilde{X}_i)^2  + i \tilde{\bar{\Psi}} \bar{\sigma}^\tau D_{\tau} \tilde{\Psi} \right\} 
\end{equation}
and,
\begin{equation} \label{eq:int}
    S_{int} = \mathrm{Tr} \oint^{R} d\tau
      \left\{  - \frac{1}{4} \frac{\lambda}{N}  \left[ \tilde{X}_i, \tilde{X}_j \right]^2 - \left(\frac{\lambda}{N}\right)^{1/2} \tilde{\bar{\Psi}} \bar{\sigma}^i \left[ \tilde{X}_i, \tilde{\Psi} \right] \right\} ,
\end{equation}
where we have rescaled the fields $X_i = (\lambda/N)^{1/2} \tilde{X}_i$ and $\Psi =  (\lambda/N)^{1/2} \tilde{\Psi}$ to obtain canonical kinetic terms. 

\FIGURE[h]{
 \centerline{\includegraphics[width=3.5in,height=1.5in]{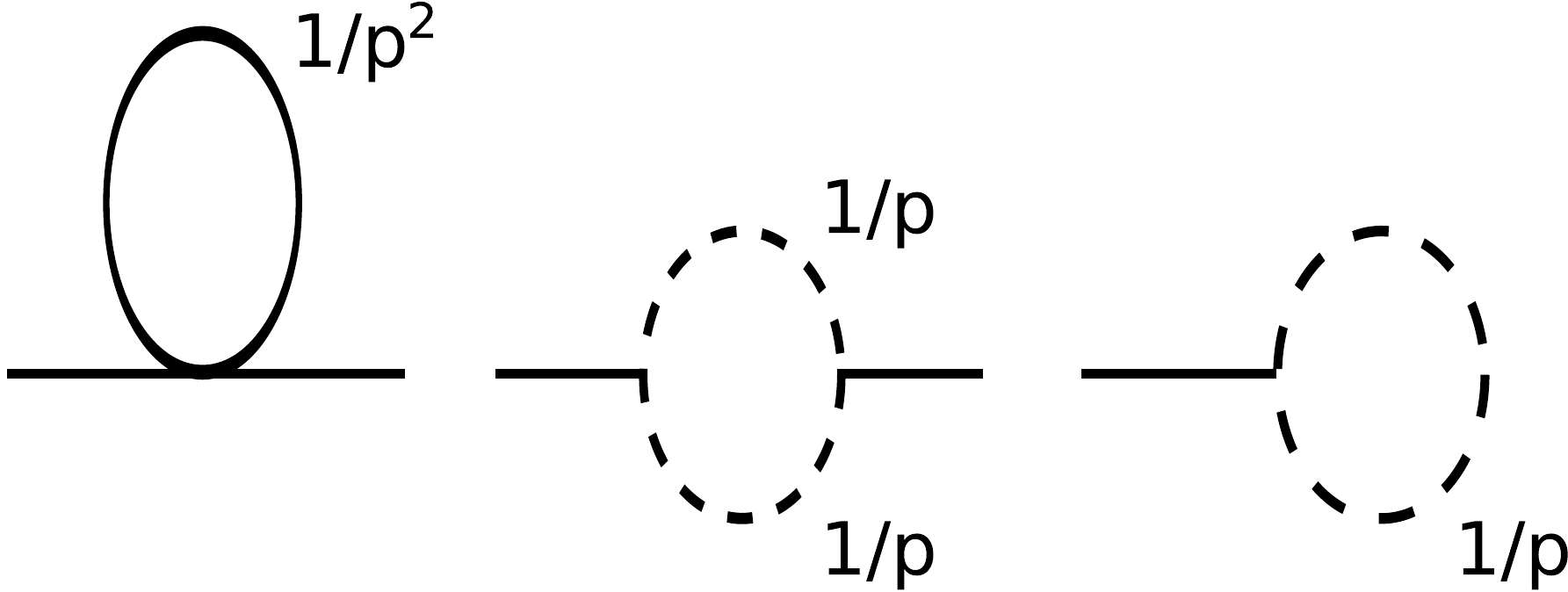}}
 \caption{Relevant diagrams in super Yang-Mills quantum mechanics perturbation theory. The heavy lines are bosonic and the dashed fermionic. The left 2 diagrams are finite mass renormalizations going as $\sim\Lambda^{-1}$. The right diagram is a fermionic tadpole loop which, while naively potentially divergent as $\sim\log{\Lambda}$, is actually also finite going as $\sim\Lambda^{-1}$, the divergence not arising due to form of the Yukawa gauge interaction.
}
   \label{fig:loops}
} 

Consider now performing perturbation theory in the coupling $\lambda$, and introducing a high energy cut-off $\Lambda$. Clearly quantum mechanics is always super-renormalizable. In general this does not preclude divergences, but simply means there can only be a finite number of them. Consider the two boson mass renormalization diagrams shown in figure \ref{fig:loops}. In both cases the loop propagators contribute $\sim 1/p^2$, and the loop integral $\int^{\Lambda} dp / p^2 \sim O(\Lambda^{-1})$ is independent of the cut-off. However for the last diagram in the figure the fermion tadpole loop has only a $\sim 1/p$ contribution from the fermion
propagator, and superficially the loop integral goes as $\int^{\Lambda} dp / p \sim O( \log \Lambda )$ and hence depends on the cut-off. The only divergences in the perturbation theory for this Yang-Mills quantum mechanics come from such fermion tadpole loops.

However, let us consider more carefully the fermion loop in the tadpole. Whilst it superficially diverges we will see the gauge and global symmetries actually render the diagram finite. We introduce $X^{A}_i = \mathrm{Tr} \,T^A X_i$ and $\Psi^{A}_{\alpha} = \mathrm{Tr} \,T^A \Psi_\alpha$ where $T^A$ are the generators of SU(N) in the adjoint representation. The generators obey $[ T^A, T^B ] = i f^{ABC} T^C$ with  antisymmetric structure constants $f^{ABC}$, and are normalized so that $\mathrm{Tr}\; T^A T^B = \delta^{AB}$. Then we may write the fermionic interaction term in \eqref{eq:int} as,
\begin{equation}\label{eq:fermint}
\mathcal{L}_{ferm} \sim  f^{ABC} \bar{\Psi}^A_{\dot{\alpha}} \bar{\sigma}^{i\dot{\alpha}\alpha} \Psi^C_{\alpha} X^B_i
\end{equation}
The fermion propagator in momentum space is given by,
\begin{equation}
\frac{1}{i \bar{\sigma}^{\tau,\dot{\alpha}\alpha} D_\tau^{AB}} = \frac{1}{\delta^{\dot{\alpha}\alpha} (i \delta^{AB} p + f^{ABC} A^C ) } = \frac{ \delta_{AB} \delta_{\dot{\alpha}\alpha}}{i p} + O(\frac{1}{p^2}) 
\end{equation}
where the subleading terms at large momentum involve the gauge matrix $A$. The tadpole loop for an incoming boson $X^C_i$ will therefore contribute an integral proportional to,
\begin{equation}
\int^{\Lambda} dp \left(  f^{ABC} \delta_{AB}  \frac{ \mathrm{Tr} (\bar{\sigma}^i )}{p} + O(\frac{1}{p^2}) \right) = O(1) \int^{\Lambda} dp \; \frac{1}{p^2} \sim O(\Lambda^{-1})
\end{equation}
where now we see the potentially divergent leading term actually vanishes identically due to both the anti-symmetry of the structure constants, $f^{ABC} \delta_{AB} = 0$, and also the fact the Pauli matrices are traceless so $\mathrm{Tr} (\bar{\sigma}^i ) = 0$. Hence we see that actually the fermion tadpole loops, and therefore all diagrams in this theory, are cut-off independent provided that the momentum regulator preserves the gauge and global symmetry and hence the form of the Yukawa interaction term above.
This implies that we can regulate our theory with impunity, and providing we have correctly maintained the degrees of freedom given by the quadratic action above, and the gauge and global symmetry, we should expect to recover the correct theory when we remove the regulator, and in particular all its supersymmetries. This applies to the naive lattice regulator we are interested in here.
  
\FIGURE[h]{
 \centerline{\includegraphics[width=2.5in,height=1.5in]{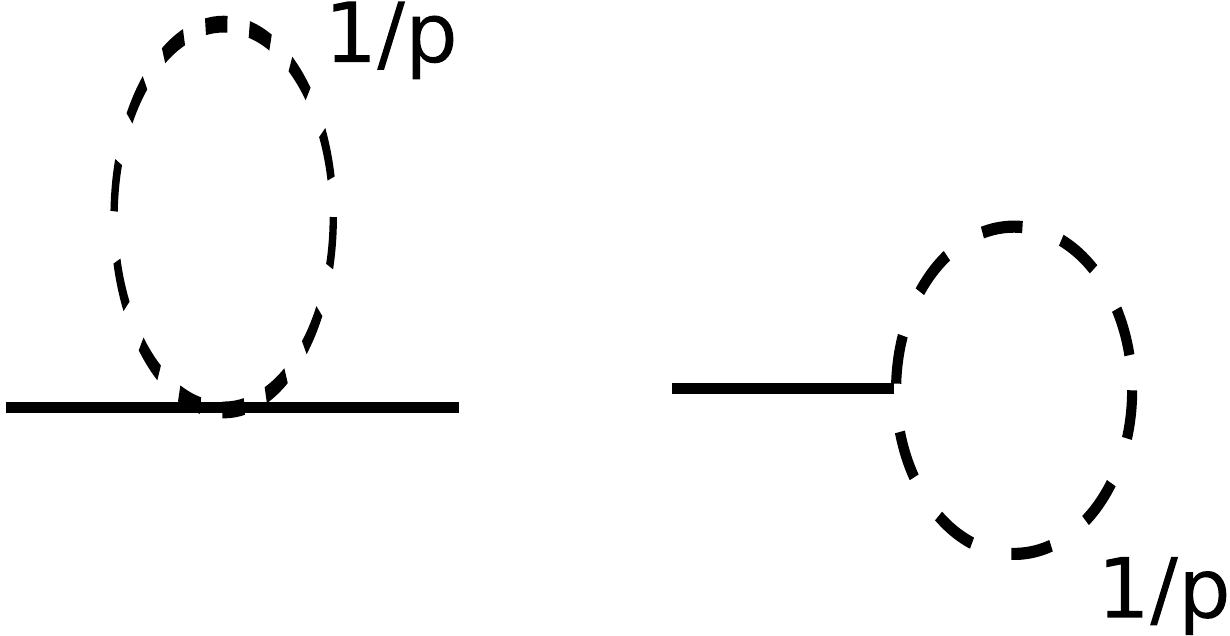}}
 \caption{Relevant diagrams in perturbation theory for Witten's supersymmetric quantum mechanics for a quartic superpotential. The left diagram, a mass renormalization, is due to the quartic superpotential term. The right diagram is a fermion tadpole loop due to the cubic term. Both are divergent going as $\sim\log{\Lambda}$.
 }
   \label{fig:loopsWitten}
}   
  
How generic is this stability against UV radiative corrections? Suppose we wished to study the 2 dimensional version of the holographic correspondence we would then have a similar 2-d Yang-Mills theory to compute with, and in this case now the mass renormalization  diagrams in figure \ref{fig:loops} above are divergent, since $\int^{\Lambda} dp^2 / p^2 \sim O( \log \Lambda )$ as the regulator is removed. Unlike the quantum mechanics case there are no internal symmetries preserved by a naive momentum regulator that mean the superficial divergence is avoided. Hence a naive discretization is not guaranteed to give a continuum limit preserving supersymmetry, and one would have to perform a lattice perturbation theory calculation to compute the potential counter-terms that would have to be added to counteract this, as for example in \cite{Elliott,Suzuki}. Notice that
the absence of such counter terms in Yang-Mills quantum mechanics
is not generic for all quantum mechanics models.
For example, Giedt et al \cite{Giedt} have studied the case of Witten's supersymmetric quantum mechanics,
\begin{equation}
S'=  \int d\tau
      \left\{ \frac{1}{2} \dot{x}^2 + h'(x)^2 +  \bar{\psi} \dot{\psi} + h''(x) \bar{\psi} \psi \right\} 
\end{equation}
Here $x$ is a bosonic field, and $\psi$ fermionic, and if we take a superpotential as $h(x) = m x^2 + \gamma x^3 + \lambda x^4$, then a mass renormalization term is generated as in figure \ref{fig:loopsWitten} from the quartic piece, and the cubic piece gives rise to the fermion tadpole 
shown in the same figure. Naive power counting shows that
both superficially diverge as $\int^{\Lambda} dp / p \sim O( \log \Lambda )$, again with no internal symmetry preserved by a naive momentum cut-off protecting from this divergence. 
Hence, when discretizing this theory naively 
one might expect to generate UV sensitive radiative corrections, and hence break supersymmetry unless the discretization preserves the supersymmetry manifestly. Indeed this was confirmed in \cite{CG, Giedt,Bergner} for the quartic term. So we see that the fact that we are able to discretize our super Yang-Mills quantum mechanics is a result not only of its low dimension, but also its interaction structure, in particular, the gauge and global symmetry which constrains the boson-fermion interaction vertices which are potentially dangerous in quantum mechanics. 

Since we are discussing quantum mechanics we may equivalently use the language of operator ordering in order to discuss the UV behaviour of the theory. Starting from a Hamiltonian operator we may construct a lattice path integral discretized in time, by using the Hamiltonian to propagate states forward by a small time step in the usual manner.
Different operator orderings in the Hamiltonian will give different discrete lattice path integrals. If there is a physical ordering ambiguity the continuum limit of the discrete path integral will yield different physical theories depending on the details of the ordering used. If there is no ordering ambiguity, any ordering will yield the same continuum physics. Hence we see that in general different naive discretizations of an action will correspond to Hamiltonians with different operator orderings, and if there is a physical ambiguity, then different naive lattice actions will give different continuum physics.

We consider again the example of Witten's supersymmetric quantum mechanics. 
Associated with the continuum interaction term $h''(x) \bar{\psi} \psi$ in the action is the interaction Hamiltonian,
\begin{equation}
\hat{H'}_{ferm} =  \hat{b}^{\dagger} \hat{b} \; h''(\hat{x})
\end{equation}
with anticommuting fermionic operators $\hat{b}, \hat{b}^\dagger$ so that $\{ \hat{b}^\dagger, \hat{b} \} = 1$. This is the only term in the Hamiltonian which is sensitive to operator ordering, as the bosonic interaction term involves only $\hat{x}$ and not its momenta.  
Indeed we see that changing the operator ordering of the fermions changes the purely bosonic interaction terms in the Hamiltonian. For the general ordering,
\begin{equation} \label{eq:wittenferm}
\hat{H'}_{ferm} =  \left( (1-\xi) \hat{b}^{\dagger}  \hat{b} - \xi \hat{b} \hat{b}^{\dagger}  \right) h''(\hat{x}) + \xi h''(\hat{x})
\end{equation}
for some $\xi$ we see we have added $\xi h''(\hat{x})$ to the bosonic interaction term. Consider the supercharges for this theory,
\begin{equation}
\hat{Q} = \left( \hat{P} + h'(\hat{x}) \right) \hat{b} , \quad \hat{Q}^{\dagger} = \hat{b}^{\dagger} \left( \hat{P} - h'(\hat{x}) \right)
\end{equation}
and so,
\begin{equation}
\hat{H'} = \left\{ \hat{Q}^{\dagger} , \hat{Q} \right\}  =  \hat{P}^2 + h'(\hat{x})^2 + h''(\hat{x}) \left[ \hat{b}^{\dagger} , \hat{b} \right] .
\end{equation}
Hence we see that supersymmetry requires a particular fermion operator ordering. From our discussion above, a naive discretization of the action will correspond to a Hamiltonian with some operator ordering, and in general it will not be the one required above by supersymmetry.
Thus naive discretizations will generically
lead to continuum theories which are not supersymmetric as the continuum bosonic potential will be incorrect. This argument
tallies with our previous considerations of 
UV behaviour in Witten's quantum mechanics from perturbation theory above.

Consider now our Yang-Mills quantum mechanics. We again introduce $X^{A}_i = \mathrm{Tr} \,T^A X_i$ and $\Psi^{A}_{\alpha} = \mathrm{Tr} \,T^A \Psi_\alpha$. This fermionic interaction in \eqref{eq:fermint} is associated to a Hamiltonian interaction term,
\begin{equation}
\hat{H}_{ferm} =  f^{ABC} \hat{b}^{\dagger A}_{\dot{\alpha}} \bar{\sigma}^{i\dot{\alpha}\alpha} \hat{b}^C_{\alpha} \hat{X}^B_i
\end{equation}
where the fermionic operators $\hat{b}^{\dagger A}_{\dot{\alpha}}, \hat{b}^{A}_{\alpha}$ anticommute as,
\begin{equation}
\quad \{ \hat{b}^{\dagger A}_{\dot{\alpha}}, \hat{b}^{B}_{\alpha} \} = \delta^{AB}  \delta_{\dot{\alpha}\alpha} .
\end{equation}
Again this is the only term in the action which might be sensitive to operator ordering, as the bosonic quadratic interaction term involves only the $X_i$ and not their momenta. However, 
\begin{equation}
\bar{\sigma}^{i,\dot{\alpha}\alpha} f^{ABC} \{ \hat{b}^{\dagger A}_{\dot{\alpha}}, \hat{b}^{C}_{\alpha} \} = f^{ABC}  \delta_{AB} \mathrm{Tr}(\bar{\sigma}^i) = 0
\end{equation}
due to both the antisymmetry of the structure constants and also the traceless property of the Pauli matrices.
We see that changing the operator ordering in this term introduces no additional bosonic terms. 
From our discussion above, since the form of the Hamiltonian is invariant under ordering, naive lattice discretizations will lead to the supersymmetric continuum physics we are interested in. Thus again we see that while the Yukawa couplings in our supersymmetric gauge quantum mechanics superficially could render a naive discretization non-supersymmetric in the continuum, the precise structure given by gauge invariance and global symmetry ensures that this does not happen.

We note that since the form of the 16 supercharge action is the same as that of the 4 supercharge theory, we expect the same argument to ensure that a naive discretization of the 16 supercharge theory will also regain full
supersymmetry in the naive continuum limit.

%
\section{Two lattice actions}
%

We now discuss the two actions we have used to simulate the 4 supercharge quantum mechanics - a naive discretization of the continuum theory
and a manifestly supersymmetric lattice action. The naive action follows the standard rules for discretizing a gauge theory. We will show that the resulting fermion determinant is positive real, which much simplifies simulation. While the lattice action preserves no supersymmetry, we have argued above that since the quantum mechanics is free from UV divergences provided gauge and global symmetries are preserved by a UV regulator, all the supersymmetries will be restored in the continuum limit - this argument also applies to the 16 supercharge theory of interest for future work. The supersymmetric action arises from discretizing
a twisted or cohomological formulation of the continuum Yang-Mills
theory in which nilpotent scalar supercharges can be
constructed as linear combinations of the original
supercharges. In this case we show
that half of the original supersymmetry of the continuum theory can be
preserved in the discrete theory.

\subsection{Naive lattice action}
\label{sec:naive}

In this section we consider a naive discretization of the 4 supercharge quantum mechanics. Firstly we rewrite the action using a dimensionless compact coordinate $\theta$, so $\tau = R \, \theta$, where $\theta$ has unit period $\theta \sim \theta + 1$. 
We create our discrete lattice choosing $M$ lattice sites at position $\theta_n =  a n$, with $a$ the discretization length $1/M$. We use $R$ to make our lattice fields dimensionless, and denote the variable at lattice site $n=0,1,\ldots,M-1$ with a subscript $X_a, \Psi_a$ so that,
\begin{equation}
X_i(R \, \theta_n) = R^{-1}  X_{i,n} , \quad \mathrm{and} \quad \Psi_\alpha(R \theta_n) =  R^{-3/2} \Psi_{\alpha,n}
\end{equation}
Since we are taking compact Euclidean time we must identify $X_0 = X_M, \Psi_0 = \pm \Psi_M$, with the sign for the fermions giving periodic or antiperiodic (thermal) boundary conditions.

An important consideration is to avoid fermion doubling on the lattice. 
This problem is encountered if one replaces
continuum derivatives by {\it symmetric} difference operators.
The resulting fermion operator possesses additional light states
-- ``doublers'' which do not decouple as the lattice
spacing is sent to zero. Theorems guarantee the appearance of
such states in any local, translationally invariant and chirally
symmetric theory. One simple way to remove
these unwanted states is to add a so-called
Wilson mass term $a r \triangle$ to the lattice
action where $a$ is the lattice spacing, $\triangle$ is the Laplacian, and $r$ is some non-zero constant. This term lifts the doublers to have mass $O(r/a)$, leaving the physical modes light. For
our naive simulations we have employed an
$r=1$ Wilson term, which in quantum mechanics,
yields the simple
Euler discretization prescription $\dot{\Psi}_a = (\Psi_{a+1} - \Psi_a)/a$. 

We introduce unitary adjoint link fields, $U_a$, to implement the gauge invariance. Gauge freedom allows us to fix the link variables on a spanning tree which in our one dimensional case means there is one unitary degree of freedom for the whole lattice and we may choose how to represent it. For the purpose of simulating the theory a convenient choice is a gauge where all the links are equal $U = U_i$. We may think of $U$ as representing the holonomy of the gauge connection about the time circle, so that the Polyakov loop is simply given as,
\begin{equation}
e^{i \oint A d\tau} = U^M .
\end{equation}
Note that the Jacobian introduced by the gauge fixing to set all the links equal is trivial. 
Now we may naively discretize the action as,
\begin{eqnarray}\label{eq:actiondis}
     S &=&  \frac{N}{\beta^3}  \sum_{a=0}^{M-1} \mathrm{Tr}\left[ \frac{1}{2 a}  \left( X_{i,a} - U X_{i,a-1} U^{\dagger} \right)^2 - \frac{a}{4} [ X_{i,a}, X_{j,a} ]^2 + \delta^{\dot{\alpha}\alpha} \bar{\Psi}_{\dot{\alpha},a} \left( \Psi_{\alpha,a} - U \Psi_{\alpha,a-1} U^{\dagger} \right)   \right. \nonumber \\
 && \qquad \qquad \qquad  \qquad
 \left. \vphantom{\frac{1}{2 a}}  - a \bar{\sigma}^{i,\dot{\alpha}\alpha} \bar{\Psi}_{\alpha,a} [ \Psi_{\alpha,a} , X_{i,a} ] \right]
\end{eqnarray}
where again we note the Euler differencing of the fermion kinetic term is free of doublers.

Let us now show that integrating out the quadratic fermion term gives rise to a positive real determinant.
As above we again introduce $X^{A}_i = \mathrm{Tr} \,T^A X_i$ and $\Psi^{A}_{\alpha} = \mathrm{Tr} \,T^A \Psi_\alpha$. Taking the generators $T^A$ to be Hermitian, they obey $[ T^A, T^B ] = i f^{ABC} T^C$ with the $SU(N)$ structure constants $f^{ABC}$ being real. Since $X_i$ are Hermitian the components $X^A_i$ are real. We then define the fermion operator from the above action so that the fermionic part is given by, $\bar{\Psi}^{A}_{\dot{\alpha},a} \mathcal{O}^{\dot{\alpha}\alpha, AB}_{ab} \Psi^{B}_{\alpha,b}$, and we have,
\begin{equation}
\mathcal{O}^{\dot{\alpha}\alpha, AB}_{ab} = \delta^{\dot{\alpha}\alpha} \left( \delta^{AB} \delta_{ab} - \delta_{a,b+1} \mathrm{Tr} T^A U T^B U^\dagger \right) - i \, a \,  f^{ABC} \bar{\sigma}^{i,\dot{\alpha}\alpha} \delta_{ab}   X^{C}_{i,a}  .
\end{equation}
Following a related argument in \cite{Ambjorn} we see that,
\begin{equation}
{\sigma}^2 \mathcal{O}^{AB}_{ab} {\sigma}^2 = \left( \mathcal{O}^{AB}_{ab} \right)^\star
\end{equation}
and hence any eigenvector $v_{\alpha A a}$ of $\mathcal{O}$ having eigenvalue $\lambda$ will be paired with another eigenvector $(\bar{\sigma}^{2,\dot{\alpha}\alpha} v_{\alpha A a})^\star$ with eigenvalue $\lambda^\star$. Since the determinant of $\mathcal{O}$ is the product of its eigenvalues, we see it is real and positive. This is very nice as it ensures that in the naive discretization we can, in principle, exponentiate the fermion determinant, and use Monte-Carlo methods to straightforwardly simulate the resulting action. Notice that we used properties of the Weyl representation, and this no longer holds for the 16 supercharge theory. Indeed, already in the 16 supercharge matrix integral, the Pfaffian obtained from integrating out the fermions is not positive \cite{Nicolai}. Thus in the 16 supercharge case, using a naive action one would likely have to take the absolute value of the Pfaffian and then use `reweighting' to simulate the phase. How effective this would be would then depend on how important this phase is in the physical regime of interest.

\subsection{Manifestly supersymmetric lattice action}

A lattice action which possesses an exact supersymmetry may be derived by
dimensional reduction of the supersymmetric 
lattice action for two-dimensional {\it twisted}
four supercharge Yang-Mills theory
described in \cite{Catterall_2d,Catterall_sims}. 
The resulting action may be written as
\beq
S=Q\sum_x{\rm Tr}
\left(\chi^\dagger_1(B_1+2D^+_1\phi^3)+\psi^\dagger_1D^+_1\phib+
\kappa^\dagger[\phi^3,\phib]+
\frac{1}{4}\eta^\dagger[\phi,\phib]+{\rm h.c}\right)\eeq
where the scalar supercharge arising from the twisting process
has the following action on the fields
\begin{eqnarray}
Q U_1&=&\psi_1\\\nonumber
Q\psi_1&=&-D^+_1\phi\\\nonumber
Q\phi^3&=&\kappa\\\nonumber
Q\kappa&=&-[\phi^3,\phi]\\\nonumber
Q\chi_1&=&B_1\\\nonumber
Q B_1&=&[\phi,\chi_1]\\\nonumber
Q\phib&=&\eta\\\nonumber
Q\eta&=&[\phi,\phib]\\\nonumber
Q\phi&=&0
\end{eqnarray}
It is straightforward to verify that $Q^2$ acts like
a gauge transformation on a generic field $f$ which is taken to
lie in the adjoint of the gauge group and of the form
$f=\sum_{i=1}^{N^2-1}f^aT^a$ where in this section we employ {\it antihermitian}
generators $T^a$. The $Q$-exact
structure of the action then guarantees the discrete action will
be supersymmetric. To avoid
fermion doubling a covariant forward difference
operator is utilized whose action on lattice scalar fields is given by
\beq
D^+_1\phi(x)=U_1(x)\phi(x+\one)-\phi(x)U_1(x)\eeq
where $U_1(x)$ is the usual Wilson gauge link.\footnote{actually
a complexified version of it since the construction requires
the lattice fields be taken complex}
Notice that several of these fields, such as $U_1(x)$, 
carry a vector index. In the continuum this is redundant for a one dimensional
theory but in this lattice construction the index
plays an important role, indicating
that such fields live on the links of the lattice and transform
under gauge transformations as
\beq
f_1(x)\to G(x)f_1(x)G^\dagger(x+\one)\eeq
where $G(x)=e^\phi(x)$.
The corresponding transformation for scalar or site fields is
\beq
f(x)\to G(x)f(x)G^\dagger(x)\eeq
These transformation reduce to the usual continuum ones in the naive
continuum limit. Notice also that the definition of the forward difference
operator when acting on a site field automatically produces a vector
or link field with the correct lattice gauge transformation properties.

Carrying out the $Q$-variation and integrating out the auxiliary field
$B_1$ we find the following lattice action
\beq
S=\sum_x {\rm Tr}\left(-\sum_{i=1}^3X^i D^-D^+X^i-\sum_{i>j}^3[X^i,X^j]^2-
\lambda^{i\dagger}_1D^+_1\rho^i-\rho^{i\dagger}D^-_1\lambda^i_1+S_{\rm
YK}\right)\eeq
where we have written $\phi=X^1+iX^2$, $\phib=X^1-iX^2$, $\phi^3=X^3$
and the covariant backward derivative $D^-_1$, which is adjoint to 
$D^+_1$, acts on link fields in the following way
\beq
D^-_1\psi_1=\psi_1(x)U^\dagger(x)-U^\dagger(x)\psi_1(x-\one)\eeq 
We have also relabeled the fermions according to
\begin{eqnarray}
\lambda_1^1&=&\psi_1\\\nonumber
\lambda_1^2&=&\chi_1\\\nonumber
\rho^1&=&\frac{\eta}{2}\\\nonumber
\rho^2&=&\kappa\\\nonumber
\end{eqnarray}
Notice that the bosonic action is real positive definite on account of
the antihermitian basis for the fields.
The Yukawa interactions take the form
\begin{eqnarray}
S_{\rm YK}&=&\rho^{2\dagger}[X^1,\rho^2]-i\rho^{2\dagger}[X^2,\rho^2]-
             \rho^{1\dagger}[X^1,\rho^1]-i\rho^{1\dagger}[X^2,\rho^1]\\\nonumber
	  &+&\lambda^{1\dagger}_1[X^1,\lambda^1_1]-i\lambda^{1\dagger}_1[X^2,\lambda^1_1]-
	     \lambda^{2\dagger}_1[X^1,\lambda^2_1]-i\lambda^{2\dagger}_1[X^2,\lambda^2_1]\\\nonumber
	  &+&\left\{\lambda^{1\dagger}_1[X^3,\lambda^{2}]
	  -\rho^{2\dagger}[X^3,\rho^1]+{\rm h.c}\right\}
\end{eqnarray}
While the bosonic action arising in this discretization is 
rather similar to the naive action described in the previous
section, differing mainly in the
definition of the
lattice derivative, 
it is clear that the fermionic action is quite different -- the fermions
can be assembled into a four component object which in the continuum is just the
usual Majorana fermion of four dimensional ${\cal N}=1$ super Yang-Mills. Thus the result of
integrating out the fermions in this twisted formulation is to produce a Pfaffian rather than
the determinant encountered with the naive discretization. Of course, in the continuum, one
can find a change of variables which allows the Pfaffian to be rewritten as a simple
determinant but this is no longer true in the lattice construction. Indeed the complex
two component spinor encountered in the naive formulation takes the
form
\beq
\left(\begin{array}{c}
\chi_1+i\eta/2\\
\psi_1+i\kappa
\end{array}\right)
\eeq
Since the different component fields carry different gauge transformation properties 
this spinor does not transform simply under lattice gauge transformations and cannot be
used to construct a gauge invariant lattice action.

A more difficult question relates to the complex nature of the fields in the supersymmetric lattice
construction. To target the correct continuum theory requires choosing the path integral
along a contour such that the imaginary parts of all fields are zero. The question then
arises as to whether the supersymmetric Ward identities are satisfied along this
path. The latter are $Q$-exact and hence by standard arguments can be evaluated
exactly in the semiclassical limit in which the fields are expanded about the
classical solution
\cite{Catterall_top}. It is not hard to show that in this limit the real and imaginary
parts of the fields decouple and the computation can be consistently truncated to the
real line. These theoretical arguments are strengthened by the results of Monte
Carlo calculations which support the existence of an exact $Q$-supersymmetry at the
quantum level \cite{Catterall_sims,Catterall_restore}.

Finally notice that this lattice action actually possesses a 
global symmetry of the
form
\begin{eqnarray}
\rho^1&\to&\rho^2\\\nonumber
\lambda^1_1&\to&\lambda^2_1\\\nonumber
X^1&\to&-X^1\\\nonumber
X^2&\to&X^2\\\nonumber
X^3&\to&X^3
\end{eqnarray}
This symmetry combined with the original $Q$-supersymmetry leads to the lattice
action possessing a {\it second} exact supersymmetry. The existence of this
second supersymmetry is consistent with the orbifold construction described
in \cite{Kaplan1}.

%
\section{Results}
%

In this section we discuss the results of lattice simulation of our two implementations. We begin by considering the quenched theory, move onto the periodic theory and end the section by discussing the thermal theory.
The results for the quenched theory agree precisely for the naive and supersymmetric actions since they differ only in their treatment of the fermions
(notice though that no gauge fixing is done in the supersymmetric
implementation). 
As we discuss, the naive and supersymmetric implementations both give compatible results for the periodic theory. In practice the naive implementation
is computationally easier since the corresponding fermion
operator is half that of the supersymmetric formulation and, as we have shown, is real and positive.
Hence the bulk of the periodic results were generated
with the naive action. 
In the case of the thermal system the situation
is more interesting - for coarse lattice spacings and a small number
of colors we observed that both codes suffered from strong lattice artifacts
most visible in the distribution of scalar eigenvalues which developed
long tails out to large eigenvalue -- a situation
quite different from the complementary runs with
periodic boundary conditions. We observe that this stems from very large
fluctuations of the scalars in the classical moduli space -- for
the periodic system the corresponding
bosonic zero mode is strongly suppressed by
its fermionic superpartner but in the thermal case the would be
fermionic zero mode
is lifted by the antiperiodic boundary conditions and
is less effective at finite $\beta$ at suppressing these
zero mode effects.
This problem appears to
be worse for the supersymmetric action and hence we have again derived
the bulk of our thermal results from the naive action runs. In the latter
case these effects appeared negligible for $N\ge 5$ and $M\ge 5$.

We have examined the following expectation values. We normalize the expectation value of the fermionic action as,
\begin{equation}
S_F = \frac{1}{2 M (N^2-1)} < S_{\mathrm{fermionic}} >
\end{equation}
and a simple scaling argument gives the Schwinger-Dyson equation $S_F =1$ which we then use to check the equilibration of our runs. 
We also compute the bosonic action as,
\begin{equation}
S_B = \frac{1}{N^2} \left( < S_{\mathrm{bosonic}} > -  S_{zero} \right)
\end{equation}
where we have subtracted the extensive zero point energy contribution, which is given explicitly as $\frac{3}{2} M (N^2-1)$ for both actions as can easily be seen in the weakly coupled high temperature limit. For the supersymmetric lattice action we have $S_B = 0$
for all $\beta$ since it is related to an
index \cite{Catterall_2d}. For the naive periodic
theory we expect $S_B$ to
approach zero as the number of lattice points $M\to\infty$. We have normalized the bosonic action by $N^2$ to ensure that $S_B$ should tend to a constant at large $N$ in a deconfined phase. 

This form of the bosonic action has an additional interpretation in the
thermal and quenched theories as
yielding a measurement of 
the dimensionless mean energy of the system, $<E>$. This dimensionless energy is given by 
the usual relation $<E> =-\frac{\partial\ln Z}{\partial \beta}$ so that,
\begin{equation}
S_B = - \frac{\beta}{3} <E> .
\end{equation}
Thus the vanishing of (the subtracted) $S_B$ in the periodic theory corresponds to the
usual requirement that the supersymmetric theory have vanishing vacuum
energy at zero temperature.

We are also interested in the behaviour of the Polyakov loop variable, and we compute the expectation value of its modulus,
\begin{equation}
P = \frac{1}{N} < | \mathrm{Tr}{e^{i \oint A d\tau}} | >
\end{equation}
and also its corresponding susceptibility
\begin{equation}
dP = \frac{1}{N}< | \left(\mathrm{Tr}{e^{i \oint A d\tau}}\right)^2| >-
                < | \mathrm{Tr}{e^{i \oint A d\tau}} | >^2 .
\end{equation}
We note the inclusion of $1/N$ in these definitions 
to ensure that at large $N$ in a deconfined phase, 
$P$, $dP$ should tend to a constant. 

We have also computed the distribution of eigenvalues of the
scalar fields, made dimensionless as $R X_i(\tau)$, and averaged over the lattice. We have seen no evidence of broken $SO(3)$ symmetry \footnote{We note that at finite $N$ such a spontaneous symmetry breaking is not possible, although it might be in the large $N$ limit.} and so assume the distribution for $i=1,2,3$ is the same, and denote it as $x(\mu)$. We observe this appears to have a well-defined large $N$-limit with a width which is controlled by 
$\beta$. From our earlier discussion of the 1-loop eigenvalue potentials we deduced that well separated dimensionless eigenvalues were attracted together until they entered a strongly coupled infra-red regime when their separation goes as $\sim \beta^{1/4}$. This therefore is the expected scale characterizing the parametric dependence of the width of the distribution $x(\mu)$, and indeed this is borne out by our results later.
For $SU(N)$  the distribution has $N$ peaks localized close to the origin, and 
as we have discussed we expect the distribution to fall off most slowly for the periodic theory, going as $x(\mu) \sim | \mu |^{-3}$ for large $\mu$. This tail appears to be universal for the periodic boundary conditions depending on $\beta$ but not $N$ \cite{Catterall_restore}. The tail means that the standard deviation is in principle ill-defined as an observable, at least for periodic boundary conditions. Therefore in order that we may characterize the quenched, periodic and thermal theory scalar distributions we define the width by the observable,
\begin{equation}
W = \beta^{-1/4} \int d\mu \; |\mu| \; x(\mu)
\end{equation}
which is well defined even for the expected tail behaviour of the periodic theory.

%
\subsection{The quenched approximation}
%

We begin by discussing the quenched theory, where the fermions are simply ignored. The lattice simulation of this bosonic theory is then a very tractable problem, and one can easily work at large $N$ and establish the 't Hooft scaling. Such gauged quantum mechanics with adjoint scalar fields was originally studied in \cite{Aharony1, Aharony2} where evidence was given that the theory undergoes a large $N$ confinement/deconfinement transition. We now review this behaviour as it indicates where the interesting dynamics in the unquenched theory is likely to occur, and how we might see this in the available observables.

The first observation we may make from the quenched theory is that the continuum limit is very easy to obtain. In the appendix \ref{app:continuum} we show the bosonic action and Polyakov loop for different numbers of lattice points $M = 5,8,12$. We see that the results we obtain for all these quantities on a lattice with $M=5$ are very close 
(within one percent) to those with $M=8,12$. Indeed we find this to be true 
for all the observables we measure in the quenched, periodic and thermal theories (with the exception of the bosonic
action for the naive periodic theory which we discuss later). 
Hence, in the main text, results are calculated using $M=5$ lattice points unless otherwise stated. Certainly, in the case of
the unquenched theory, the statistical errors in most of
our measurements are larger than
the systematic discretization error
incurred by using the small lattice for the range of $\beta$ studied.

\FIGURE[h]{
\centerline{ \includegraphics[width=3.5in,height=2.5in]{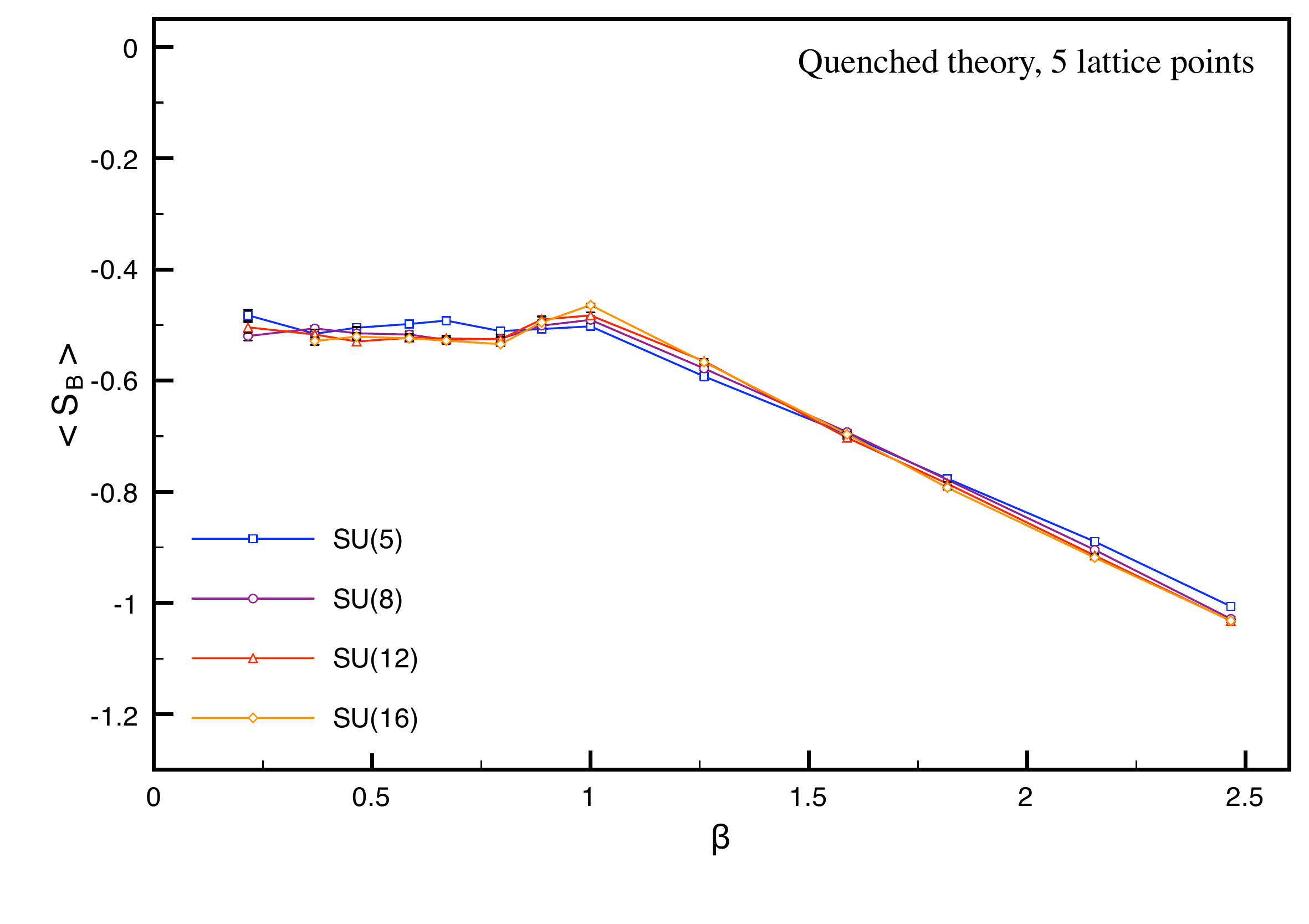}\includegraphics[width=3.5in,height=2.5in]{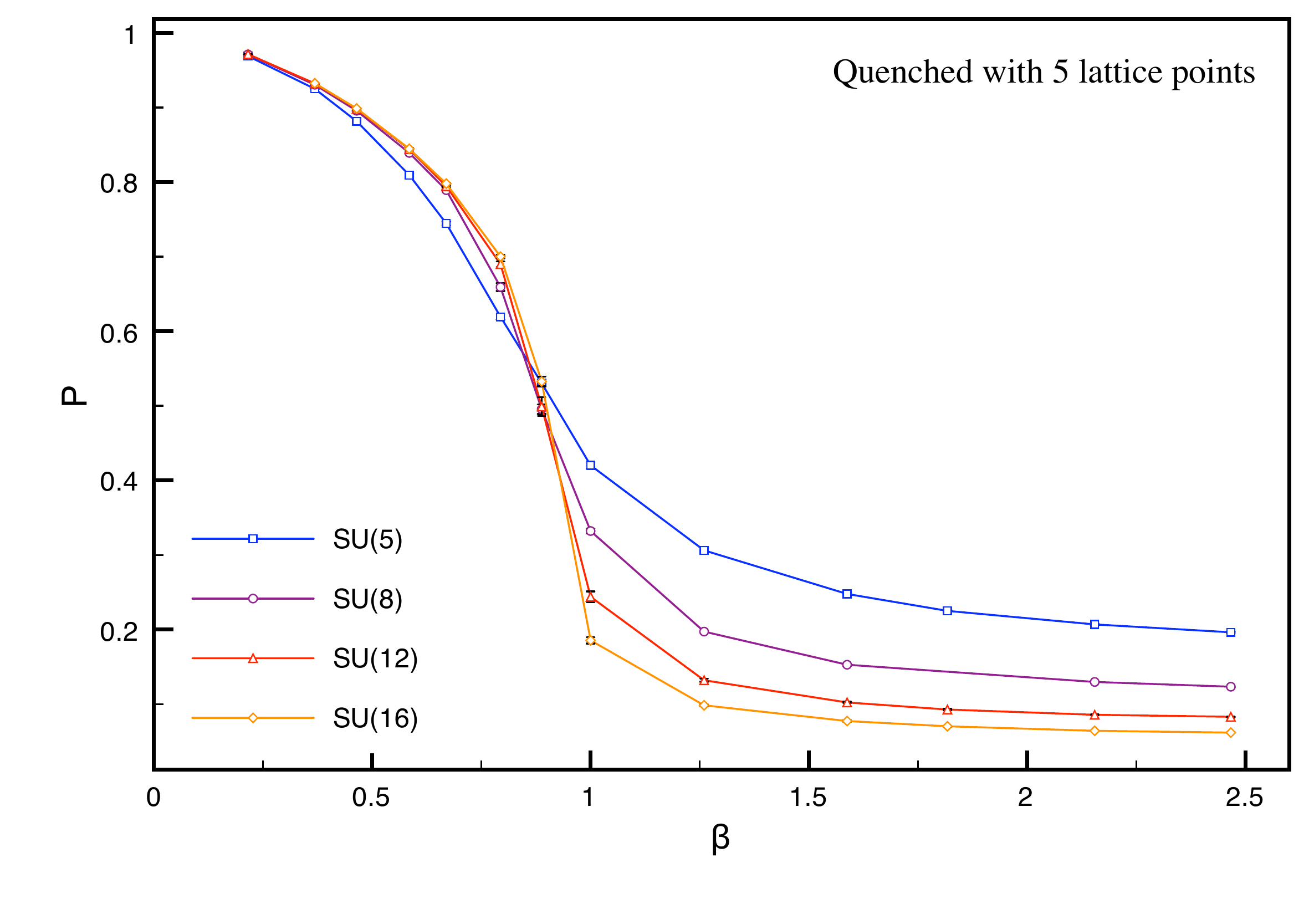}  }
\centerline{ \includegraphics[width=3.5in,height=2.5in]{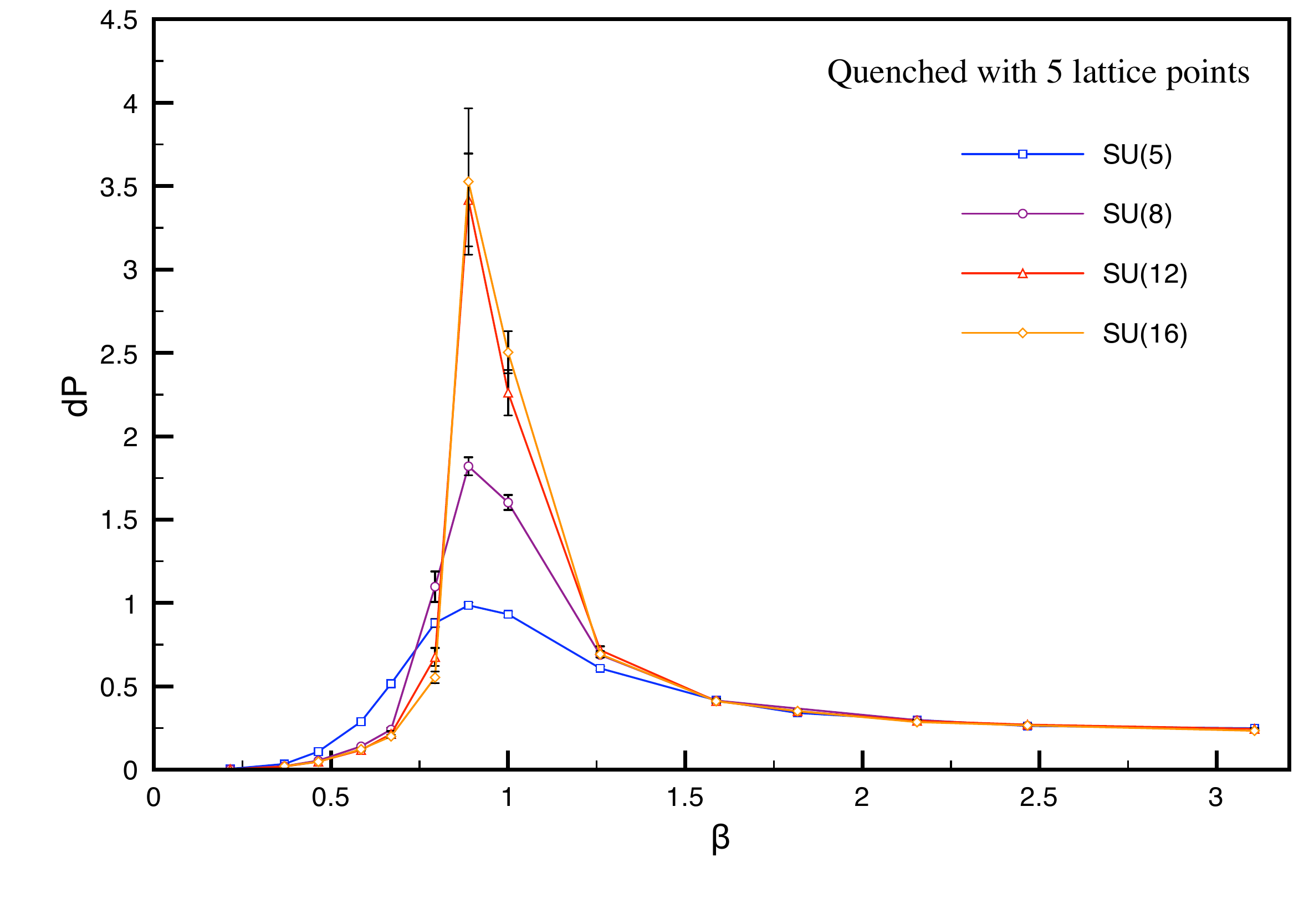}  \includegraphics[width=3.5in,height=2.5in]{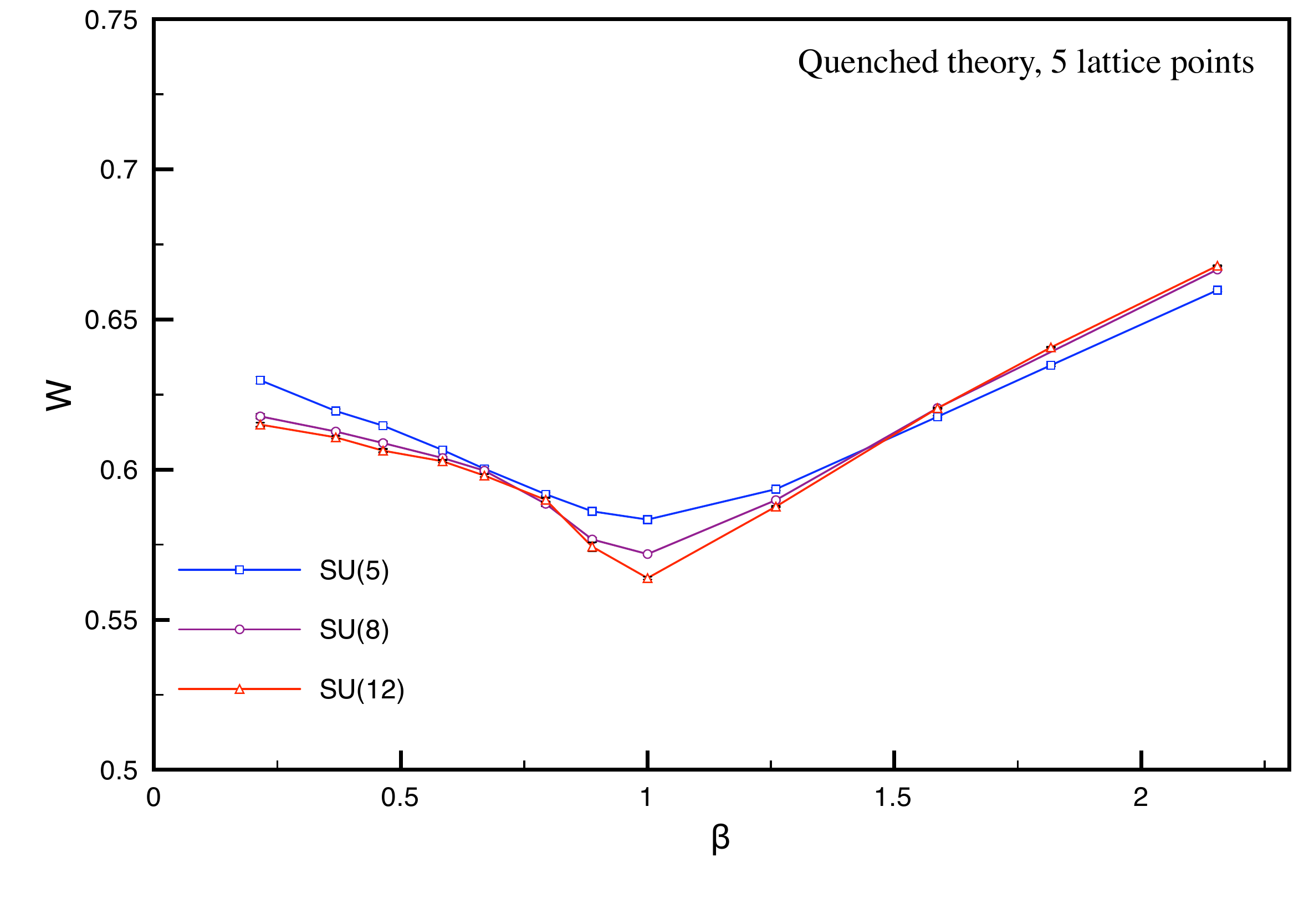}  }
      \caption{Plots of action $S_B$, Polyakov loop $P$, its susceptibility $dP$ and the scalar width $W$ for the quenched theory for varying $N$, and 5 lattice points.}
   \label{fig:bos1}
}

In  figure~\ref{fig:bos1} we plot the expectation value of the bosonic
action $S_B$, the expectation value of the modulus of the Polyakov loop, $P$, and its corresponding susceptibility $dP$
for various numbers of colors $N$ up to $16$. 
Indeed it would be easy to compute at larger $N$ but this is not our objective here. As observed earlier in \cite{Aharony1, Aharony2}
we see that there appears to 
be a sharp, probably first order large $N$ phase transition in behaviour at $\beta = \beta_c \simeq 0.85$, signaled by the bosonic action $S_B$, the Polyakov loop $P$, its 
susceptibility $dP$ and also the scalar eigenvalue distribution width $W$. In particular, the Polyakov loop variable $P$ remains finite for $\beta < \beta_c$, but appears to be consistent with
zero at large $N$ for $\beta > \beta_c$ (see \cite{Aharony1, Aharony2} for data with $N$ up to $30$). Furthermore, the mean energy $<E>$
appears to suffer a discontinuity
around the critical point.

Notice also that the linear regime observed in $S_B$ at large $\beta$
is consistent with a non-zero vacuum energy  -- the latter being given
by the slope of $S_B = -\beta/3<E>$ with respect to $\beta$
at large $\beta$. This is, of course, consistent
with what one would
expect for a non-supersymmetric theory. Conversely the observed
$\beta$-independence of $S_B$ at high temperature (small $\beta$)
is consistent with
the expected classical thermal behavior $<E>\sim N^2T$ for a theory possessing $N^2$
deconfined gluons.

%
\subsection{Full supersymmetric theory: Periodic}
%

We now focus on the theory with dynamical fermions having periodic boundary conditions - ie. the theory in finite volume and zero temperature. Our task is to compare the results obtained from our two independent implementations - the naive and supersymmetric - and show they are consistent, and produce the correct supersymmetric continuum physics.

\FIGURE[h]{
\centerline{ \includegraphics[width=3in,height=2in]{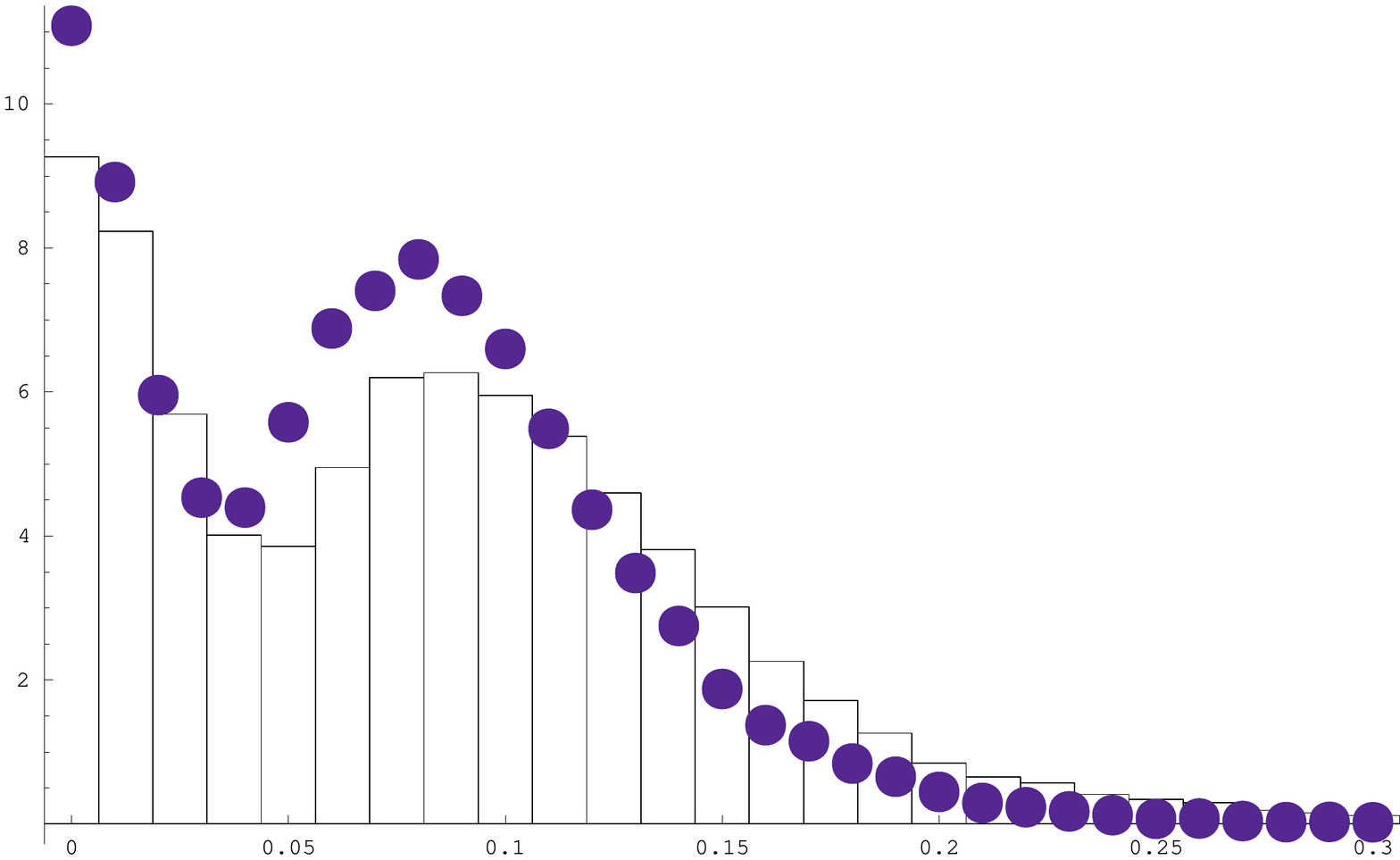}    \includegraphics[width=3in,height=2in]{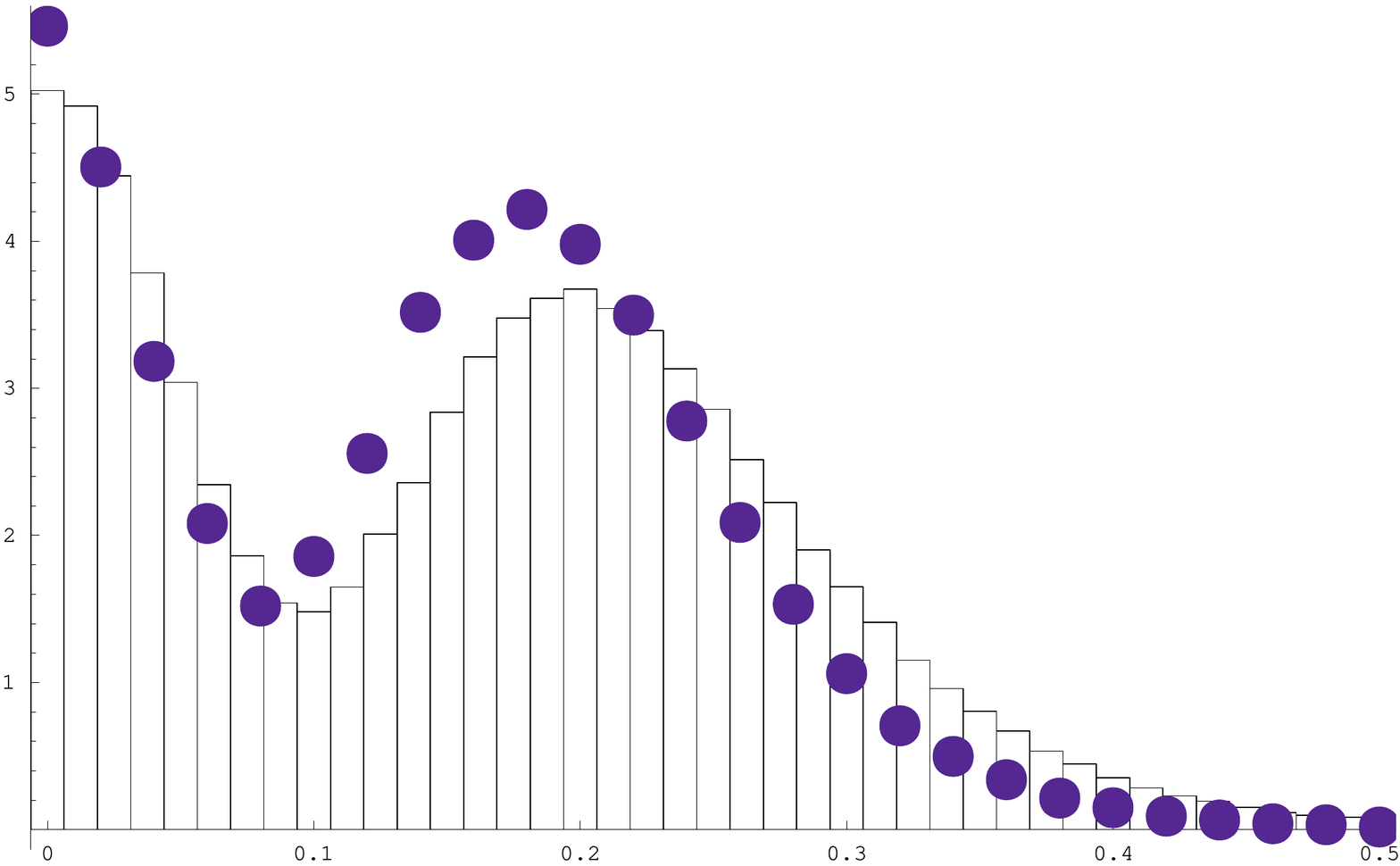}  }
\centerline{ \includegraphics[width=3.5in,height=2.5in]{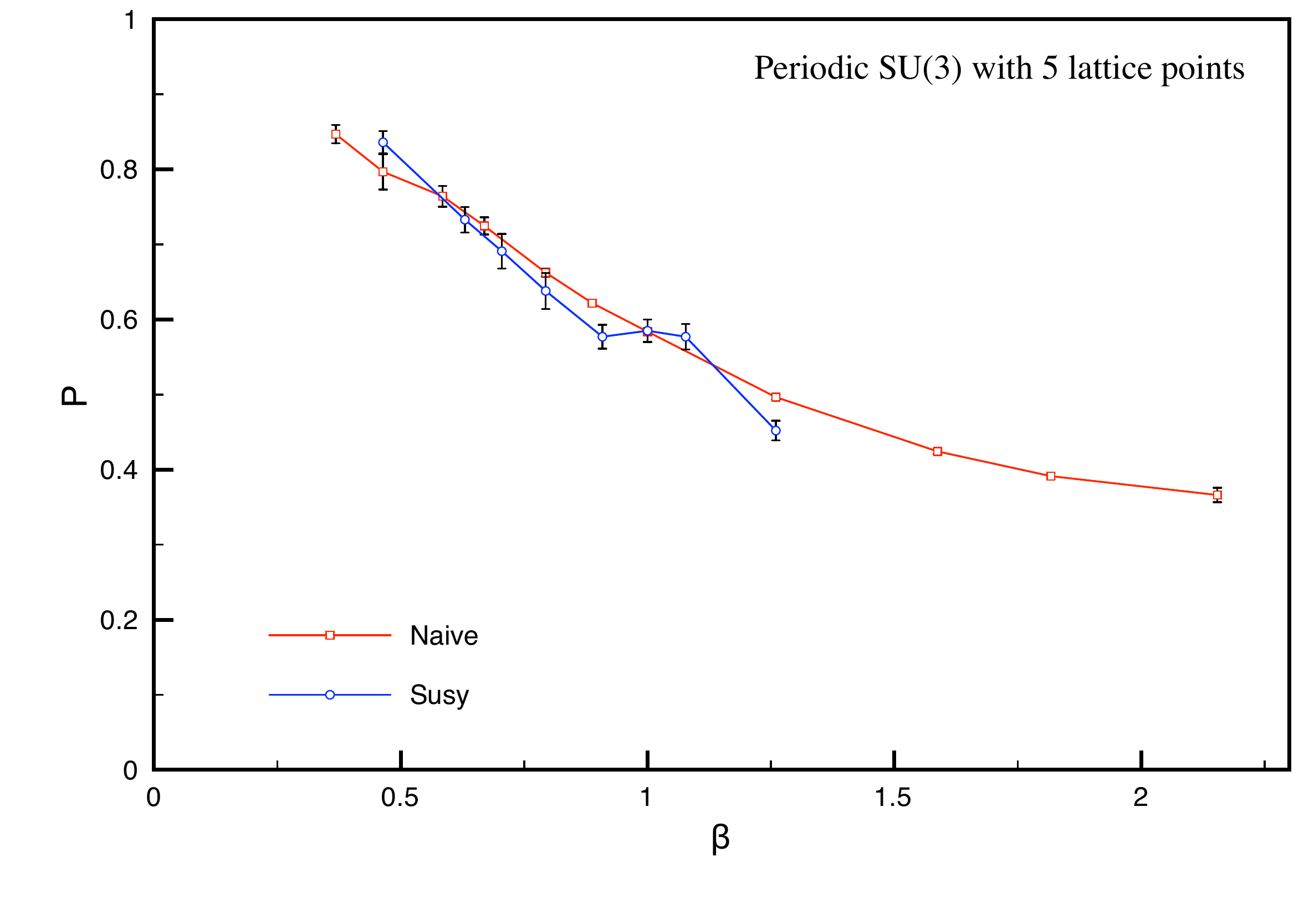}    \includegraphics[width=3.5in,height=2.5in]{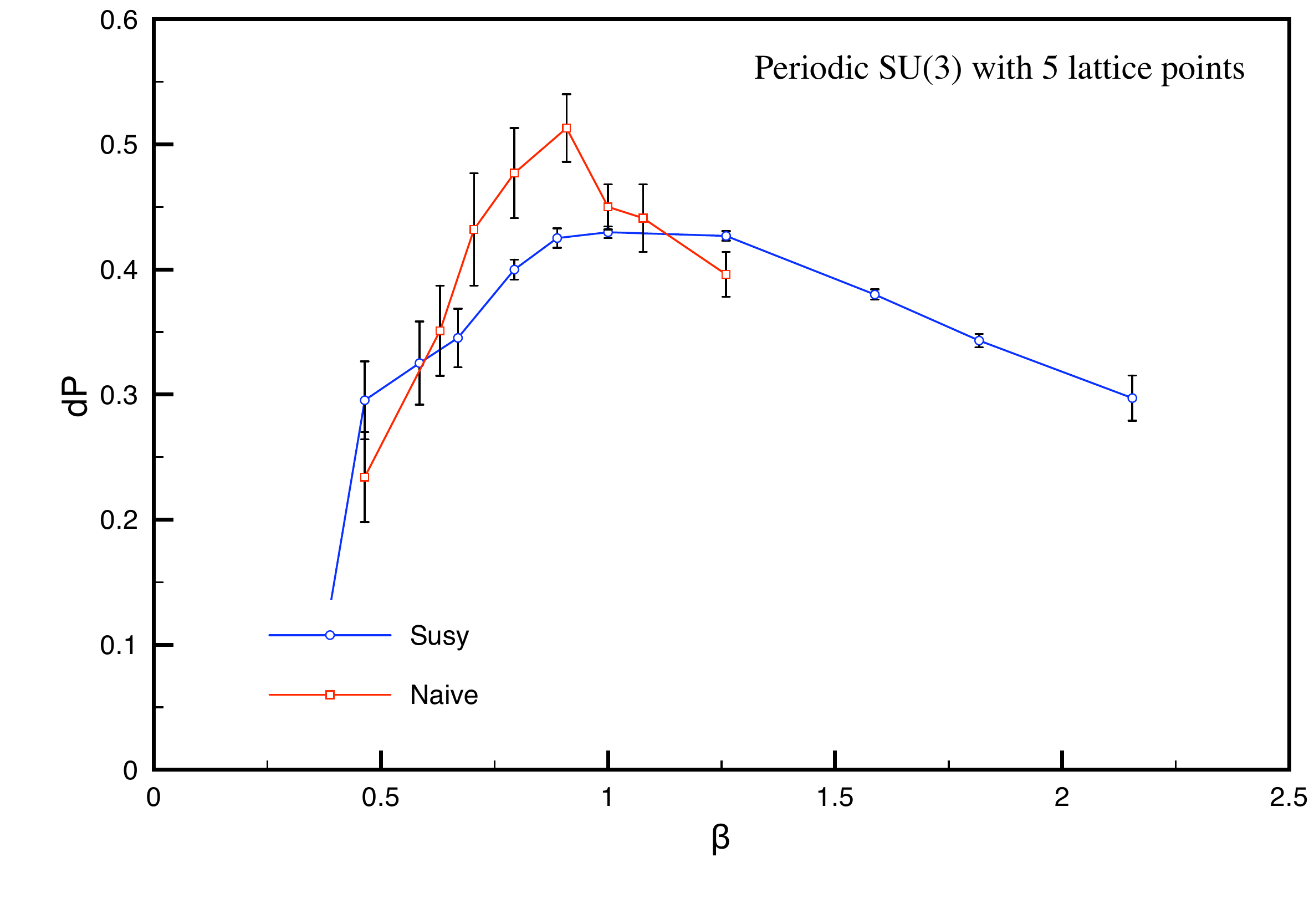}  }
      \caption{Top: Plot comparing scalar field distributions for the 2 implementations with periodic fermions (bars for naive, dots for supersymmetric; left $\beta^3 = 0.1$, right $\beta^3 = 2.0$). Bottom: the Polyakov loop $P$ and its susceptibility
      $dP$ for $N=3$ and $5$ lattice points.}
   \label{fig:periodic1}
}

Figure \ref{fig:periodic1} shows a comparison of the Polyakov loop observable $P$ and its susceptibility $dP$ for the two lattice
implementations in the case of $N=3$ and $M=5$ lattice points. We see the results are in in rather close agreement, in fact within statistical error, for $P$, $dP$. The close agreement is a good check on both implementations
as it indicates that
lattice spacing effects appear to be small and supports the
claim that the $M=5$ runs with naive action yield results which are
already decent approximations to the continuum.
The figure also shows a detailed comparison
of the scalar eigenvalue distributions for the naive and
supersymmetric actions and again reasonable agreement is
seen.  
We study the continuum behaviour of the naive periodic 
theory in appendix \ref{app:continuum}, and confirm that the observables $P$
and $dP$ are close to their continuum values by simulating at $M=8$ and $12$ lattice points. Given constraints on resources we have not been able to compute for larger $N$ in the supersymmetric implementation and therefore cannot check the agreement there.

\FIGURE[h]{
 \centerline{ \includegraphics[width=3.5in,height=2.5in]{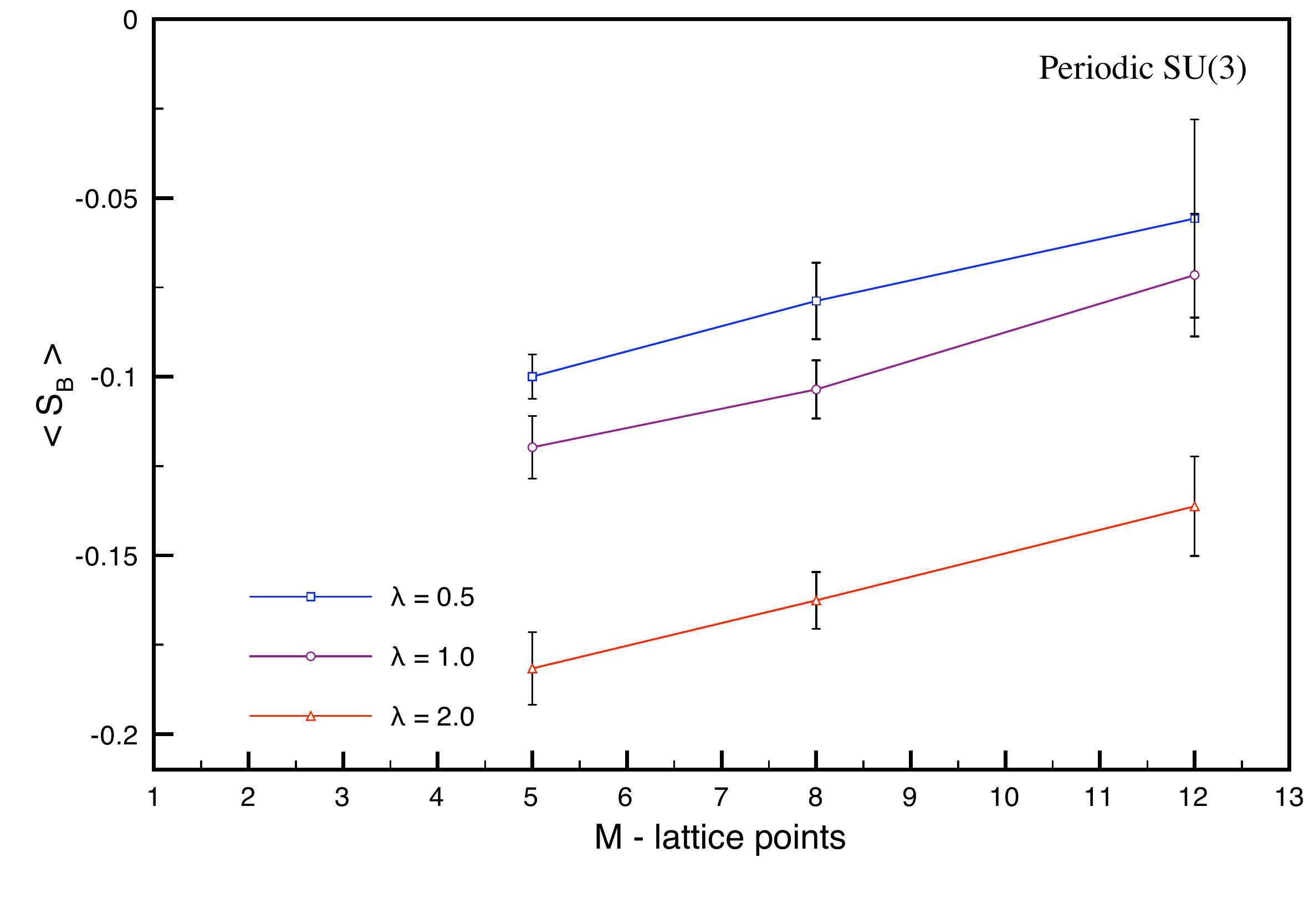} \includegraphics[width=3.5in,height=2.5in]{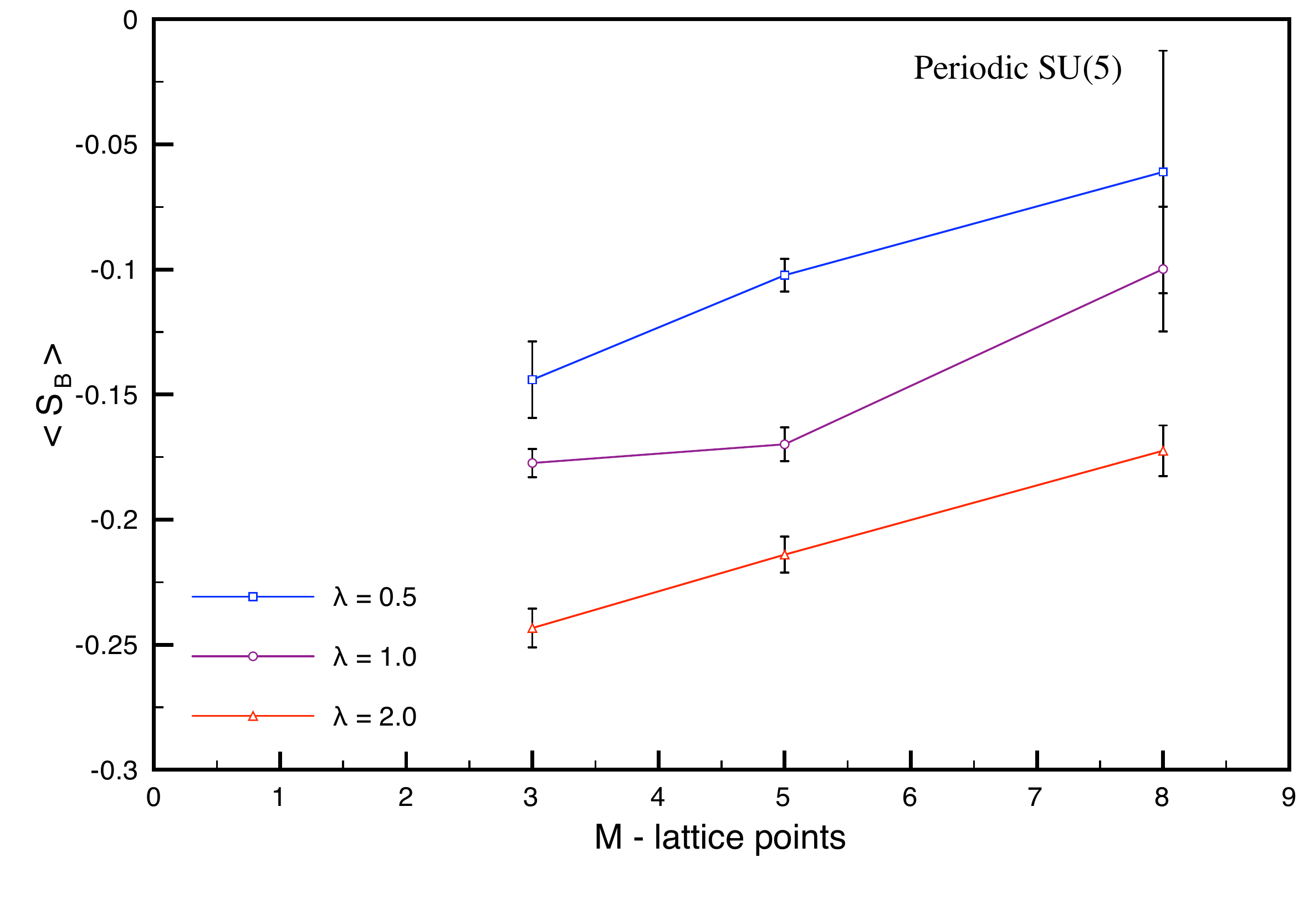}}
        \caption{Plots of the bosonic action $S_B$ for the naive discretization with periodic fermion boundary condition for $N=3$ (left) and $N=5$ (right) for increasing lattice size $M=5,8,12$ for 3 values of $\beta^3 = 0.5,1,2$.}
   \label{fig:periodic2}
}

We show the action $S_B$ for the supersymmetric implementation in figure \ref{fig:periodicX} for $N=5$ and $5$ lattice points and see that it is consistent with zero, as we expect for the
vacuum energy of a system with 
exact supersymmetry.

\FIGURE[h]{
 \centerline{ \includegraphics[width=3.5in,height=2.5in]{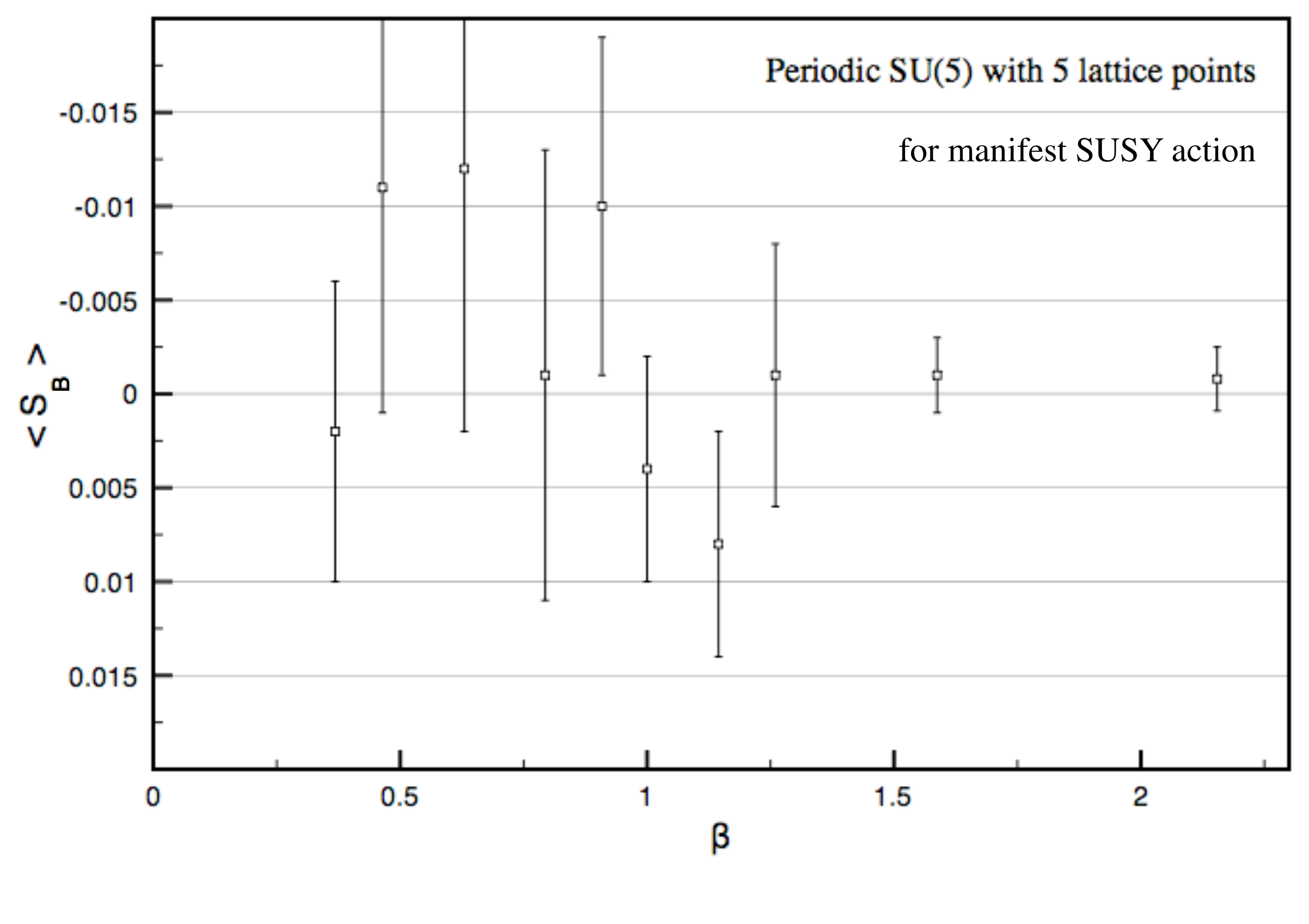}}
        \caption{Plot of the bosonic action $S_B$ for the supersymmetric discretization with periodic fermion boundary condition for $N=5$ with $M = 5$ lattice points. We see it is consistent with zero as we expect.}
   \label{fig:periodicX}
}

Figure \ref{fig:periodic2} shows the bosonic action $S_B$  
for the naive theory at three
representative values of $\beta^3$ with $N=3,5$ as a function of increasing numbers of
lattice points. While by construction this vanishes in the supersymmetric theory, in the naive theory we only expect it to vanish in the limit
$M\to\infty$. We see, both for $N=3$ and $5$ and
fixed $\beta^3$, that
the numerical results
are indeed consistent with $S_B$ decreasing to zero as
$M$ increases. 
The time taken to compute larger lattice sizes with reasonable statistical errors has prohibited extending these plots to higher numbers of lattice points. While the approach to the continuum value of the action is rather slow, appearing to go as $\sim 1/M^p$ with $p\simeq \frac14-\frac12$, even for the modest number of lattice points $M=5$  its actual value is already rather small compared to the quenched or thermal theory (eg. see figures \ref{fig:bos1} and \ref{fig:thermal1}) for these values of  $\beta$.

\FIGURE[h]{
 \centerline{ \includegraphics[width=3.5in,height=2.5in]{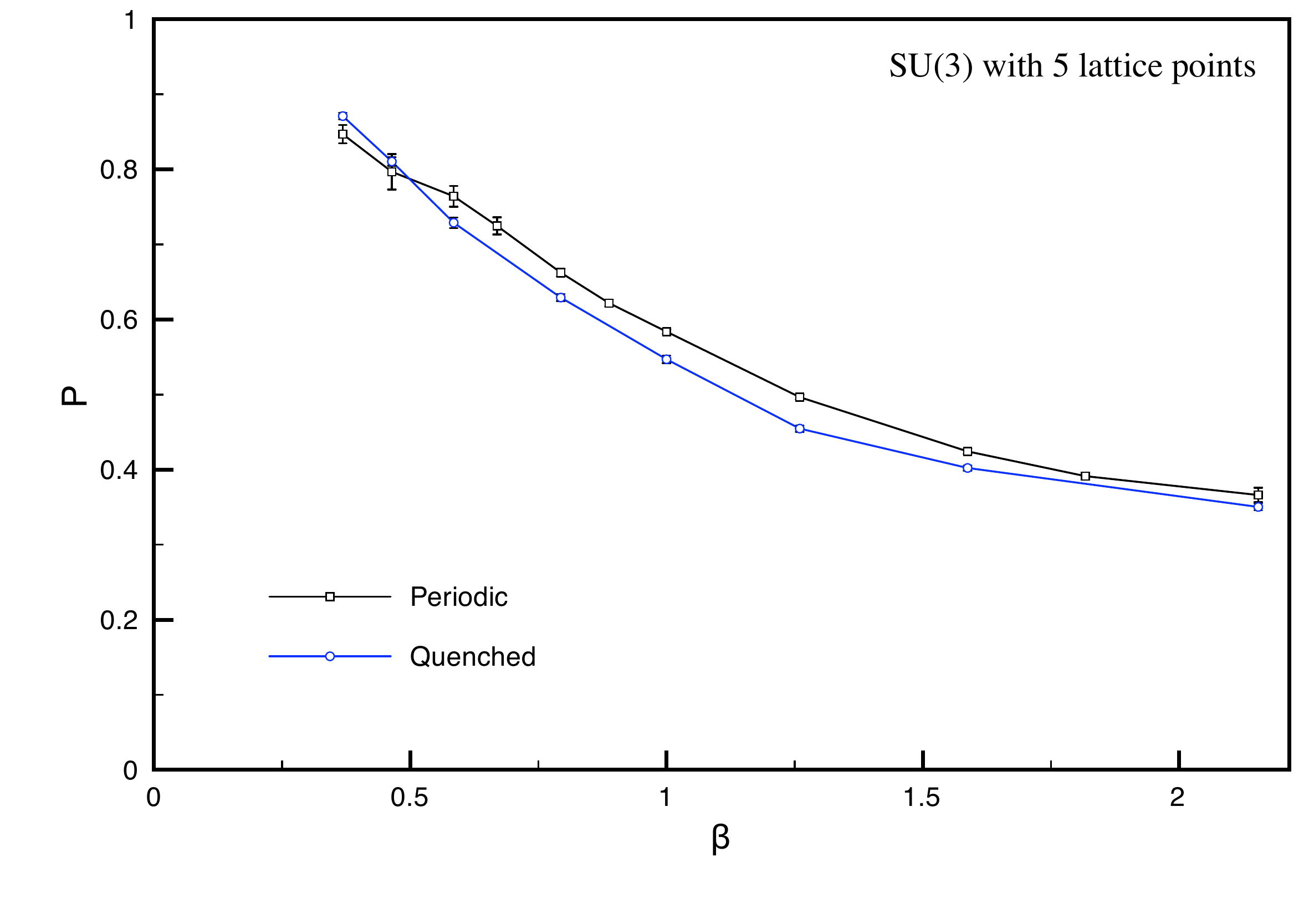} \includegraphics[width=3.5in,height=2.5in]{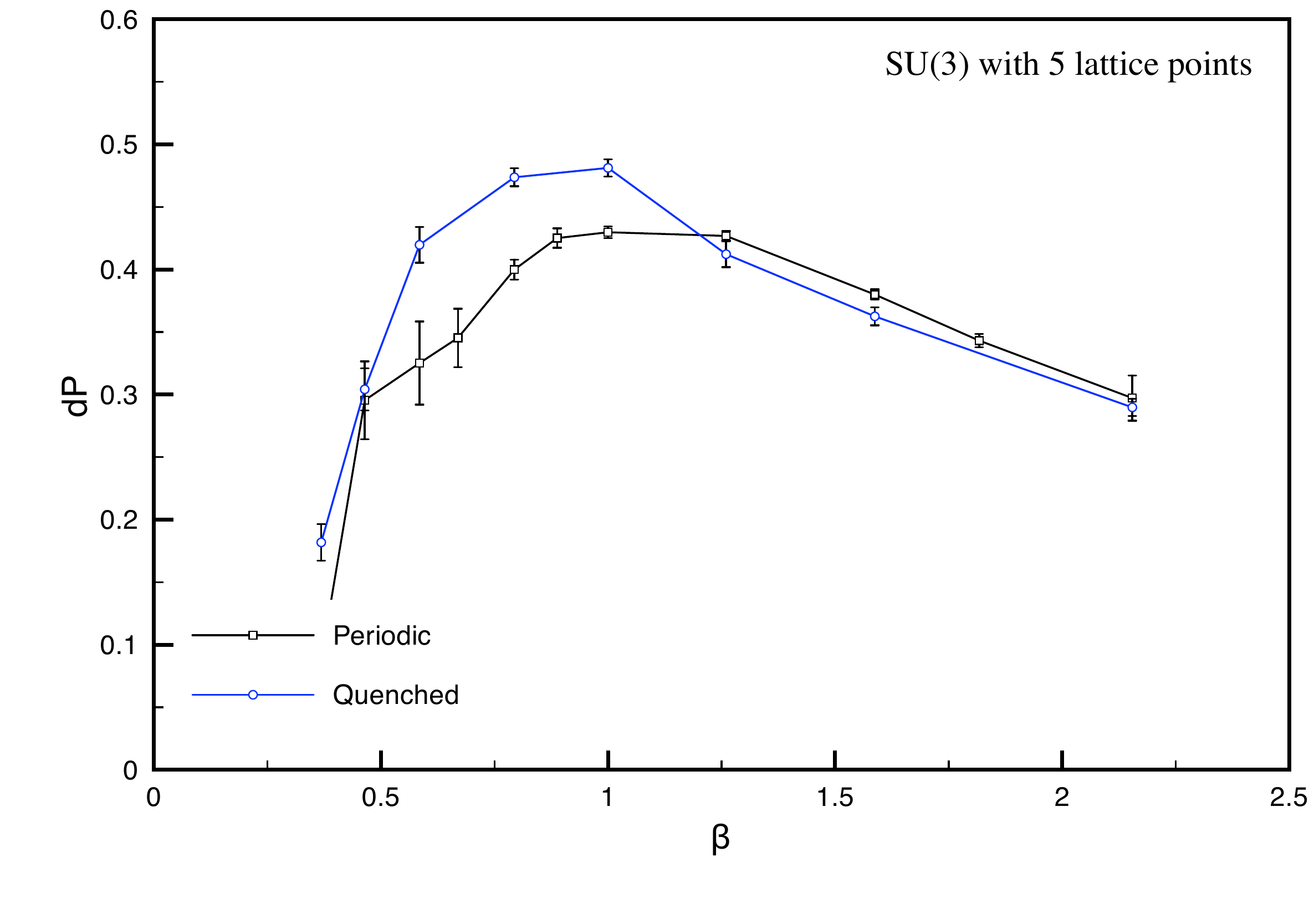}}
 \centerline{  \includegraphics[width=3.5in,height=2.5in]{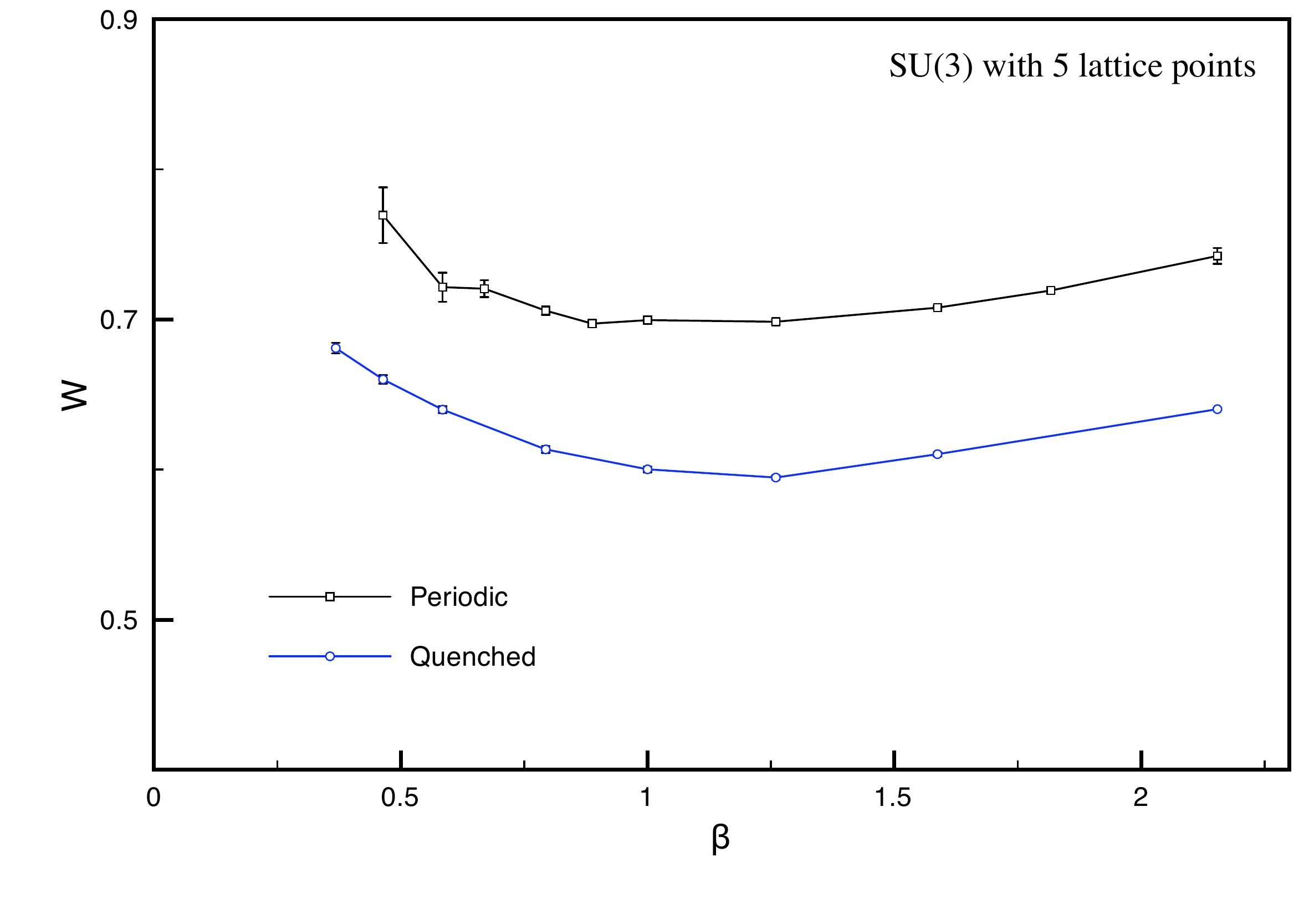}}
        \caption{Plots of the Polaykov loop $P$, its susceptibility $dP$ and the scalar width $W$ for the periodic theory for $N=3$ with 5 lattice points, also compared to the quenched theory.}
   \label{fig:periodic3}
}

In figure \ref{fig:periodic3} we plot the Polyakov loop $P$, its susceptibility $dP$ and the scalar eigenvalue distribution width $W$ for $N=3$ compared with those of the quenched theory. Interestingly, while the action and scalar distribution width are different to the quenched theory, the Polyakov loop behaves rather similarly. In figure \ref{fig:thermal1} we also see the same is true for $N=5$. We have currently not simulated the periodic theory for $N>5$ and therefore cannot confirm this correspondence with the periodic theory occurs in the large $N$ limit.
If it does, it leads to the interesting conclusion that while
the bosonic action $S_B$ is a constant in the periodic continuum theory 
independent of volume and coupling, the behaviour of the Polyakov loop may nevertheless not be smooth in the large $N$ limit. We note that the scalar eigenvalue width is broader for the periodic theory than for the quenched. This is consistent with the 1-loop calculations of the potentials on the classical moduli space where we expect the periodic theory will have power law tails in the eigenvalue distributions, whereas the quenched will not.

Let us summarize these results. We have compared the supersymmetric and naive lattice actions for dynamical fermions with periodic boundary conditions. The results are consistent, and both actions yield the expected behaviour with the supersymmetric version giving a vanishing bosonic action and the naive version giving a continuum limit consistent with vanishing bosonic action. Hence our earlier analytic claims that a naive discretization of the action will give the correct supersymmetric continuum physics appears to be borne out in practice.

%
\subsection{Full supersymmetric theory: Finite temperature results}
%

We now turn to the most interesting part of our
results: the supersymmetric theory at finite temperature. If we
were studying the $16$ supercharge theory
these lattice simulations would be dual to a 
computation of the thermal properties of $N$ D0-branes, and should
reproduce the thermodynamics of black holes described earlier at low
temperature. In the case of the $4$ supercharge model studied in this
paper no such correspondence exists. Nevertheless, we might expect
on the basis of the one loop calculations described earlier that this
model lies in a similar universality class and so understanding how to
extract continuum results from this model should stand us in good stead
for a future simulation of the thermal $16$ supercharge theory.

We have observed that the thermal theory exhibits large lattice artifacts
for small values of $N$ and the number of lattice points $M$. These lead to an apparent instability in the scalar eigenvalues. The problems seem most acute with the supersymmetric action and so we
have concentrated on using the naive implementation for the bulk of our thermal runs. 
In this case to avoid these strong artifacts we require
$N\ge 5$, $M\ge 5$ for the range of $\beta$ we are studying, $\beta^3 \le 10.0$, and then a good lattice continuum limit is seen. We discuss these effects in more detail in the appendix \ref{app:continuum}, and it is an interesting direction for future research to better understand how to more stably implement the supersymmetric lattice action at finite temperature.

\FIGURE[h]{
 \centerline{ \includegraphics[width=3.5in,height=2.5in]{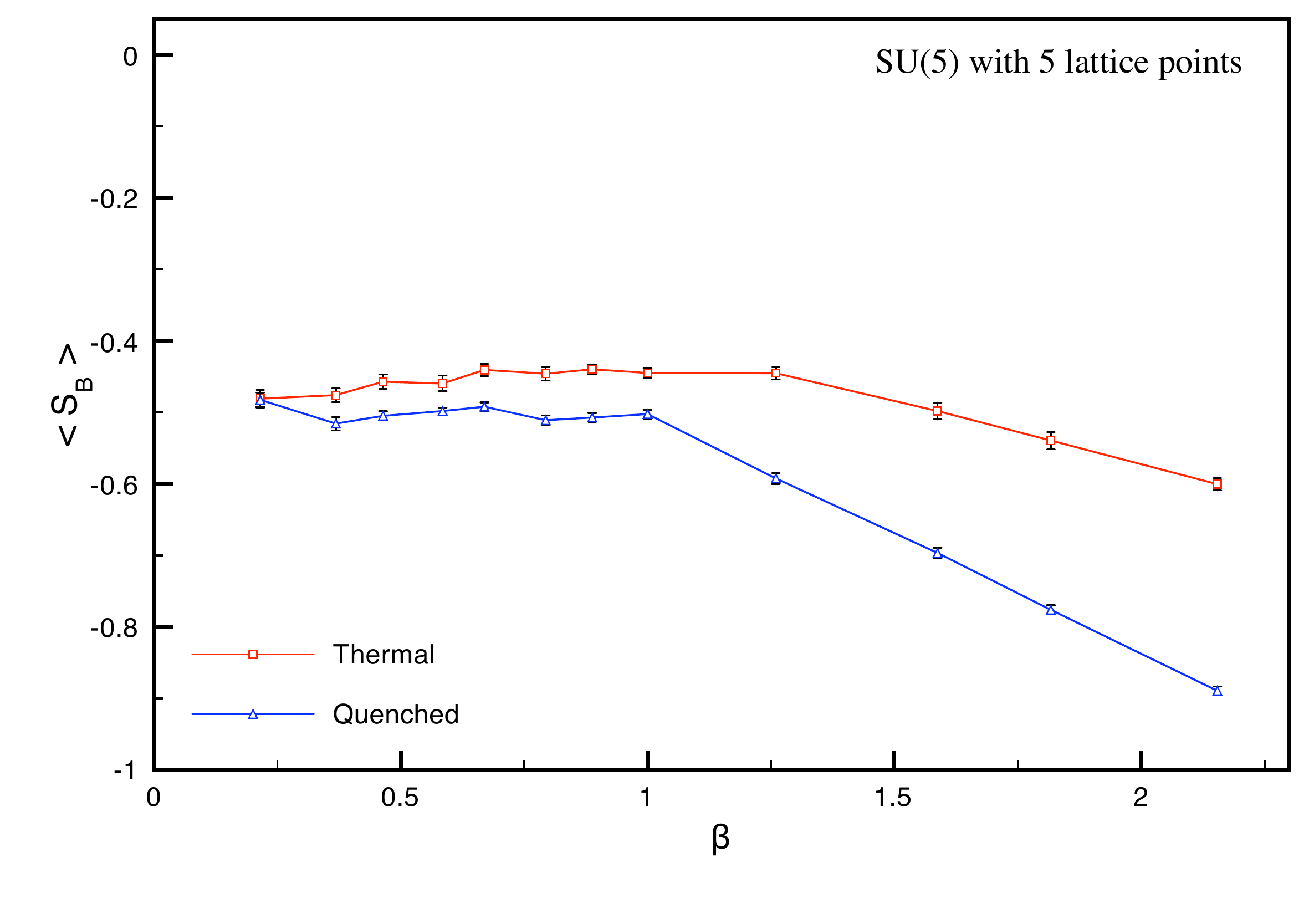}\includegraphics[width=3.5in,height=2.5in]{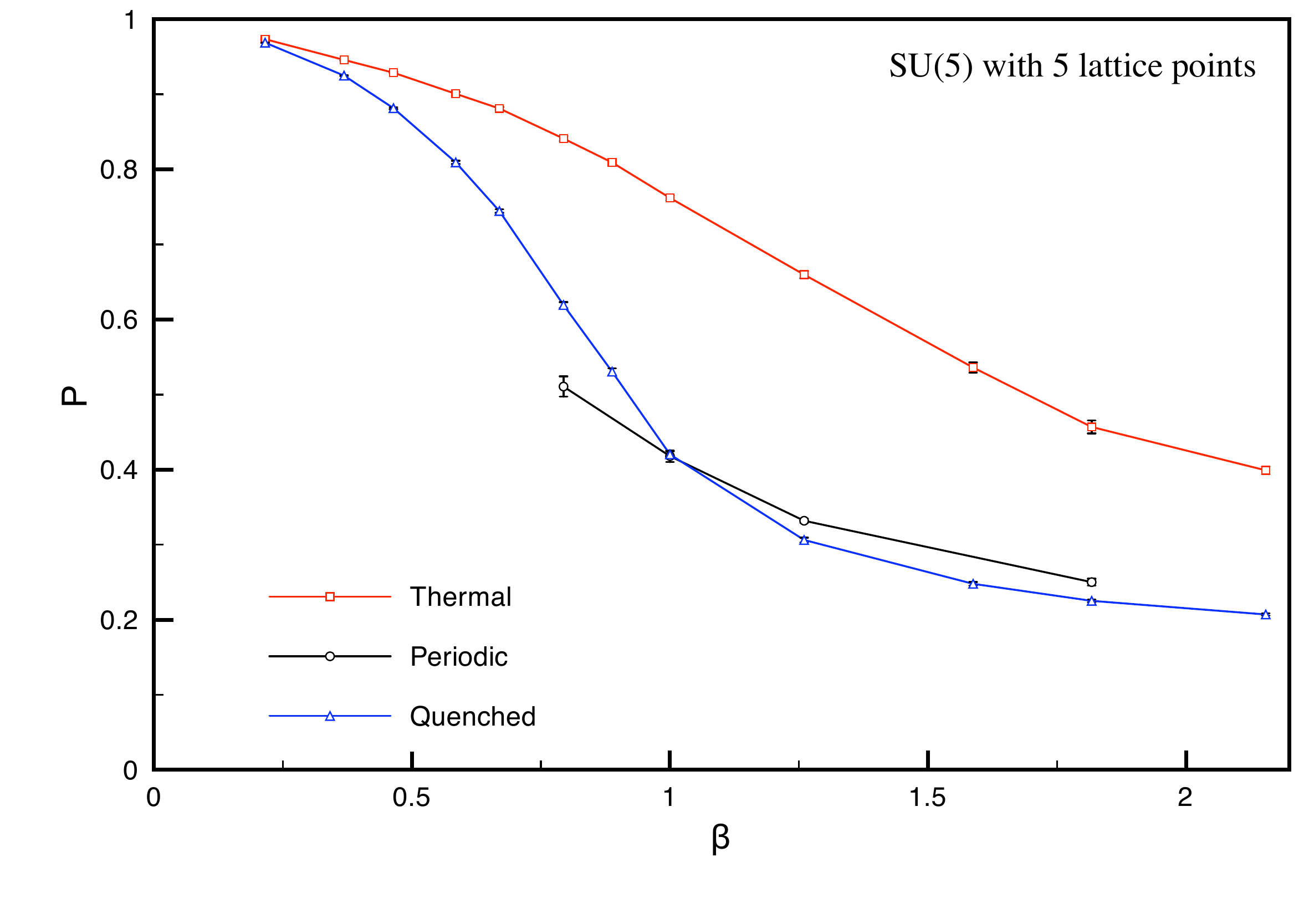}}
  \centerline{\includegraphics[width=3.5in,height=2.5in]{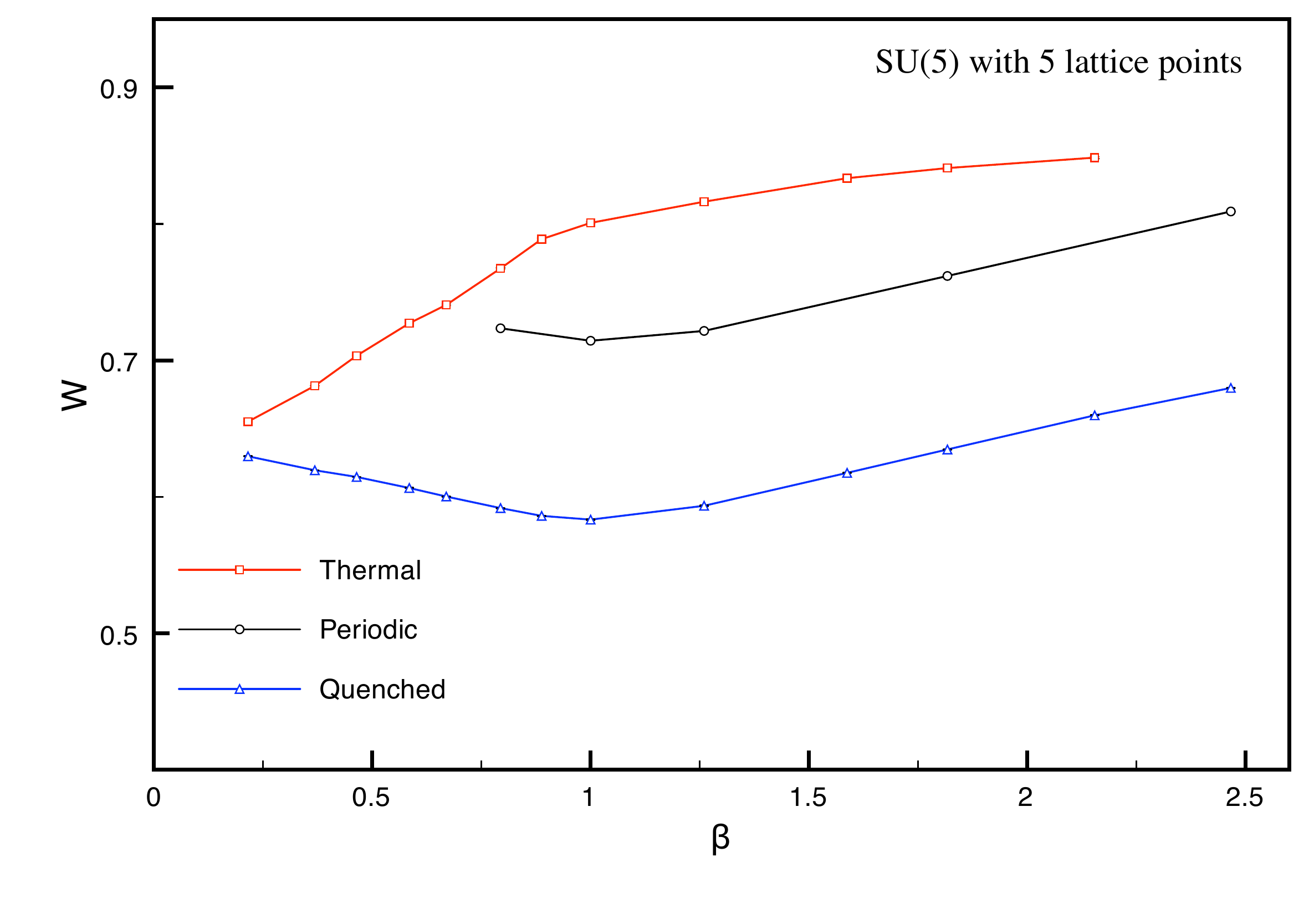}}
        \caption{Plots of the bosonic action $S_B$ for the quenched and thermal theories, and the Polaykov loop $P$ and scalar width $W$ for the quenched, periodic and thermal theories for $N=5$ and 5 lattice points. These quantities are expected to be close to their continuum values.}
   \label{fig:thermal1}
}

We begin in figure \ref{fig:thermal1} by comparing this thermal theory with the previously discussed quenched and periodic theories for $N=5$ and $M=5$ lattice points. In appendix \ref{app:continuum} we show these quantities for increasing lattice sizes, showing that already these $5$ lattice point data capture the continuum reasonably accurately. 

Our expectation for this theory is that at small $\beta$ (high temperature) the thermal theory will behave like the quenched theory since the fermions are lifted out of the dynamics by their thermal mass. On the other hand in the absence of spontaneous supersymmetry
breaking, in the large $\beta$ limit (low temperature) we expect the energy of the theory to go to zero and the behaviour to coincide with the periodic theory. 

We clearly see from the data that at small $\beta$ the action $S_B$, Polyakov loop $P$ and scalar eigenvalue distribution width $W$ do coincide with the quenched behaviour as expected. Conversely, at large $\beta$ we see the same observables depart from the corresponding quenched quantities and approach the periodic results.
The variation of the bosonic action, $S_B$, is flatter and apparently smoother than that of the quenched theory and appears to have a small linear slope for large $\beta$, which would imply a non-zero energy and hence supersymmetry breaking if it were to continue to large values of $\beta$. 
However, given we have data only to inverse temperatures $\beta \sim 2.2$ it is difficult to say whether
we are really seeing asymptotic behavior -- any sublinear
behavior would imply supersymmetry is restored at
zero temperature. And the apparent asymptotics may also be
influenced by discretization effects. Indeed, to address the possibility of supersymmetry
breaking one should first extrapolate the data to zero lattice spacing
at fixed $\beta$ and then examine the $\beta$ dependence of the extrapolated
curve. 
This extrapolation
is beyond our current resources. Nevertheless, it should be noted that
the asymptotic behavior of this quantity appears
to be rather different than
that expected for the 16 supercharge theory where we see from equation \eqref{eq:stringprediction} that holography predicts $S_B \sim \beta^{-9/5}$.
 
Independent of the final conclusion concerning supersymmetry breaking there
is little indication of a discontinuity in the mean energy as a function of
$\beta$ as $N$ increases and our numerical results
are consistent with the existence of
just a single phase for all $\beta$. In figure \ref{fig:thermal4} we show the bosonic action $S_B$, Polyakov loop variable $P$, its susceptibility $dP$ and the scalar eigenvalue width $W$ for the thermal theory for $N=5,8,12$. We see the results confirm the 't Hooft scaling - recall we expect the observables $S_B, P, W$ to tend to a constant at fixed $\beta$ in the large $N$ limit, as we see confirmed in the data. Whilst in the quenched theory there is a large $N$ transition, with $P$ vanishing at large $N$ for $\beta > \beta_c$, and the susceptibility diverging at $\beta = \beta_c$, we see no such behaviour here for the thermal theory in the range of $\beta$ studied. Instead $P$ appears smooth and the susceptibility $dP$ reduces for increasing $N$ over most of the range of $\beta$. Together with the apparently smooth bosonic action $S_B$ and scalar width $W$ we conclude that the thermal theory
is always in one, presumably deconfined, phase with non-vanishing $P$ and with bosonic energy $O(N^2)$.

\FIGURE[h]{
 \centerline{\includegraphics[width=3in,height=2.5in]{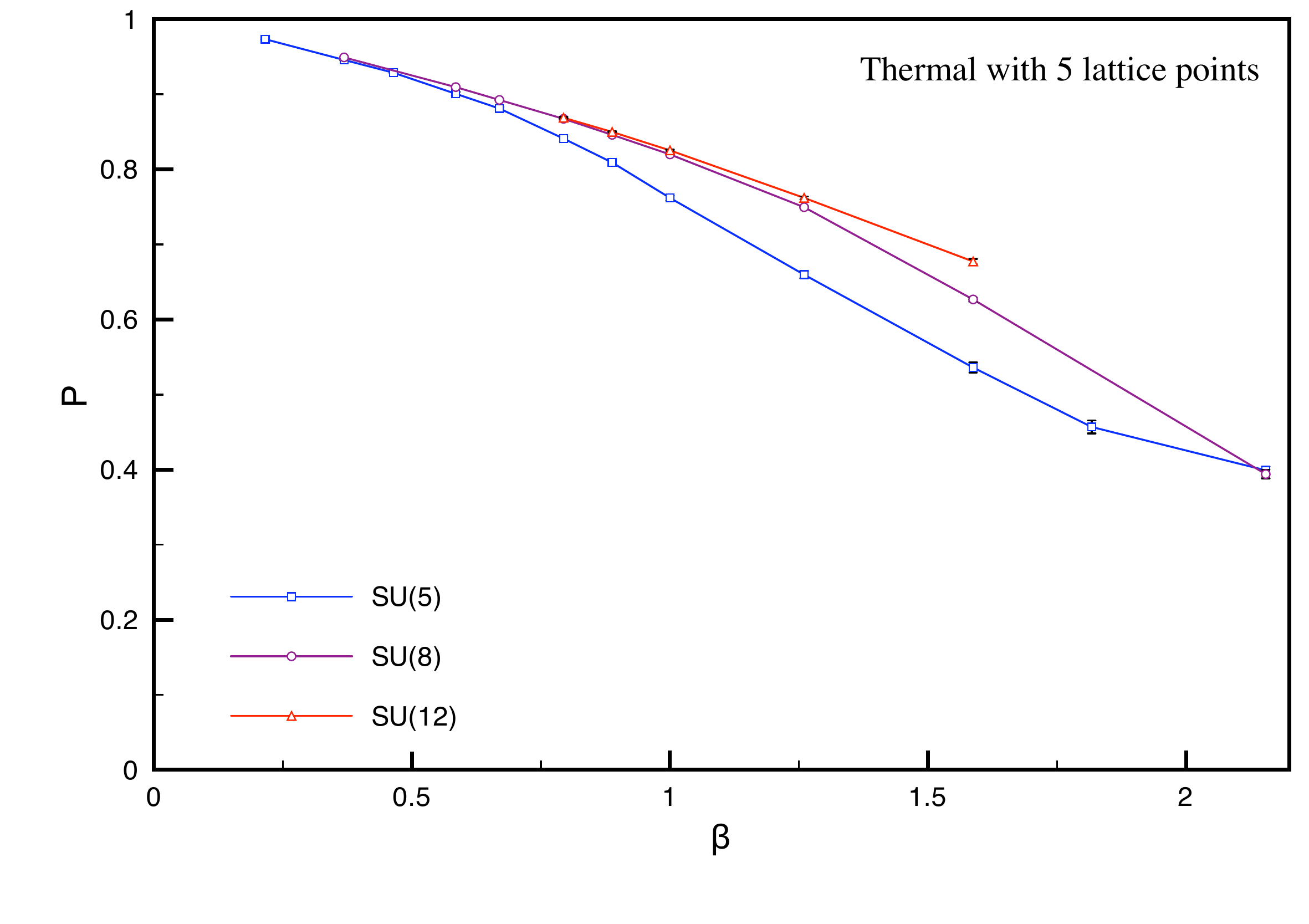}\includegraphics[width=3in,height=2.5in]{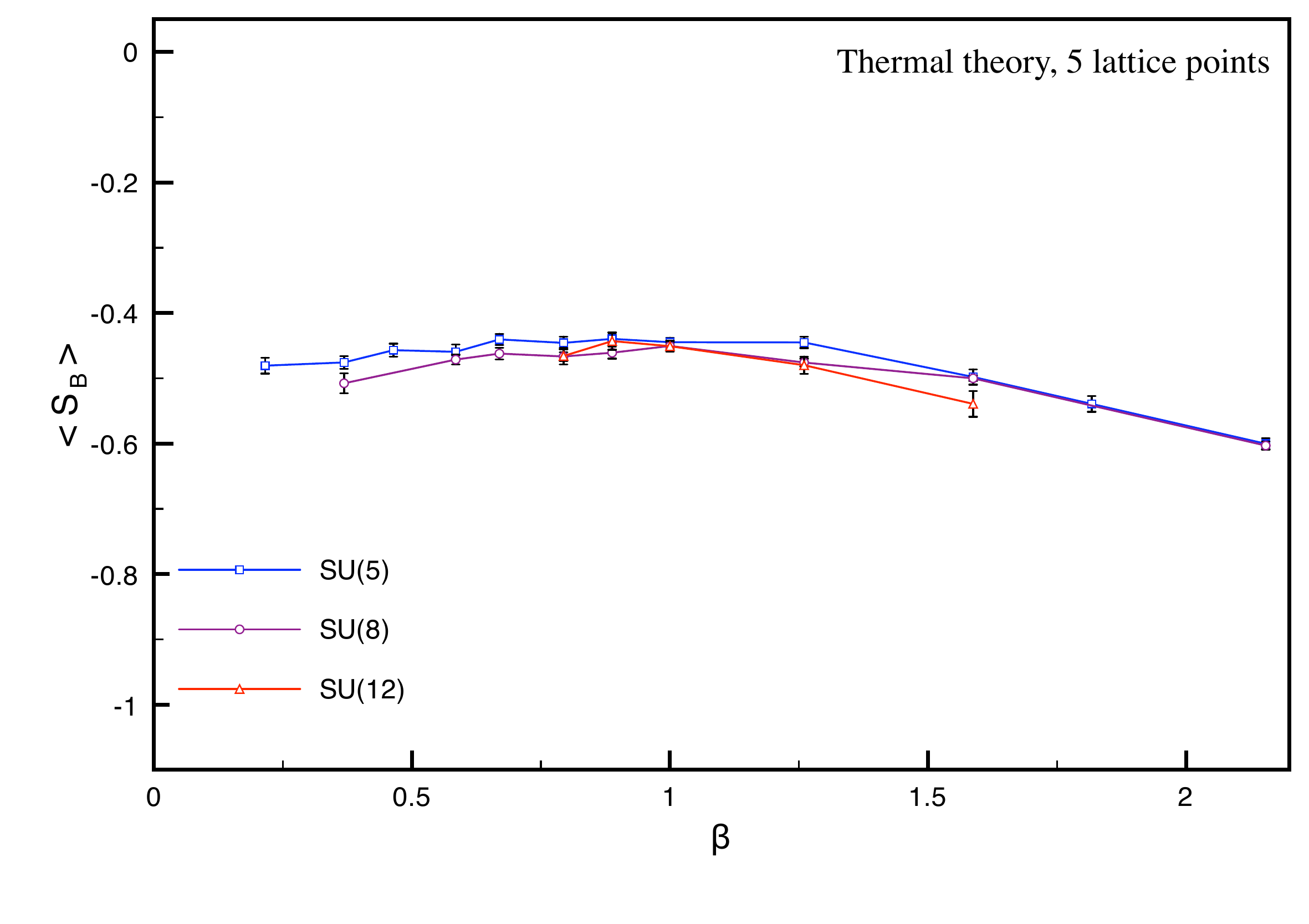}}
  \centerline{\includegraphics[width=3in,height=2.5in]{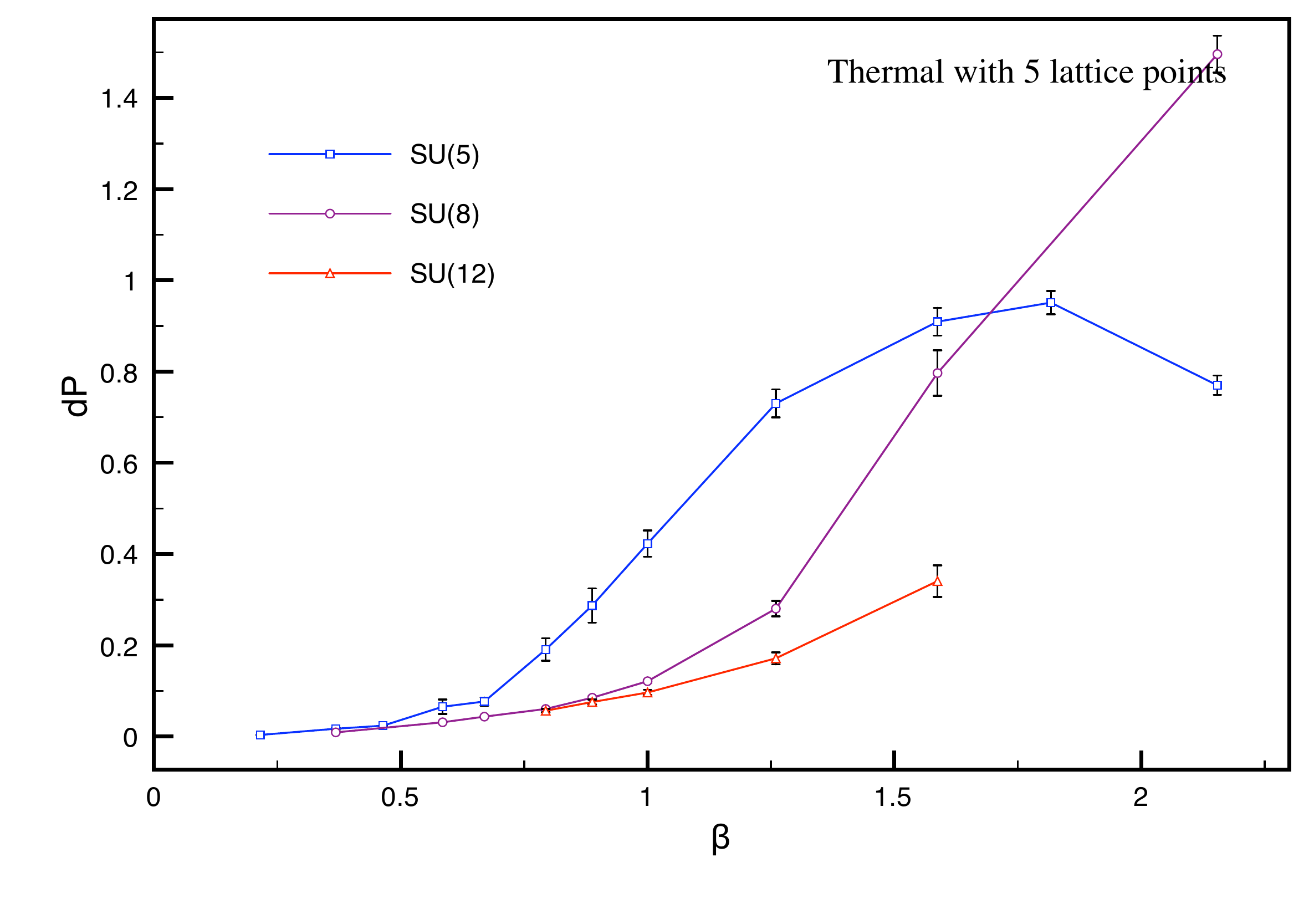}\includegraphics[width=3in,height=2.5in]{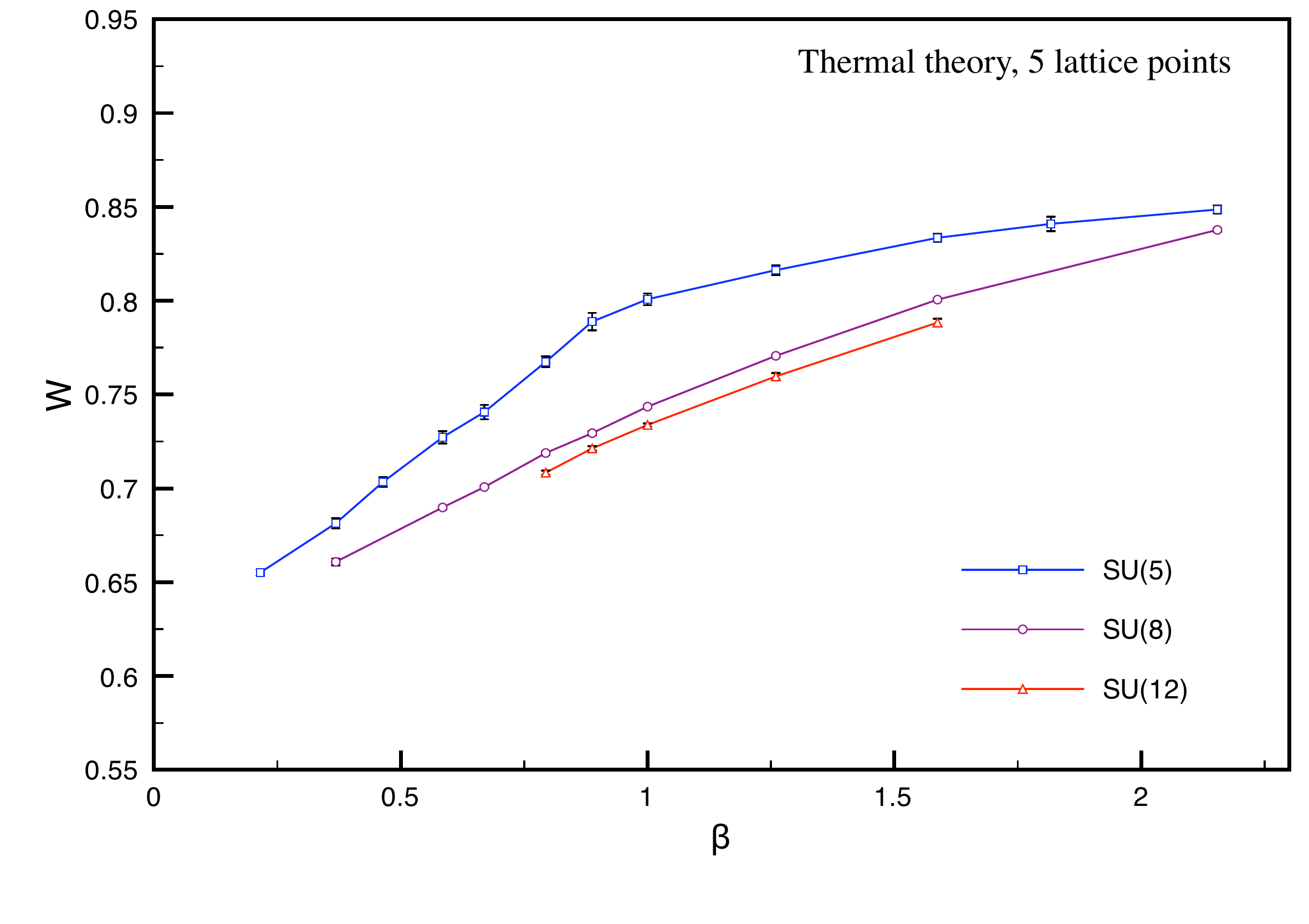}}
        \caption{Plots of the bosonic action $S_B$, Polaykov loop $P$, its susceptability $dP$ and the scalar field width $W$ for the thermal theory for increasing $N=5,8,12$ with 5 lattice points. These quantities are expected to be close to their continuum values and show no indication of large $N$ phase transitions. }
   \label{fig:thermal4}
}

\section{Discussion}

This paper is devoted to a study of four supercharge Yang Mills quantum mechanics at large $N$. This work is motivated
by the idea that a related model -- namely the sixteen supercharge theory, should be dual to IIA string theory at
least at sufficiently low energies that it can be approximated by a supergravity theory. In this regime
the Yang-Mills model is strongly coupled and hence we have developed lattice discretizations of
the model which allow for Monte Carlo simulation. 
Two such discretizations have been studied -- a so-called
naive action in which supersymmetry is broken classically by terms of the order the
lattice spacing and a manifestly supersymmetric action which arises from a discretization of a twisted
form of the continuum theory. 

We have analytically argued that the quantum mechanics is independent of a UV momentum regulator provided it preserves the gauge and global symmetries and hence our naive discretization should give the correct continuum physics. 
The mean energy and the Polyakov loop computed with these
two discretizations agree well for the case when supersymmetry preserving periodic
boundary conditions are used for the fermions, and the mean energy of the naive discretization is indeed consistent with vanishing in the continuum as required by supersymmetry. 

The theory has a large classical bosonic moduli space. We have analytically computed the 1-loop effective action for the bosonic moduli in this theory -- both for
periodic and antiperiodic (thermal) fermion boundary conditions and find that the classical
moduli space is lifted and the theory does not suffer from infrared divergences. These analytical
results are in agreement with our numerical work which shows that the eigenvalues  of the scalar fields
are localized close to the origin in field space where the low energy behaviour is strongly coupled.

We have studied the thermodynamic behaviour of the thermal theory for $N \le 12$ and observe the expected 't Hooft scaling of thermodynamic quantities. In contrast to the quenched theory, the behaviour of the thermal theory with dynamical fermions appears smooth and we find no evidence of a large $N$ phase transition, the theory always appearing deconfined. This is similar to holographic expectations for the 16 supercharge theory, although we find that the temperature dependence of the
free energy in the 4 supercharge model is rather different from that
expected for 16 supercharges.

It would be very interesting to extend these calculations to the case of 16 supercharges. This
is computationally more challenging with the main problem being how effectively the Monte
Carlo procedures can handle the phase of
the Pfaffian arising after integration over the fermions in that case. This problem is currently
being studied. If calculations with this Pfaffian prove possible it would be extremely
interesting as then the lattice model could be used to further test the duality between
gauge theories and gravity and perhaps learn more about the nature of the gravitational
theory in regimes of high temperature where stringy corrections are not small.

\acknowledgments

We would like to thank Jo Minahan, Shiraz Minwalla, Savdeep Sethi, Nemani Suryanarayana and David Tong for useful discussions.
SC is supported in part by DOE grant
DE-FG02-85ER40237. TW is partly supported by a PPARC advanced
fellowship and a Halliday award. Simulations were performed on the
LQCD supercomputer at Fermilab.

\appendix

%
\section{Holographic dual of 16 supercharge quantum mechanics}
%
\label{app:dual}

Following Itzhaki et al \cite{Itzhaki}, we consider the ``decoupling'' limit of $N$ coincident D0-branes. We take $N$ large with $N g_s$ fixed, where $g_s$ is the string coupling. The decoupling limit is then defined by considering excitations of these D0-branes with fixed energy while sending the string length scale to zero so $\alpha' \rightarrow 0$. In this limit the degrees of freedom of the system split up into those localized near the branes - the `near horizon' excitations - and those living far from the brane which we are not interested in here.

Depending on $N g_s$ there are two perturbative descriptions of the degrees of freedom living near the branes. For $N g_s << 1$ the D0-branes decouple from the ambient 10-d gravity and the degrees of freedom are well described simply by the worldvolume theory of the D0-branes whose degrees of freedom are the open strings ending on the branes, with dynamics governed by the 16 supercharge $SU(N)$ Yang-Mills quantum mechanics. Since we consider fixed energy excitations and $\alpha' = l_s^2 \rightarrow 0$, the action is just the conventional two derivative one, the higher derivative $\alpha'$ corrections being irrelevant. The gauge coupling is then found to be $g_{YM}^2 = g_s \alpha'^{-3/2} / (2\pi)^2$ and in this $N g_s << 1$ regime it is small. 

For $N g_s >>1$ the D0-branes couple strongly to gravity and the appropriate perturbative description is given by the target spacetime supergravity solution for $N$ D0-branes. The string frame metric is 
\begin{eqnarray}
ds^2_{IIA} & = & - \frac{f}{h^{1/2}} dt^2 + h^{1/2} \left( \frac{1}{f} dr^2 + r^2 d\Omega_{(8)}^2 \right) , \nonumber \\
f(r) & = & 1 - \left( \frac{r_0}{r} \right)^7 , \qquad h(r) = 1 + \left( \frac{R}{r} \right)^7 
\end{eqnarray}
and the dilaton and Ramond-Ramond (RR) gauge field are given as,
\begin{equation}
e^{\phi} = g_s h^{\frac{3}{4}} , \qquad A_0 = - \frac{1}{2} \left( \frac{1}{h} - 1 \right) .
\end{equation}
The solution describes a black hole with temperature $T$ and Bekenstein-Hawking entropy $S$,
\begin{equation}
T = \frac{7}{2 r_0} \sqrt{1 + \frac{R^7}{r_0^7}} \; , \quad S = \frac{2}{(2 \pi)^7 g_s^2 l_s^8}\Omega_8 \left( e^{-2 \phi} h^2 r^8 \right)_{r=r_0}
\end{equation}
using the conventions of \cite{Itzhaki} with $\Omega_8$ the area of a unit 8-sphere. Each brane carries RR charge, and hence the number of D0-branes is computed from the charge of the black hole by $N = \int_{S^8} \star F$ with $F$ the RR field strength, and one finds,
\begin{equation}
N g_s = \frac{R^7}{b \; l_s^7} \left( 1 + \frac{r_0^7}{R^7} \right)^{1/2} .
\end{equation}
Hence in the limit $N g_s$ is large the characteristic curvature radius of the solution, estimated by $R$, is much larger than the string length and therefore the string worldsheet theory is weakly coupled, and supergravity is a good approximation. 
Geometrically this solution has two regions. One is asymptotically flat 10-d spacetime, the other is the geometry describing the `near-horizon' region of the branes. Finite energy excitations in these two regions are separated from each other by a potential barrier. To take the decoupling limit and focus on the excitations only in the `near-horizon' region we must fix our physical energy scales of interest and take $\alpha' \rightarrow 0$. We identify the energy scales we wish to fix as $U$ and $U_0$, where
\begin{equation}
U = \frac{r}{\alpha'} , \qquad U_0 = \frac{r_0}{\alpha'} 
\end{equation}
and so $U\ge U_0$. Then taking $\alpha'  \rightarrow 0$ keeping $U, U_0$ fixed we see that $N g_s \rightarrow \frac{R^7}{b \; l_s^7}$ and the near-horizon geometry becomes,
\begin{eqnarray}
ds^2 & = & \alpha' \left( \frac{U^{\frac{7}{2}}}{2 \pi \sqrt{b \lambda}} (-f dt^2) + 2 \pi \sqrt{b \lambda} \left( U^{-\frac{7}{2}} \frac{dU^2}{f} + U^{-\frac{3}{2}} d\Omega^2\right) \right) \nonumber \\
f(U) & = & 1 - \frac{U_0}{U}
\end{eqnarray}
with $\lambda = N g_{YM}^2$. In this decoupling limit the entropy becomes,
\begin{equation}\label{eq:YMent}
S = \frac{1}{28 \sqrt{15} \pi^{7/2}} N^2 \left( U_0 / \lambda^{1/3} \right)^{9/2} ,
\end{equation}
and the temperature is given by,
\begin{equation}\label{eq:YMtemp}
T/\lambda^{1/3} = \frac{7}{16 \sqrt{15} \pi^{7/2}} \left( U_0 / \lambda^{1/3} \right)^{5/2} .
\end{equation}

The key statement of Maldacena's duality is that both these perturbative desciptions - the Yang-Mills for $N g_s << 1$ and the stringy black hole for $N g_s >> 1$ - are not limited to the regime of $N g_s$ where they are perturbatively good. In particular the Yang-Mills description remains well defined for all $N g_s$. The string black hole description also persists for finite $N g_s$ away from $N g_s >> 1$, and for $N g_s$ finite one must take into account stringy $\alpha'$ corrections to the description of the black hole in supergravity. 

To summarize the correspondence, the claim is that Yang-Mills with coupling $g_{YM}$ at finite temperature $T$ taken in the large $N$ limit is dual to string theory with target space given in the large $N g_s$ limit by the near horizon geometry of $N$ D0-branes as above, given by fixing our thermal energy scale $U_0$ and taking $\alpha' \rightarrow 0$. The dimensionless effective coupling at finite temperature in the Yang-Mills is given by  $\beta^3 = \lambda / T^3$. The equations \eqref{eq:YMent} and \eqref{eq:YMtemp} then relate the Yang-Mills quantities to the string theory.

Let us further explore the corrections to the supergravity description of the string theory above.
It is crucial that all curvatures and the dilaton are small in order that the above supergravity solution is valid. The curvature radius $\rho$ at energy scale $U$ is characterized by the radius of the sphere in the above geometry, so that in string units,
\begin{equation}
\frac{\rho}{\alpha'^{1/2}} \sim \left( \frac{\lambda}{U^3} \right)^{1/4}
\end{equation}
and the dilaton at the radius $U$ is,
\begin{equation}
e^{\phi} \sim  \frac{1}{N} \left( \frac{\lambda}{U^3} \right)^{7/4}
\end{equation}
We also require the temperature $T$ to be large enough that Euclidean winding modes are not present at the horizon.

Hence we see that provided $\lambda/U^3, \lambda/U_0^3$ and $N$ are large the supergravity solution above is a good approximation. In particular the dilaton condition shows we must take the 't Hooft limit, first taking $N$ to infinity with $\lambda/U^3, \lambda/U_0^3$ fixed, and then take these large.

If we take the black hole to have a high energy/temperature with $\lambda/U_0^3 \sim 1$ we reach the Horowitz-Polchinski correspondence region where the IIA supergravity breaks down even at the horizon due to $\alpha'$ curvature corrections becoming important. One requires full IIA string theory to describe the horizon region. Conversely for a black hole with ultra low energy/temperature, so $U_0^3/\lambda \sim N^{-4/7}$, which is outside the 't Hooft scaling limit, the dilaton becomes large near the horizon and the string theory becomes strongly coupled there. It may then be resolved by lifting to M-theory where again one finds 11-d supergravity is valid. However, we emphasize that provided we remain within the 't Hooft scaling regime, we cannot access such low temperatures.

%
\section{Potentials on classical moduli space}
%
\label{app:moduli}

We now derive the effective 1-loop actions for the classical zero modes. We compute first the potential for the quenched theory, and derive the effective coupling controlling the 1-loop integration. Then we proceed to the thermal theory, and finally the periodic theory.

\subsection{Quenched case}

The results in this subsection are a special case studied in Aharony et al \cite{Aharony2}. We expand our bosonic fields in fluctuations about the classical zero-modes,
\begin{eqnarray}
A_{ab}(\tau) & = & A^a {\delta}_{ab}  + \delta A_{ab}(\tau)  \nonumber \\
X_{i,ab}(\tau) & = & x_i^a {\delta}_{ab}  +  \delta X_{i,ab}(\tau)
\end{eqnarray}
where the the classical bosonic moduli are $A^a,x^a_i$. We take the fluctuations $\delta A, \delta X_i$ to have no constant component in their diagonal terms. We will shortly expand these fluctuations in harmonics so,
\begin{eqnarray}
 \delta A_{ab}(\tau)  & = & \frac{1}{\sqrt{2 \pi}} \sum_{m=-\infty}^{\infty} \delta A^{(m)}_{ab} e^{\frac{2 \pi}{R} i m \tau} \nonumber \\
 \delta X_{i,ab}(\tau) & = & \frac{1}{\sqrt{2 \pi}} \sum_{m=-\infty}^{\infty}  \delta X^{(m)}_{i,ab} e^{\frac{2 \pi}{R} i m \tau}
\end{eqnarray}
and thus $\delta A^{(0)}_{ab}, \delta X^{(0)}_{i,ab}$ vanish when $a=b$. Since $A_{ab}(\tau),X_{i,ab}(\tau)$ are Hermitian fields, we have $(\delta A^{(m)}_{ab})^\star = \delta A^{(-m)}_{ba}$ and $(\delta X^{(m)}_{i,ab})^\star = \delta X^{(-m)}_{i,ba}$.
Since the gauge group is $SU(N)$, the sums $\sum_a A^a, \sum_a x_i^a$ vanish. We define the dimensionless $\Delta A^{ab} = R (A^a - A^b)$ and $\Delta x_i^{ab} = R (x_i^a - x_i^b)$, and the following quantities,
\begin{eqnarray}
\delta \Phi_{\mu,ab} & = & \left( \delta A_{ab}, \delta X_{i,ab} \right) \nonumber \\
D^{ab}_\mu & = &  \left( R \partial_\tau + i \Delta A^{ab},  i \Delta x^{ab}_i \right) 
\end{eqnarray}
We may now  write the action to quadratic order in the perturbations, obtaining,
\begin{equation} \label{eq:varyL}
\mathcal{S}^{quad} = \frac{N}{\lambda R^2} \int d\tau  \sum_{a<b}  \delta \Phi^{\star}_{\mu,ab}  M^{ab}_{\mu\nu} \delta \Phi_{\nu,ab} 
\end{equation}
where,
\begin{equation}
M^{ab}_{\mu\nu} =  \frac{1}{2} \left( D_\mu^{ab} D_\nu^{ab} - \delta_{\mu\nu} (D_\rho^{ab} )^2 \right) .
\end{equation}
We must take care with the action since we have not fixed a gauge. The gauge transformation generates a bosonic fluctuation $\delta \Phi_{\mu,ab} = D_{\mu}^{ab} \lambda_{ab}$ for arbitrary functions $\lambda_{ab}$ of $\tau$ and correspondingly we see,
\begin{equation}
M^{ab}_{\mu\nu} D_{\nu}^{ab} = 0 \quad \forall \; a,b
\end{equation}
We choose to use the gauge freedom to fix the gauge $\delta A_{ab}(\tau) = 0$, and hence the gauge field is diagonal and constant in $\tau$. This gauge fixing introduces a Jacobian factor of $\prod_{a,b}\det{D_{\tau}^{ab} }$ from the path integral measure. In the case of the constant modes on the circle, this is just the familiar Vandermonde determinant.
Now the quadratic action becomes,
\begin{equation} 
S^{quad} = \frac{N}{\lambda R^2}  \int d\tau  \sum_{a<b}  \left( 
 \delta X^{\star}_{i,ab} M^{ab}_{ij}  \delta X_{j,ab} \right) - \sum_{a,b} \ln |\det D_{\tau}^{ab}| 
\end{equation}
We may now expand in Fourier modes to obtain,
\begin{equation} \label{eq:quad}
S^{quad} = \frac{N}{\lambda R}  \sum_{a<b} \sum_{m=-\infty}^{\infty}  \left( 
 \delta X^{\star (m)}_{i,ab} M^{(m)ab}_{ij}  \delta X^{(m)}_{j,ab} \right) - \sum_{a < b} \sum_{m=-\infty}^{\infty} \ln (D_\tau^{(m)ab})^2
\end{equation}
where $D_\alpha^{(m)ab} = ( 2 \pi i m + i \Delta A^{ab},  i \Delta x^{ab}_i )$, and 
\begin{equation}
M^{(m)ab}_{\mu\nu} =  \frac{1}{2} \left( D_\mu^{(m)ab} D_\nu^{(m)ab} - \delta_{\mu\nu} (D_\rho^{(m)ab} )^2 \right) .
\end{equation}
For a given mode $m$, and colour indices $ab$, the eigenvalues of the $3\times3$ matrix $M^{(m)ab}_{ij}$ are $-(D_\tau^{(m)ab})^2$ once, and $-\sum_{\alpha} (D_\alpha^{(m)ab})^2$ repeated twice. Now our quadratic action is positive after our gauge fixing, we may integrate out the fluctuations yielding, up to terms independent of the moduli,
\begin{eqnarray} 
S^{quad} & = & 2 \sum_{a<b} \sum_{m=-\infty}^{\infty}  \ln | D_\alpha^{(m)ab}|^2 \nonumber \\
& = & 2 \sum_{a<b} \sum_{m=-\infty}^{\infty}  \ln \left( (2 \pi m + \Delta A^{ab} )^2 + | \Delta x^{ab} |^2 \right)^2
\end{eqnarray}
where $| \Delta x^{ab} |^2 = \sum_i (\Delta x^{ab}_i)^2$. We note that the constant $m=0$ modes have an enhanced $SO(4)$ target spacetime global symmetry. This is explicitly broken by our gauge choice, but the result should have this full invariance, and we indeed see that this is the case.
We also note that the effective action takes the form of a pair-wise interaction between the moduli with colour labels $a$ and $b$.

The infinite sum over Fourier modes must be regulated in the quenched case. A Pauli-Villars regulator can be utilized, giving a regularized 1-loop action,
\begin{eqnarray} 
S^{reg} & = & 2 \lim_{\Omega \rightarrow \infty} \sum_{a<b} \sum_{m=-\infty}^{\infty}  \ln \left( \frac{ (2 \pi m + \Delta A^{ab} )^2 + | \Delta x^{ab} |^2 }{ (2 \pi m)^2 + \Omega^2 } \right)^2 \nonumber \\
& = & 2 \sum_{a<b} \ln \left( \cosh{ | \Delta  x^{ab} |  } - \cos{ \Delta A^{ab} } \right)
\end{eqnarray}
where in the second line we have suppressed the trivial divergence going as $R \Omega$.

When was this 1-loop integration valid? As with any dimensional reduction on a circle, the strongest coupled modes are the ones that are constant on the circle. We should compare our quadratic action \eqref{eq:quad} for these constant modes with the interaction terms,
\begin{equation}
\mathcal{S}^{int} = - \frac{1}{2} \frac{N}{\lambda} R \; \mathrm{Tr} \left( \delta\Phi^{\star(0)}_{\mu}  \delta \Phi^{(0)}_{\nu}  \delta\Phi^{\star(0)}_{\mu}  \delta \Phi^{(0)}_{\nu} - \delta\Phi^{\star(0)}_{\mu}  \delta \Phi^{(0)}_{\nu}  \delta\Phi^{\star(0)}_{\nu}  \delta \Phi^{(0)}_{\mu}   \right)
\end{equation}
In order to obtain a canonical kinetic term in \eqref{eq:quad} we should rescale the constant fluctuation fields,
\begin{equation}
\delta \tilde{\Phi}^{(0)}_{\mu,ab} = \sqrt{ \frac{N |D_\alpha^{(0)ab}|^2}{R \lambda} } \delta {\Phi}^{(0)}_{\mu,ab}
\end{equation}
Our interaction term now takes the schematic form,
\begin{equation}
\mathcal{S}^{int} \sim \frac{\lambda R^3}{|D_\alpha^{(0)ab}|^4}  \frac{1}{N^3}  (\delta \tilde{\Phi}^{(0)})^4  = g_{eff} \frac{1}{N^3} (\delta \tilde{\Phi}^{(0)})^4
\end{equation}
where we have suppressed the obvious index and matrix structure.
The factors of $N$ are associated with the 't Hooft limit, leaving the effective coupling for the constant modes,
\begin{equation}
g_{eff} = \frac{\beta^3}{\left( (\Delta A^{ab} )^2 + | \Delta x^{ab} |^2 \right)^2 }
\end{equation}
Since $R A^a$ should be thought of as an angular variable the effective coupling is really characterized by just the non-compact bosonic moduli, 
\begin{equation}
g_{eff} \sim \frac{\beta^3}{| \Delta x^{ab} |^4}
\end{equation}
and hence we expect  the 1-loop approximation is good provided,
\begin{equation}
 |\Delta x^{ab}|  >> \beta^{3/4} .
\end{equation}

\subsection{Thermal case}

Again this case is straightforward, and closely follows the quenched calculation above. The results in this section are a special case studied in Aharony et al \cite{Aharony2}. We expand our fields as before, now including the fermions with the appropriate boundary conditions, so,
\begin{eqnarray}
A_{ab}(\tau) & = & A^a {\delta}_{ab}  + \frac{1}{\sqrt{2 \pi}} \sum_{m=-\infty}^{\infty}  \delta A^{(m)}_{ab} e^{ \frac{2 \pi}{R} i m \tau} \nonumber \\
X_{i,ab}(\tau) & = & x_i^a {\delta}_{ab}  + \frac{1}{\sqrt{2 \pi}} \sum_{m=-\infty}^{\infty}  \delta X^{(m)}_{i,ab}e^{\frac{2 \pi}{R} i m \tau} \nonumber \\
\Psi_{\alpha,ab}(\tau) & = & \frac{1}{\sqrt{2 \pi}} \sum_{m=-\infty}^{\infty}  \delta \Psi^{(m)}_{\alpha,ab} e^{\frac{2 \pi}{R} i (m +\frac{1}{2}) \tau} 
\end{eqnarray}
We see that due to the antiperiodicity the fermions can have no constant mode on the circle. As before we take $\delta A^{(0)}_{ab}, \delta X^{(0)}_{i,ab}$ and now also $\delta \Psi^{(0)}_{\alpha,ab}$ to vanish when $a=b$. We gauge fix the action in the same manner, and now arrive at a similar action to the quenched one above in equation \eqref{eq:quad}, but with a fermionic piece too,
\begin{equation} \label{eq:quad2}
S^{quad} = \frac{N}{\lambda R}  \sum_{m=-\infty}^{\infty} \left[ \sum_{a<b}   \left( 
 \delta X^{\star (m)}_{i,ab} M^{(m)ab}_{ij}  \delta X^{(m)}_{j,ab} \right) + \sum_{a,b} \left( 
 \delta \bar{\Psi}^{(m)}_{ab}  i \bar{\sigma}^{\rho} D_\rho^{(m + \frac{1}{2})ab} \delta \Psi^{(m)}_{ab} \right) - \sum_{a < b}  \ln |D_\tau^{(m)ab}|^2 \right]
\end{equation}
Now performing the fermionic integral yields a determinant equal to $\prod_{m=-\infty}^{\infty} \prod_{a,b} (D_{\alpha}^{(m + \frac{1}{2})ab})^2$. The bosonic fluctuation integral yields the same result as for the quenched theory. Putting this together we obtain the effective action for the bosonic zero modes,
\begin{eqnarray} 
S^{quad} & = & 2 \sum_{a<b} \sum_{m=-\infty}^{\infty}  \ln |D_\alpha^{(m)ab}|^2 - \sum_{a,b} \sum_{m=-\infty}^{\infty}  \ln |D_\alpha^{(m+ \frac{1}{2})ab}|^2\nonumber \\
& = & 2 \sum_{a<b} \sum_{m=-\infty}^{\infty}  \ln \left(  \frac{ (2 \pi m + \Delta A^{ab} )^2 + | \Delta x^{ab} |^2 }{ (2 \pi m + \pi + \Delta A^{ab} )^2 + | \Delta x^{ab} |^2} \right)^2 \nonumber \\
& = & 2 \sum_{a<b} \ln \left( \frac{ \cosh{ | \Delta  x^{ab} |  } - \cos{ \Delta A^{ab} } }{ \cosh{ | \Delta  x^{ab} |  } + \cos{ \Delta A^{ab} } } \right)
\end{eqnarray}
where no regulator is required to evaluate the infinite sum in the last line.

\subsection{Periodic case}

The results in this section are new, but follow straightforwardly from previous work in Aoki et al and Aharony et al \cite{Aoki,Aharony2}. Again we expand our fields,
\begin{eqnarray}
A_{ab}(\tau) & = & A^a {\delta}_{ab}  + \frac{1}{\sqrt{2 \pi}} \sum_{m=-\infty}^{\infty}  \delta A^{(m)}_{ab} e^{ \frac{2 \pi}{R} i m \tau} \nonumber \\
X_{i,ab}(\tau) & = & x_i^a {\delta}_{ab}  + \frac{1}{\sqrt{2 \pi}} \sum_{m=-\infty}^{\infty}  \delta X^{(m)}_{i,ab}e^{\frac{2 \pi}{R} i m \tau} \nonumber \\
\Psi_{\alpha,ab}(\tau) & = & \xi_{\alpha}^a {\delta}_{ab}  + \frac{1}{\sqrt{2 \pi}} \sum_{m=-\infty}^{\infty}  \delta \Psi^{(m)}_{\alpha,ab} e^{\frac{2 \pi}{R} i m \tau} 
\end{eqnarray}
but note that now with periodic boundary conditions we must include fermion zero modes $\xi^a_{\alpha}$. 
As before we take $\delta A^{(0)}_{ab}, \delta X^{(0)}_{i,ab},\delta \Psi^{(0)}_{\alpha,ab}$ to vanish when $a=b$, and the $SU(N)$ colour symmetry implies that the sums $\sum_a A^a, \sum_a x_i^a$ and also $\sum_a \xi_\alpha^a$ vanish.
Using the notation $\Delta \xi^{ab} = \xi^a - \xi^b$ for the fermion zero modes we may now write the action to quadratic order in the perturbations, obtaining,
\begin{equation} 
S^{quad} = \frac{N}{\lambda R}  \sum_{a<b} \sum_{m=-\infty}^{\infty} 
 \left( \begin{matrix} \delta \Phi^{\star(m)}_{\mu,ab} &  \quad \delta \bar{\Psi}^{(m)}_{\dot{\alpha},ab} \end{matrix} \right) 
 \left( \begin{matrix} M^{(m)ab}_{\mu\nu} & \Delta \bar{\xi}^{ab}_{\dot{\alpha}} \bar{\sigma}^{\mu,\dot{\alpha}\alpha}  \\  \bar{\sigma}^{\nu,\dot{\alpha}\alpha}  \Delta{\xi}^{ab}_{\alpha} & i \bar{\sigma}^{\rho,\dot{\alpha}\alpha} D_\rho^{(m)ab} \end{matrix} \right) 
 \left( \begin{matrix} \delta \Phi^{(m)}_{\nu,ab} \\  \quad \delta {\Psi}^{(m)}_{\alpha,ab} \end{matrix} \right) 
\end{equation}
We see now that there are off-diagonal terms that mix the fermion and boson fluctuations, which are coupled together by the presence of the fermion zero modes. These can be removed by the following transformation,
\begin{equation}
\delta \Psi^{(m)}_{ab} = \delta \Lambda^{(m)}_{ab} - i \frac{1}{(D_\rho^{(m)ab} )^2} (\sigma^\alpha D_\alpha^{(m)ab} ) (\bar{\sigma}^\beta \Delta\xi^{ab} ) \delta \Phi^{(m)}_{\beta,ab}
\end{equation}
which we note is a simple shift in the fermion fluctuation, and hence does not alter the path integral measure. Now we find the quadratic fluctuation action is diagonal,
\begin{equation} 
S^{quad} = \frac{N}{\lambda R}  \sum_{a<b} \sum_{m=-\infty}^{\infty} 
 \left( \begin{matrix} \delta \Phi^{\star(m)}_{\mu,ab} &  \quad \delta \bar{\Lambda}^{(m)}_{ab} \end{matrix} \right) 
 \left( \begin{matrix} M^{(m)ab}_{\mu\nu} + S^{(m)ab}_{\mu\nu} & 0  \\  0 & i \bar{\sigma}^{\rho} D_\rho^{(m)ab} \end{matrix} \right) 
 \left( \begin{matrix} \delta \Phi^{(m)}_{\nu,ab} \\  \quad \delta {\Lambda}^{(m)}_{ab} \end{matrix} \right) 
\end{equation}
where,
\begin{equation}
S^{(m)ab}_{\mu\nu} = - i \frac{1}{(D_\rho^{(m)ab} )^2} ( \Delta \bar{\xi}^{ab} \bar{\sigma}^\mu ) (\sigma^\alpha D_\alpha^{(m)ab}) (\bar{\sigma}^\nu \Delta\xi^{ab}) .
\end{equation}
It follows from $\delta \Phi^{(m)}_{\mu,ab} = \delta \Phi^{(-m)\star}_{\mu,ba}$, and $D^{(m)ab}_\alpha = - D^{(-m)ba}_\alpha$ that the action actually projects only onto the antisymmetric component $S^{(m)ab}_{[\mu\nu]} = \frac{1}{2} \left( S^{(m)ab}_{\mu\nu} - S^{(m)ab}_{\nu\mu} \right)$. This antisymmetric part can be written as,
\begin{equation}
S^{(m)ab}_{[\mu\nu]} = - i \epsilon^{\mu\nu\alpha\beta} \frac{D_\alpha^{(m)ab}}{(D_\rho^{(m)ab} )^2} ( \Delta \bar{\xi}^{ab} \bar{\sigma}^\beta \Delta\xi^{ab}) ,
\end{equation}
We may now write the quadratic action as,
\begin{eqnarray} \label{eq:quadpbc}
S^{quad} & = & \frac{N}{\lambda R}  \sum_{m=-\infty}^{\infty} \left( \sum_{a<b}  P + \sum_{a,b}  i \delta \bar{\Lambda} \bar{\sigma}^{\rho} D_\rho \delta {\Lambda} \right) \nonumber \\
P & = & \delta \Phi^{\star}_{\mu} \left( D_\mu D_\nu -  \delta_{\mu\nu} D_\alpha^2 + \epsilon_{\mu\nu\alpha\beta} D_\alpha J_\beta \right) \delta \Phi_{\nu} 
\end{eqnarray}
where,
\begin{equation}
J^\alpha = - 2 i  \frac{1}{(D_\rho)^2} ( \Delta \bar{\xi} \bar{\sigma}^\alpha \Delta\xi)
\end{equation}
and in these last two equations we have suppressed the Fourier index $m$, and colour indices $a,b$ for clarity. Again the bosonic operator in this action has a zero eigenvalue corresponding to the gauge freedom. Correspondingly we see,
\begin{equation}
( M^{(m)ab}_{\mu\nu} + S^{(m)ab}_{[\mu\nu]} ) D_{\nu}^{(m)ab} = 0 .
\end{equation}
Performing the same gauge fixing as before we have,
\begin{eqnarray} 
S^{quad} & = & \frac{N}{\lambda R}  \sum_{m=-\infty}^{\infty} \left( \sum_{a<b}  P^{g.f.} + \sum_{a,b}  i \delta \bar{\Lambda} \bar{\sigma}^{\rho} D_\rho \delta {\Lambda} \right) - \sum_{a<b}  \ln | \det D_\tau|^2  \nonumber \\
P^{g.f.} & = & \delta \Phi^{\star}_{i} \left( D_i D_j -  \delta_{ij} D_\alpha^2 + \epsilon_{ij\alpha\beta} D_\alpha J_\beta \right) \delta \Phi_{j} 
\end{eqnarray}
and now integrating over the bosons yields the determinant of the $3 \times 3$ matrix in $ij$ above, giving the elegant result 
\begin{equation} 
S^{quad}  =  \sum_{m=-\infty}^{\infty} \sum_{a<b} \ln \left( (D_\mu^2 )^2 + (D_\mu^2)(J_\nu^2) - (D_\mu J^\mu)^2 \right) + \frac{N}{\lambda R}  \sum_{m=-\infty}^{\infty} \sum_{a,b}  i \delta \bar{\Lambda} \bar{\sigma}^{\rho} D_\rho \delta {\Lambda} 
\end{equation}
where we again see that the $SO(4)$ spacetime symmetry for the $m=0$ constant modes is restored after the gauge fixing had broken it. Performing the fermion integration then yields the 1-loop action,
\begin{equation} 
S^{quad} = \sum_{m=-\infty}^{\infty} \sum_{a<b} \ln \left( 1 + \frac{1}{(D_\mu^2)^2} \left( (D_\mu^2) (J_\nu^2) - (D_\mu J^\mu)^2 \right)  \right)
\end{equation}
and we immediately see that if it where not for the fermion zero modes contributing to the presence of $J_\mu$, the action would vanish. Hence there has been almost total cancellation between the boson and fermion determinants. Now since $J^\mu \sim \Delta \bar{\xi} \bar{\sigma}^\mu \Delta\xi$ and each spinor $\xi_\alpha$ has only 2 components, the action can contain at most quadratic terms in $J_\mu$. Hence we may expand the logarithm above to give,
\begin{eqnarray} 
S^{quad} & = & + \sum_{m=-\infty}^{\infty} \sum_{a<b} \frac{4}{\left((D_\rho^{(m)ab})^2\right)^4} \left( D_\alpha^{(m)ab} D_\beta^{(m)ab} - (D_\mu^{(m)ab})^2 \delta_{\alpha\beta} \right) ( \Delta \bar{\xi}^{ab} \bar{\sigma}^\alpha \Delta\xi^{ab})( \Delta \bar{\xi}^{ab} \bar{\sigma}^\beta \Delta\xi^{ab}) \nonumber \\
& = & - \sum_{m=-\infty}^{\infty} \sum_{a<b} \frac{24}{\left((D_\rho^{(m)ab})^2\right)^3} ( \Delta \bar{\xi}^{ab}_1 \Delta\bar{\xi}^{ab}_2 \Delta \xi^{ab}_1\Delta\xi^{ab}_2)
\end{eqnarray}
where we have written the expression out fully. The partition function is then given by the functional integral over the bosonic and fermionic zero modes of this action,
\begin{eqnarray}
Z & \simeq & \int dA dX_i d\xi d\bar{\xi} e^{-S^{quad}} \nonumber \\
& = & \int dA dX_i d\xi d\bar{\xi} \prod_{m=-\infty}^{\infty} \prod_{a<b} \left( 1 + \frac{24}{\left((D_\rho^{(m)ab})^2\right)^3} ( \Delta \bar{\xi}^{ab}_1 \Delta\bar{\xi}^{ab}_2 \Delta \xi^{ab}_1\Delta\xi^{ab}_2) \right) \nonumber \\
& = & \int dA dX_i d\xi d\bar{\xi} \prod_{a<b} \left( 1 + \sum_{m=-\infty}^{\infty}  \frac{24}{\left((D_\rho^{(m)ab})^2\right)^3} ( \Delta \bar{\xi}^{ab}_1 \Delta\bar{\xi}^{ab}_2 \Delta \xi^{ab}_1\Delta\xi^{ab}_2) \right)
\end{eqnarray}
where we will not keep track of the overall normalization of $Z$. It is implicit in the measure that the gauge group is $SU(N)$ and hence there is the constraint $\sum_a \xi^a_\alpha,\sum_a \bar{\xi}_{\dot{\alpha}^a} = 0$, and this is important in giving the form of the expression below. For example, if we had instead taken the gauge group $U(N)$, the partition function would vanish as the trace of the adjoint fermion matrices that are constant on the circle decouple and since give fermion zero modes at 1-loop. After integrating these, the partition function will vanish.

Following Aoki et al \cite{Aoki}, we may perform the integral over the fermionic zero modes to give an effective 1-loop action for the bosonic moduli. In their analysis they express the answer in terms of sums over maximal trees with various valences as they consider 4,8 and 16 supercharge matrix models. However as they note, the case of 4 supercharges is somewhat simpler than the others, and the `tree' machinery is somewhat redundant. The answer can in fact simply be written as,
\begin{equation}
Z = \int dA dX_i \sum_{(a_1,a_2,\ldots,a_N) \in P} M^{a_1a_2}M^{a_2a_3} \ldots M^{a_{N-2}a_{N-1}} M^{a_{N-1}a_{N}}
\end{equation}
where $P$ is the set of permutations of $(1,2,\ldots,N)$ and,
\begin{equation}
M^{ab} = \sum_{m=-\infty}^{\infty} \frac{1}{\left( (m + \Delta A^{ab})^2+(\Delta x_i^{ab})^2 \right)^3} .
\end{equation}
This gives rise to the effective action given in the main text. We note that the infinite sum over Fourier modes is finite, and can simply be evaluated explicitly, although the expression obtained is unilluminating.

%
\section{Simulation details}
%
\label{app:details}

Both naive and supersymmetric lattice actions take the form
\beq
S=\kappa\left(S_B(X)+S_F(\psib,\psi,X)\right)\eeq
where
\beq
S_B=\sum_{x=0}^M 
-\frac{1}{2}\sum_i^3 (D^+X^i(x))^2-\frac{1}{2}\sum_{i>j}^3[X^i(x),X^j(x)]^2\eeq
has the same form in both cases except for a different definition of the
covariant derivative. Notice that the scalar fields in the above
action are expanded on the traceless {\it antihermitian} matrix basis of
$SU(N)$. The lattice coordinates $x$ are equally distributed with spacing
$a$ on a circle of length $R=Ma$ and $R$ can be identified
with $\frac{1}{T}$ in the case of non-zero temperature.
The fermion action takes the generic form
\beq
S_F=\sum_{x,y}\psib(x) {\cal M}(X)_{x,y}\psi(y) \eeq
where the fermion operator ${\cal M}(X)$ depends on the discretization.
An immediate question arises as to how to scale $\kappa$ as the
lattice spacing is reduced and the continuum limit approached.
Clearly the relevant dimensionless parameter in the continuum is
$\lambda R^3$ where the 't Hooft
coupling $\lambda=g^2_{\rm phys}N$ is used to access the
large $N$ limit. 
Equating the inverse of this parameter to the
corresponding lattice quantity $\frac{\kappa}{M^3}$ yields the
needed scaling of $\kappa$ 
\begin{equation}
\kappa =  \frac{N M^3}{\lambda R^3}=\frac{NM^3}{\beta^3}  
\end{equation}

To simulate this system we must first integrate out the Grassmann
fields yielding either ${\rm det}(M)$ or ${\rm det}^{\frac{1}{2}}(M)$
for naive or supersymmetric actions. In practice this is
accomplished by introducing complex commuting {\it pseudofermion fields} $F$
with the same quantum numbers as the fermions $\psi$ and modifying the
action as
\beq
S=\kappa\left(S_B+F^\dagger ({\cal M}^\dagger {\cal M})^{-p} F\right)\eeq
where $p=\frac{1}{2}$ or $p=\frac{1}{4}$ in the two cases.
Notice that this requires the determinants be real and positive definite.

Thus we now require an algorithm which can efficiently handle fields
coupled through a fractional inverse power of the fermion operator --
realized as a $P\times P$ matrix with dimension $P=2(N^2-1)M$ 
in the naive discretization (and twice this in the supersymmetric case).
To proceed further requires that we use an approximation for
the fractional power. Most effective is a partial fraction
realization of a rational function
approximation in some interval
\beq
x^{p}\sim a_0+\sum_{i=1}^Q \frac{a_i}{x+b_i}\quad \epsilon<x<1\eeq
The optimal coefficients $\{a_i,b_a\}$ can be determined offline
using the remez algorithm which seeks to minimise the absolute
value of the relative error for fixed $(Q,\epsilon)$.
In practice we have used approximations with $Q=10-15$ and intervals
ranging from $\epsilon=10^{-12}-10$ which conservatively covers
the range needed and yields relative errors $O(10^{-4}-10^{-8})$.
The latter systematic error is far below the statistical errors of
the Monte Carlo calculation and can thus be ignored.
The action we have simulated thus takes the form
\beq
S=\kappa\left(S_B(\phi)+a_0 F^\dagger F+\sum_{i=1}^Q
a_iF^\dagger\frac{1}{{\cal M}^\dagger {\cal M}+b_i}F\right)
\eeq
which resembles the contribution of a number of doublets of degenerate
fermions each with different mass parameters.    

This is still a non-local action and to simulate it efficiently
requires the use of an auxiliary classical dynamics. To be precise
one replaces the original partition function $Z=\int Dq e^{-S(q)}$
depending on a generic set of fields denoted by $\{q\}$ by another
comprising $\{q,p\}$ with partition function $Z^\prime=\int DqDp e^{-H}$
where
\beq 
H=S+\sum \frac{1}{2}p^2\eeq
Clearly expectation values derived from $Z^\prime$ and $Z$ are
identical. Furthermore (approximate) Hamiltonian evolution in this
phase space can be used to generate a series of global moves on
$\{q,p\}$ which, when subjected to the usual metropolis test, will
generate the canonical ensemble needed to simulate $Z^\prime$.\footnote{
This requires that the finite time step classical evolution be reversible
and area preserving which is true for the leapfrog integrator used here} The resulting
algorithm is termed {\it Hybrid Monte Carlo} \cite{hmc}.

In detail one starts from some initial set of
coordinates $q$, draws new momenta $p$ from a gaussian distribution and
then evolves the fields $\{p,q\}$ according to Hamilton's equations
\begin{eqnarray}
\frac{\partial q}{\partial t}&=&p\\\nonumber
\frac{\partial p}{\partial t}&=&-\frac{\partial S}{\partial q}=f
\end{eqnarray}
In practice a leapfrog integration with finite timestep $dt$ is used
to advance the fields along a classical trajectory of length $\tau$
\begin{eqnarray}
p_{n+1/2}&=&p_n+\frac{\Delta t}{2}f_n\\\nonumber
q_{n+1}&=&q_n+\Delta t p_{n+1/2}\\\nonumber
p_{n+1}&=&p_{n+1/2}+\frac{\Delta t}{2}f_{n+1}
\end{eqnarray}
In practice we have used this update $n_T$ times
corresponding to trajectory lengths $\tau=n_T\Delta t=0.05-1$. After this
we expect the lattice Hamiltonian to be conserved to $O(\Delta t^2)$.
We can then remove this $\Delta t$ dependence by treating the
final configuration as a potential global update in a metropolis
simulation and accept the new configuration with probability
$e^{-\Delta H}$. Subsequently the momenta are refreshed and
a new trajectory commenced.

In our case $q=\{X,U_1,\psi\}$ and the major inputs to this algorithm
are the forces $\frac{\partial S}{\partial \phi}$, $\frac{\partial S}{\partial
U_1}$ and $\frac{\partial S}{\partial F}$. Of these the most
computationally expensive are the pseudofermion terms. For example
the force term $\frac{\partial S_{PF}}{\partial X}$ is given by
\beq
\frac{\partial S_{PF}}{\partial X}=
-\sum_{i=1}^Q a_i\chi^{i\dagger}\frac{\partial ({\cal M}^\dagger {\cal
M})}{\partial X}\chi^i
\eeq
where $\chi^i$ is the solution of the auxiliary problem
\beq
({\cal M}^\dagger {\cal M}+b_i)\chi^i=F\eeq
The final trick required to render this approach computationally
feasible is to utilize a {\it multi-mass solver} to solve
this set of $Q$ equations iteratively and with a computational
cost similar to the case when $Q=1$. We have implemented a multimass
CG-solver for our work \cite{multimass}. Implementing the rational
approximation for the fractional fermion operator
in conjunction with the multimass solver in the way
described is termed the {\it Rational Hybrid Monte Carlo} algorithm
\cite{rhmc}.

%
\section{Continuum limits}
%
\label{app:continuum}

In this section we show the bosonic action observable $S_B$ and the Polyakov loop observable $P$ for increasing lattice sizes using the naive implementation for the quenched (figure \ref{fig:continuum1}), periodic  (figure \ref{fig:continuum2}) and thermal theories  (figure \ref{fig:continuum3}). We see that in all cases, over the range of $\beta$ of interest, 5 lattice points already appears to give results that are close to the continuum limit. Note that for the periodic theory, 
the continuum limit of the bosonic action is discussed in the main text in the results section.

\FIGURE[h]{
\centerline{ \includegraphics[width=3.5in,height=2.5in]{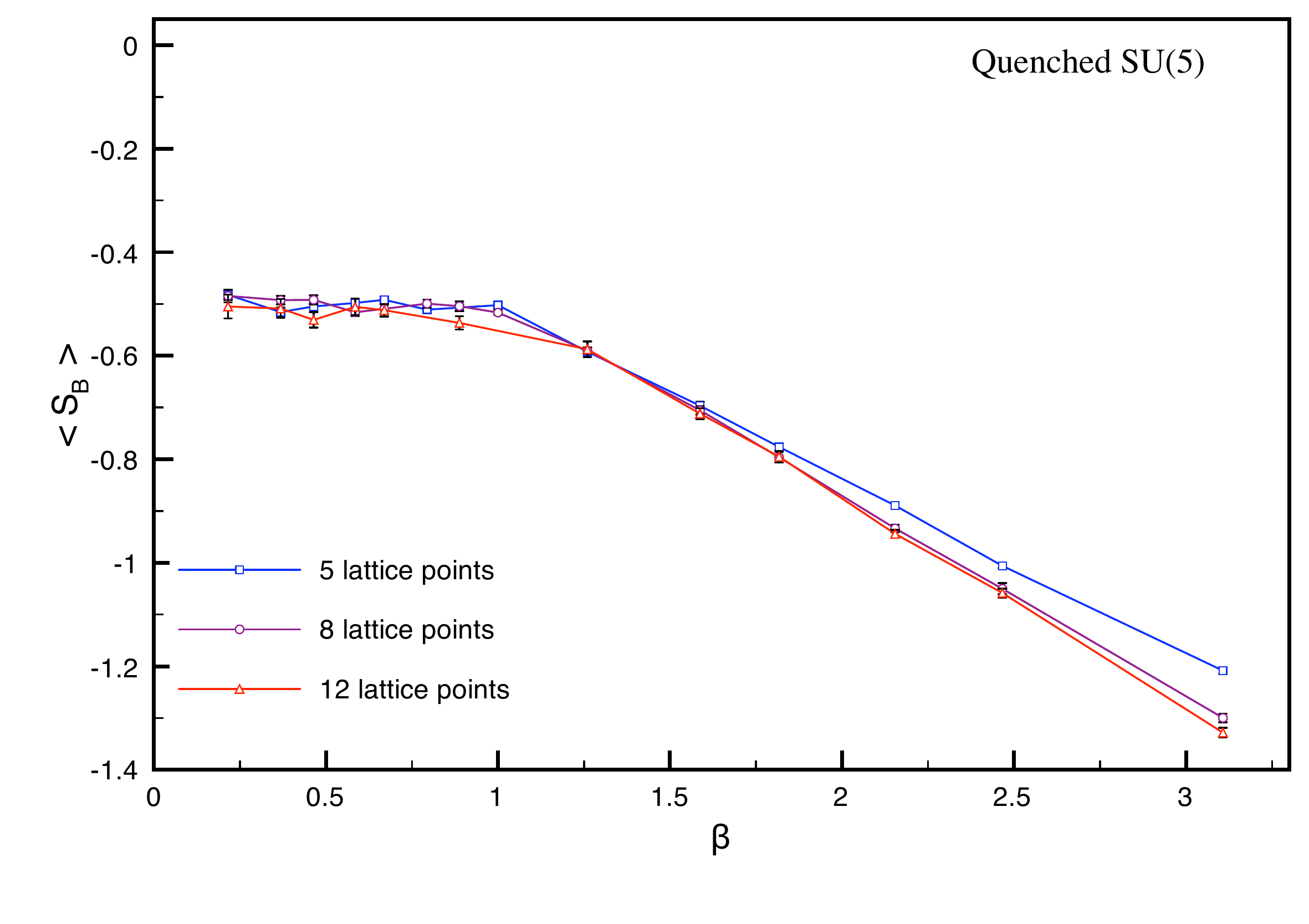}\includegraphics[width=3.5in,height=2.5in]{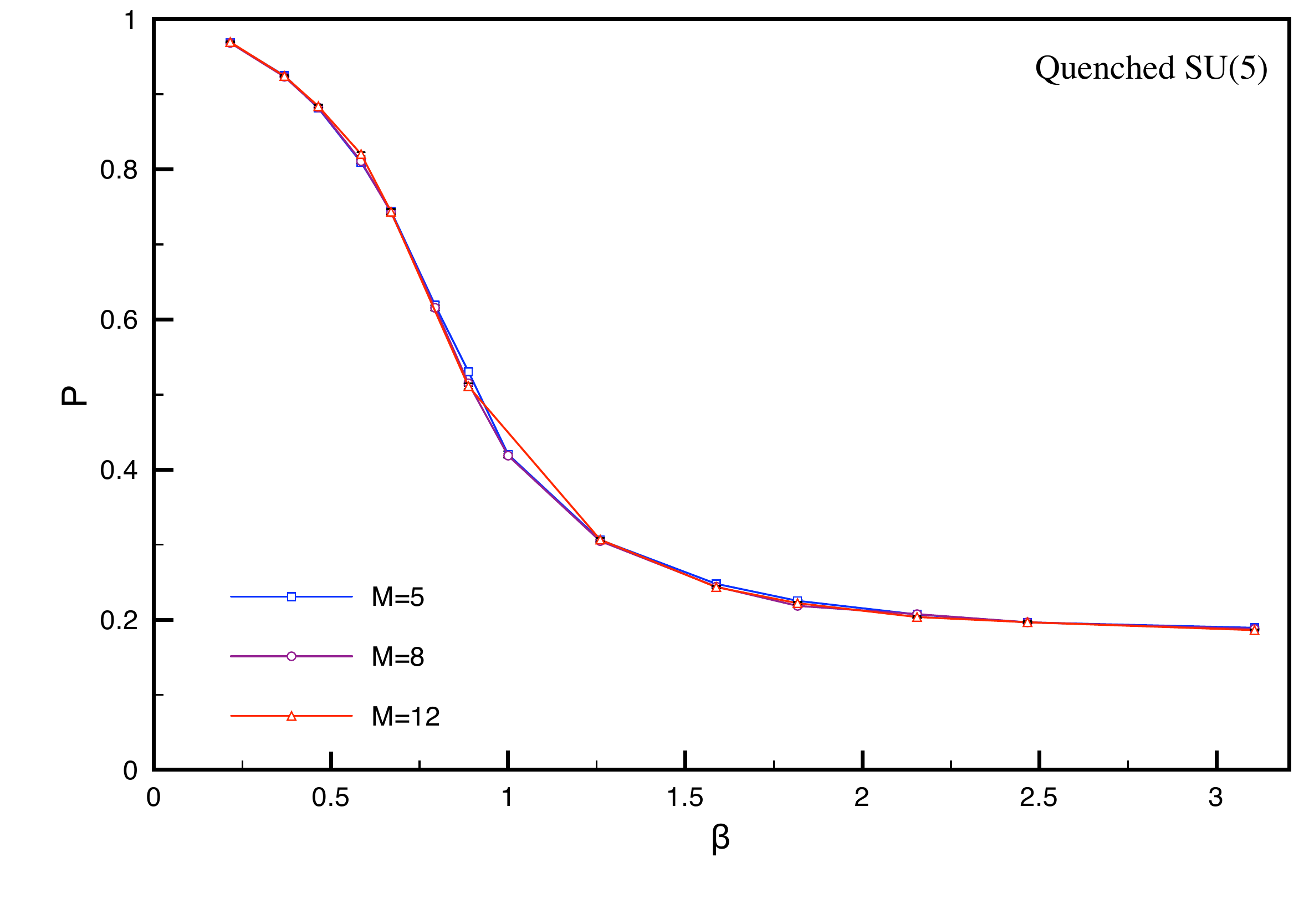} }
\centerline{ \includegraphics[width=3.5in,height=2.5in]{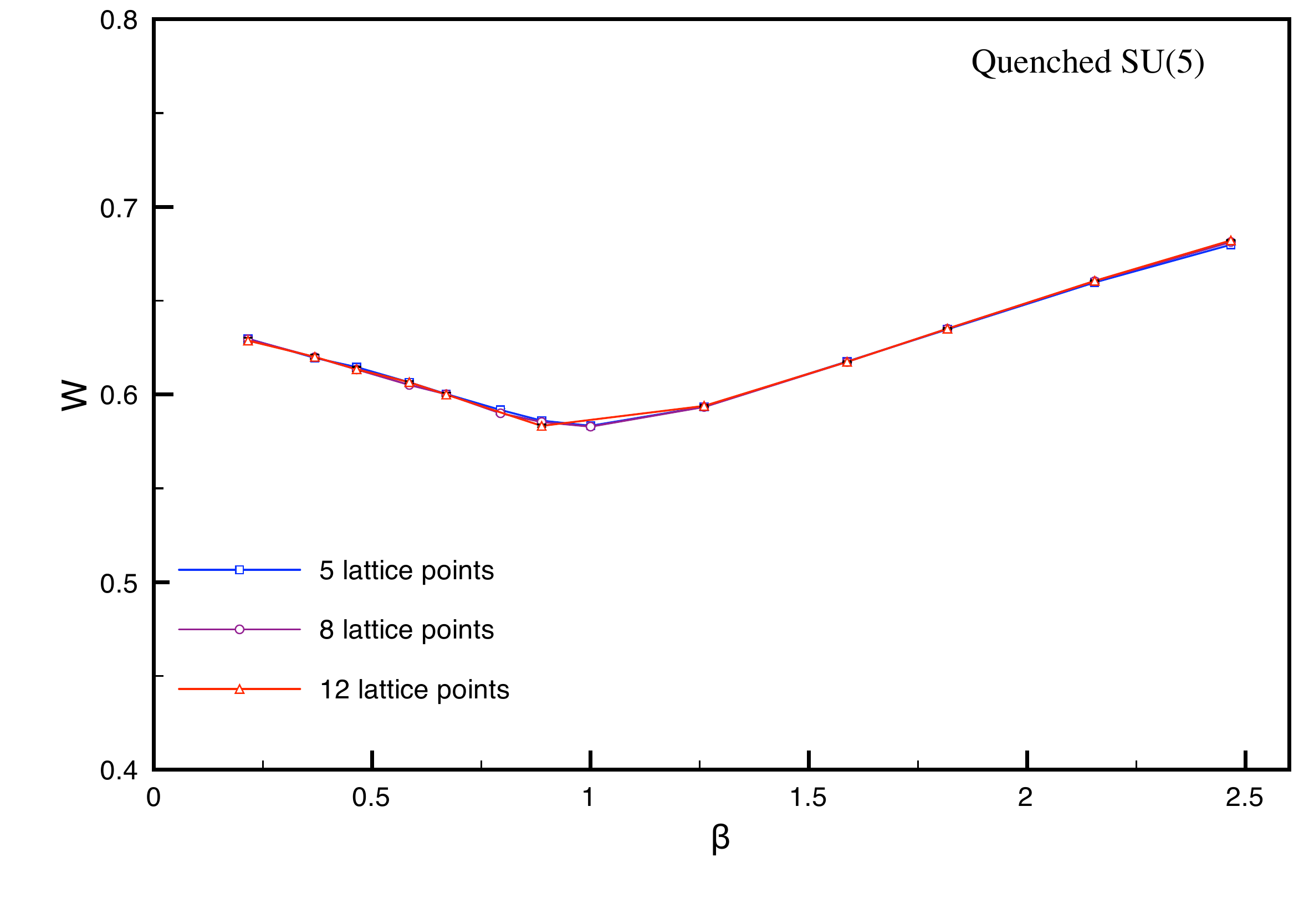}  }
     \caption{Plots of quenched theory action $S_B$, Polyakov loop $P$ and scalar width $W$ for $N=5$ with varying lattice points $M=5,8,12$.}
   \label{fig:continuum1}
}

\FIGURE[h]{
 \centerline{ \includegraphics[width=3.5in,height=2.5in]{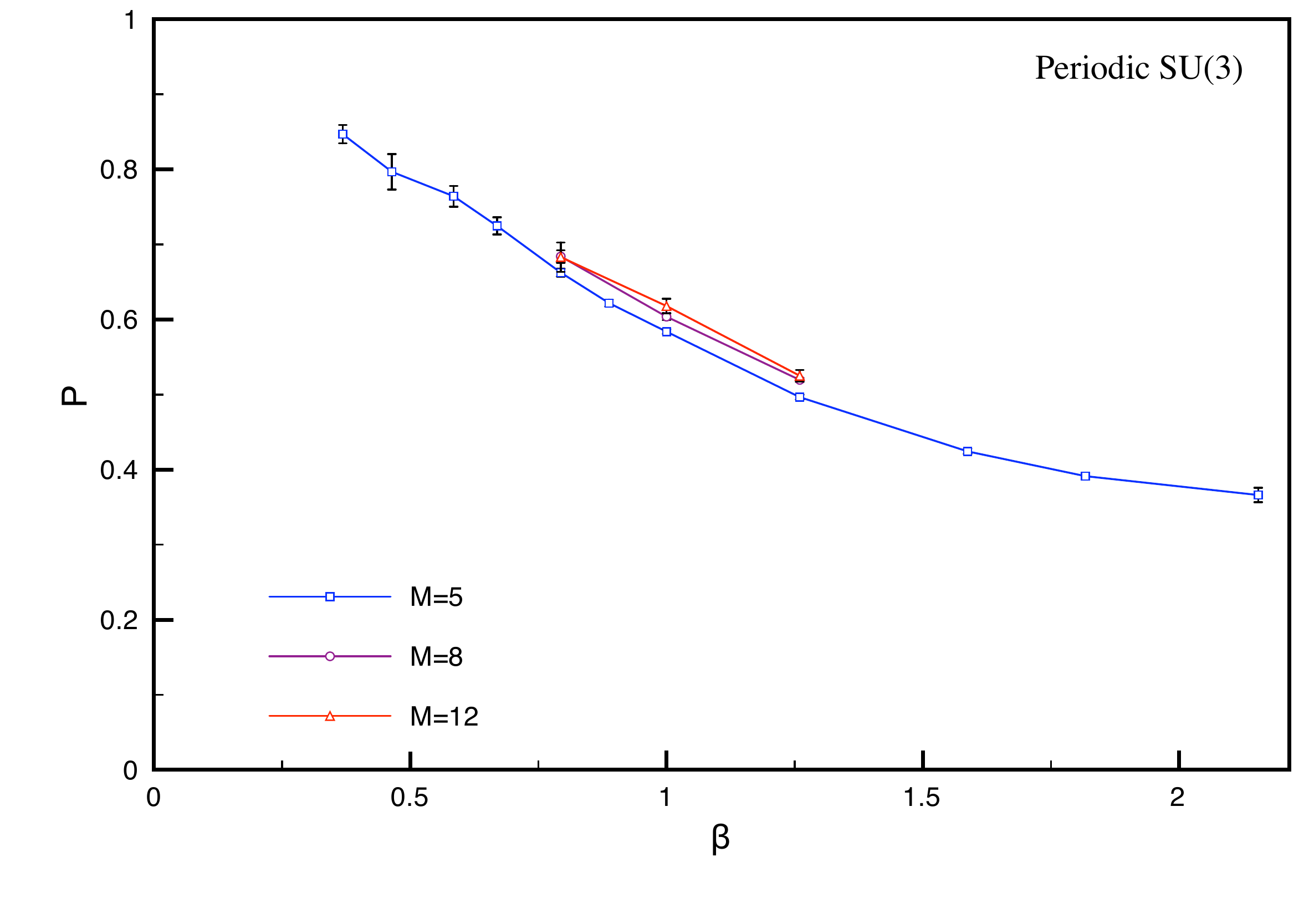} \includegraphics[width=3.5in,height=2.5in]{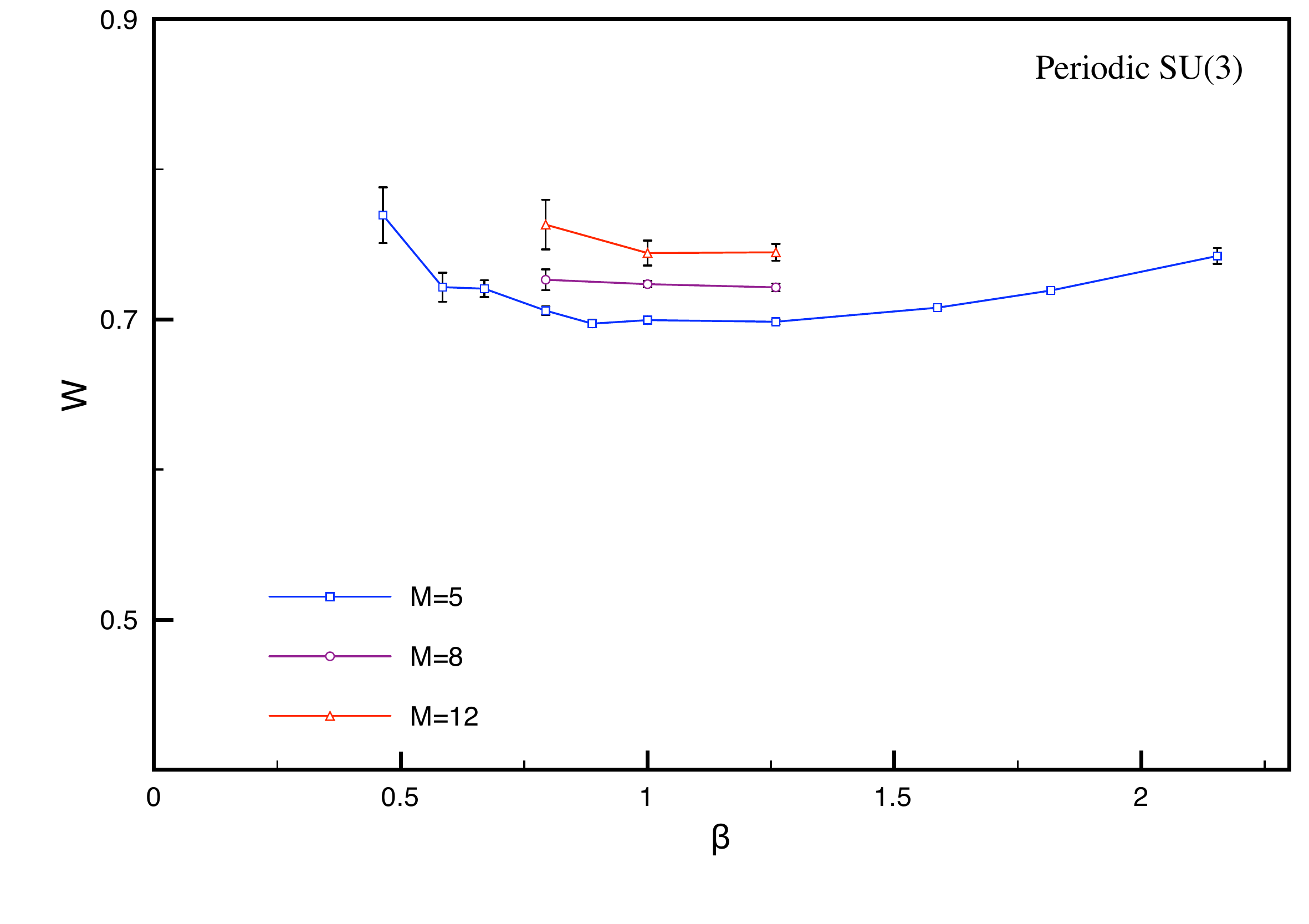} }
        \caption{Plots of periodic theory Polyakov loop $P$ and scalar width $W$ for $N=3$ with varying lattice points $M=5,8,12$.}
   \label{fig:continuum2}
}

\FIGURE[h]{
 \centerline{ \includegraphics[width=3.5in,height=2.5in]{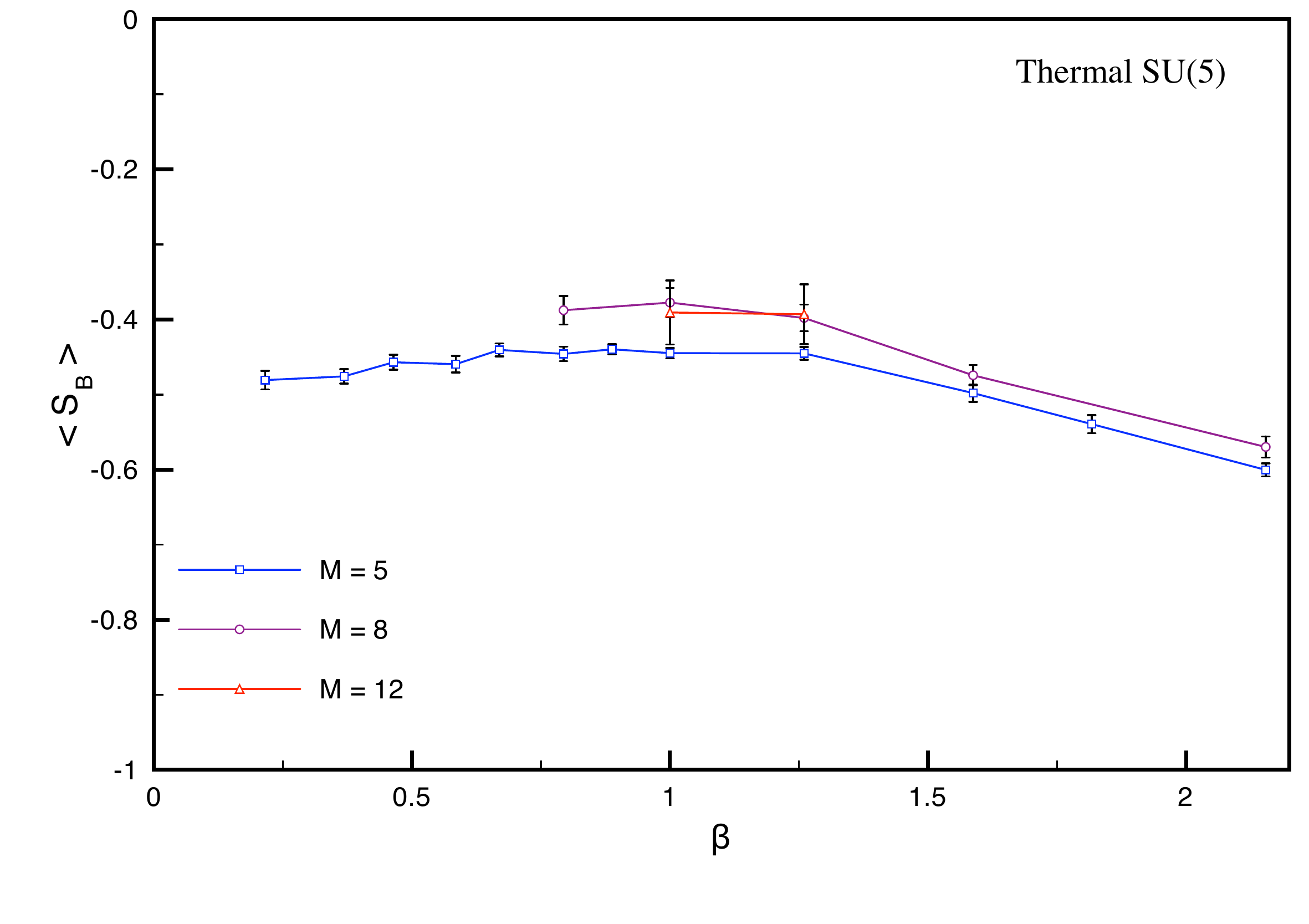} \includegraphics[width=3.5in,height=2.5in]{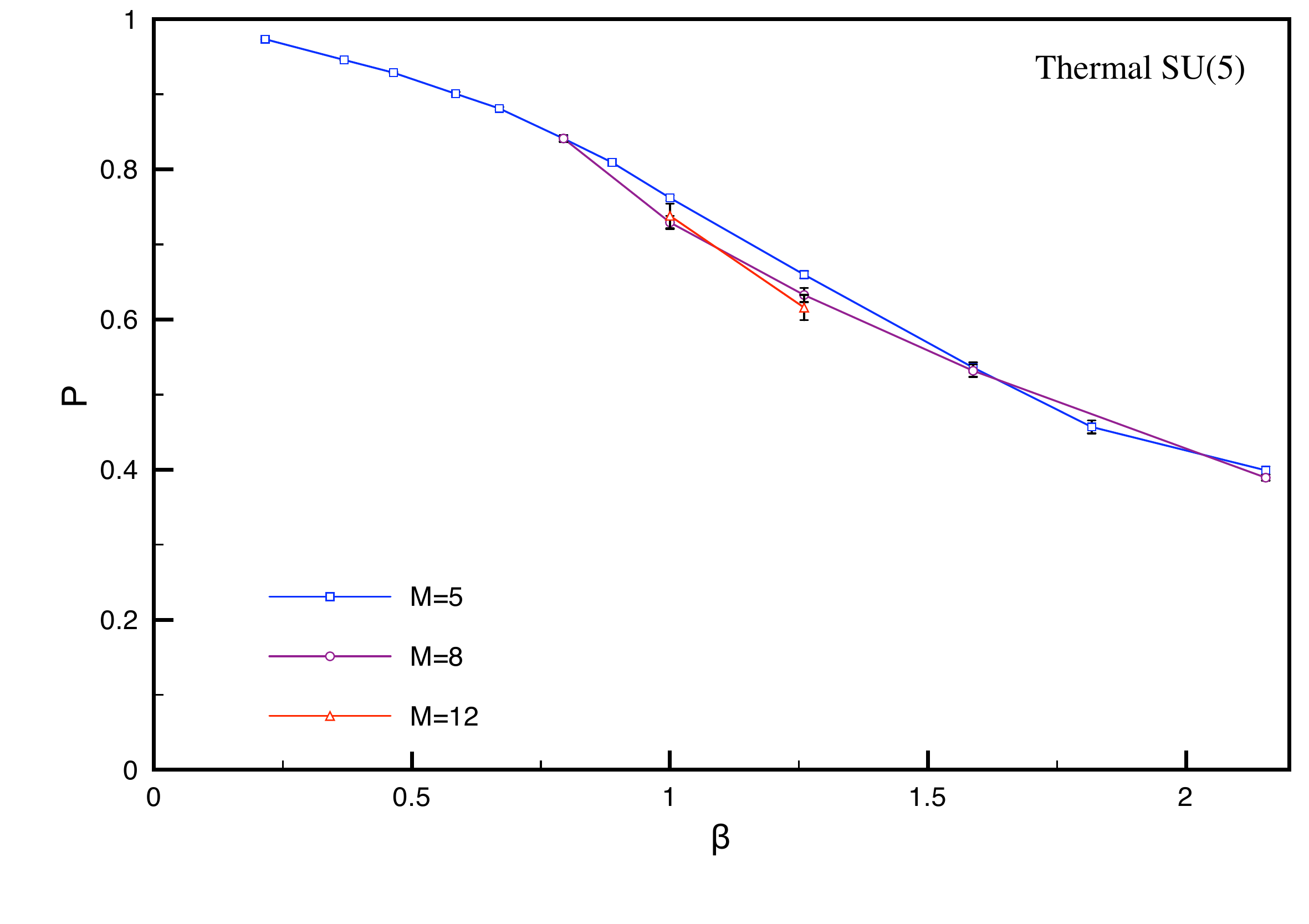}}
 \centerline{ \includegraphics[width=3.5in,height=2.5in]{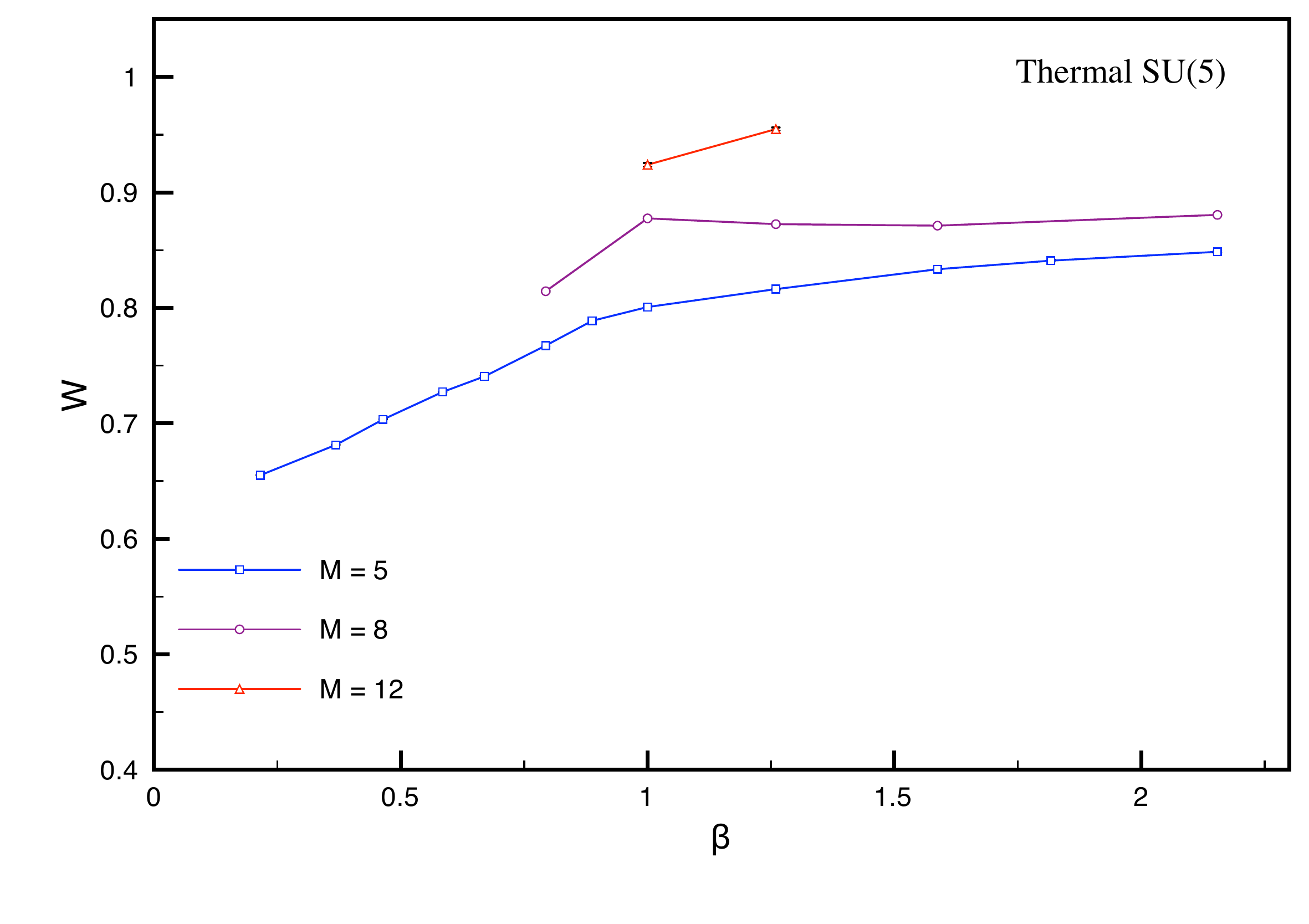}}
        \caption{Plots of thermal theory action $S_B$, Polyakov loop $P$ and scalar width $W$ for $N=5$ with varying lattice points $M=5,8,12$.}
   \label{fig:continuum3}
}

We have observed that the thermal theory exhibits large lattice artifacts
for small values of $N$ and the number of lattice points $M$. This is true
both for the naive and supersymmetric actions. To illustrate this  figure~\ref{fig:su2apbc} shows a plot of the Monte Carlo history of the maximal
fermion eigenvalue $\lambda_{\rm max}$ for the $SU(2)$ theory
at $\beta^3=10.0$
obtained using the supersymmetric implementation. The
plot includes data for lattice sizes $M=5,10,20$. The data is shown as a function of
physical time measured in units of $\tau=1$ RHMC trajectories. Data is plotted
every $100$ units of time with short time fluctuations
being removed by plotting a running average obtained using a temporal window of
$10$ units of time. This fermion
eigenvalue is
strongly correlated with the rms value of the scalar eigenvalues. Very large
fluctuations are seen with extremely large autocorrelation times for
$M=5$ lattice points. These fluctuations are not manifest in
the bosonic action $S_B$ and appear to derive from large fluctuations
of the scalars in the classical moduli space. These motions are
strongly suppressed in the periodic theory on account of the
presence of superpartner fermion zero modes. Of course as the
number of lattice sites is increased the masses of these
would be zero modes are lowered and once again
they act so as to inhibit fluctuations of the scalars in the
zero mode directions. We see this in the data which
shows a marked reduction in the amplitudes of these
fluctuations as the number of lattice points
$M$ increases. Of course the dimension of the moduli space
varies like $N$ and so we would also expect these effects to be suppressed
at large $N$ which is also observed.

The problems seem most acute with the supersymmetric action and so for data presented in the main text we
have concentrated on using the naive implementation for the bulk of our thermal runs. 
In this case to avoid these strong artifacts we require
$N\ge 5$, $M\ge 5$ for the range of $\beta$ we are studying, $\beta^3 \le 10.0$, and as we see in the figure \ref{fig:continuum3} we obtain a good continuum limit.

\FIGURE[h]{
 \centerline{\includegraphics[width=4.5in,height=3.5in]{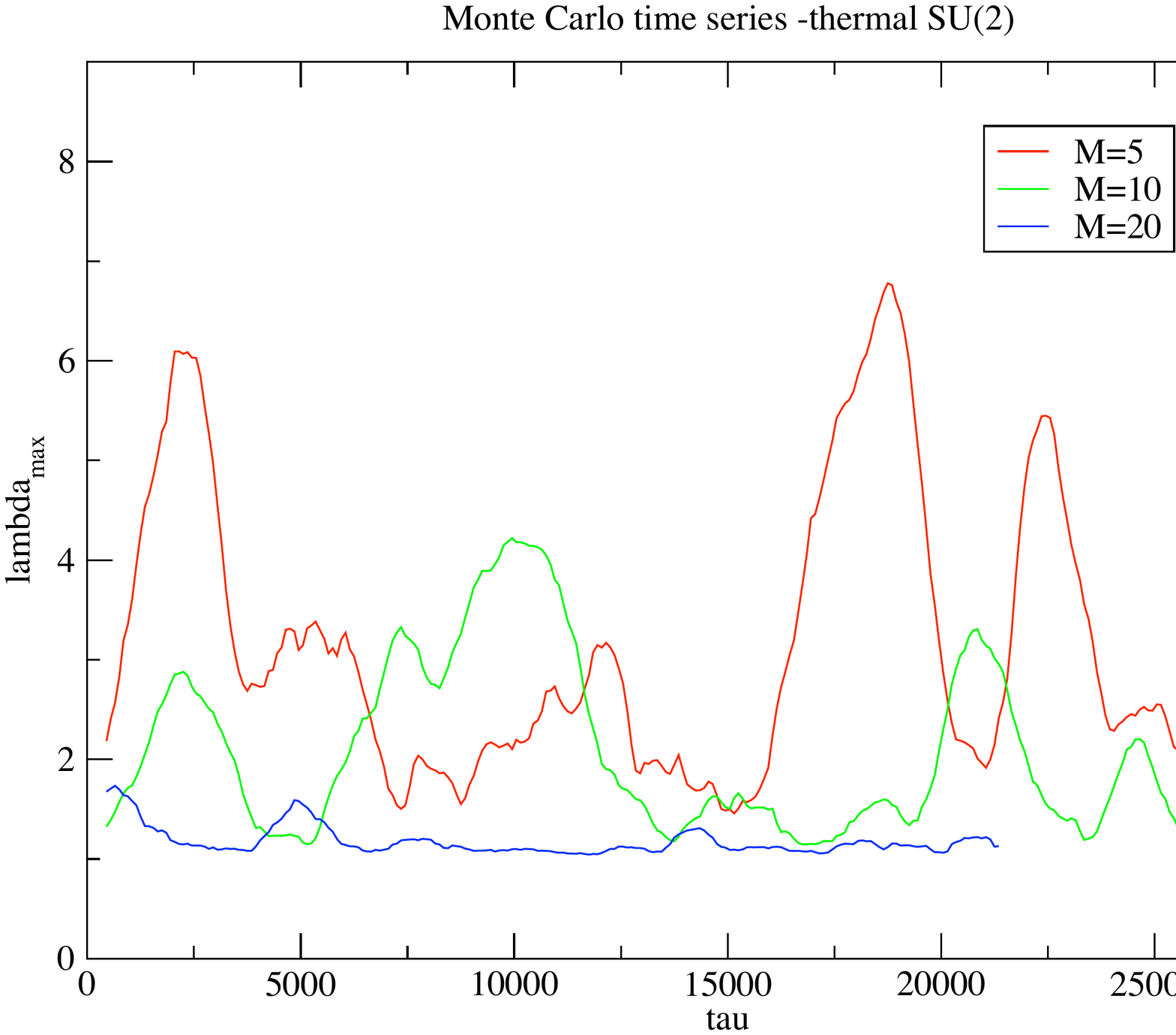}}
 \caption{Monte Carlo history of maximal fermion eigenvalue for
 $SU(2)$ and supersymmetric action}
   \label{fig:su2apbc}
}

\FIGURE[h]{
 \centerline{ \includegraphics[width=3.5in,height=2.5in]{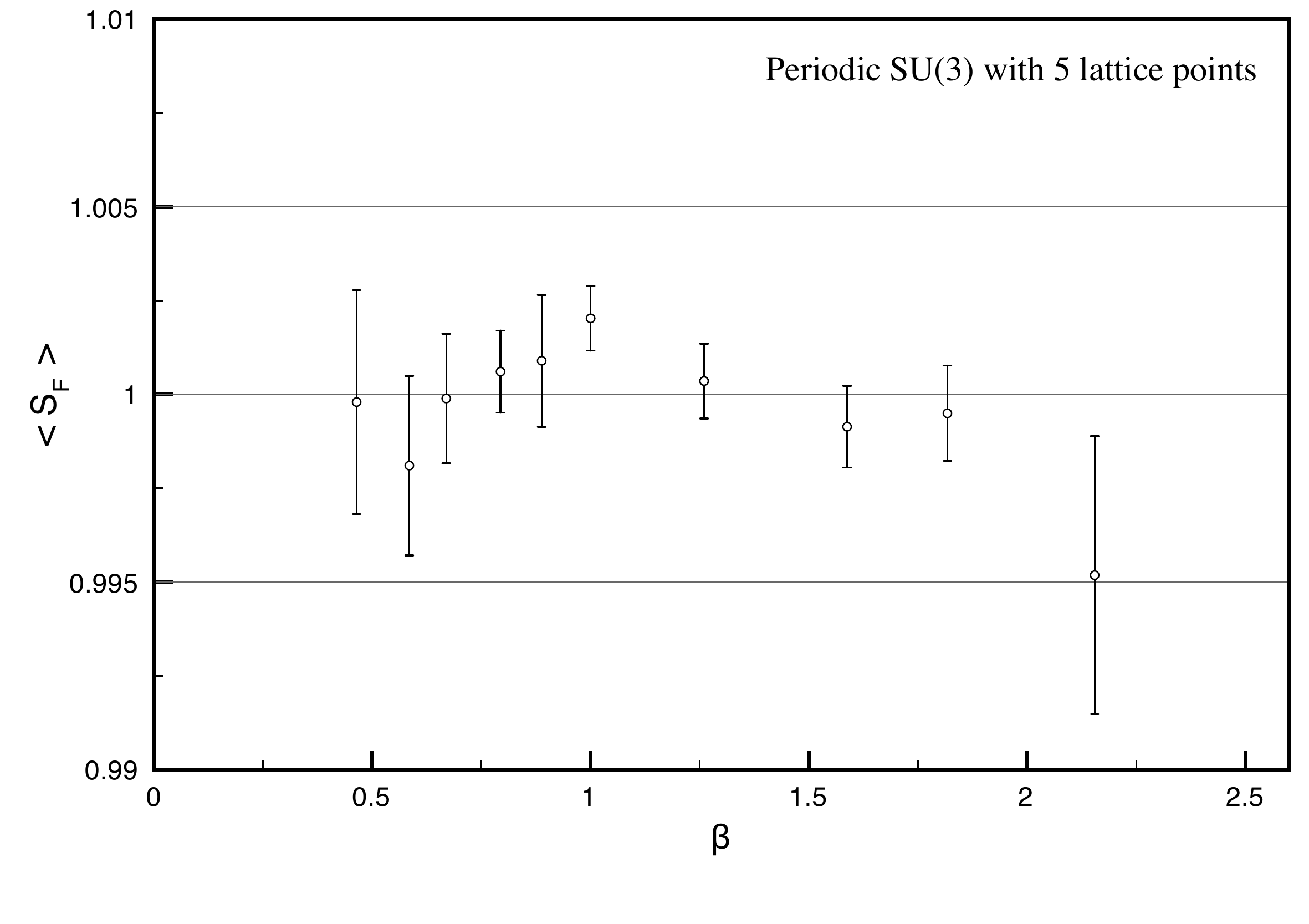} \includegraphics[width=3.5in,height=2.5in]{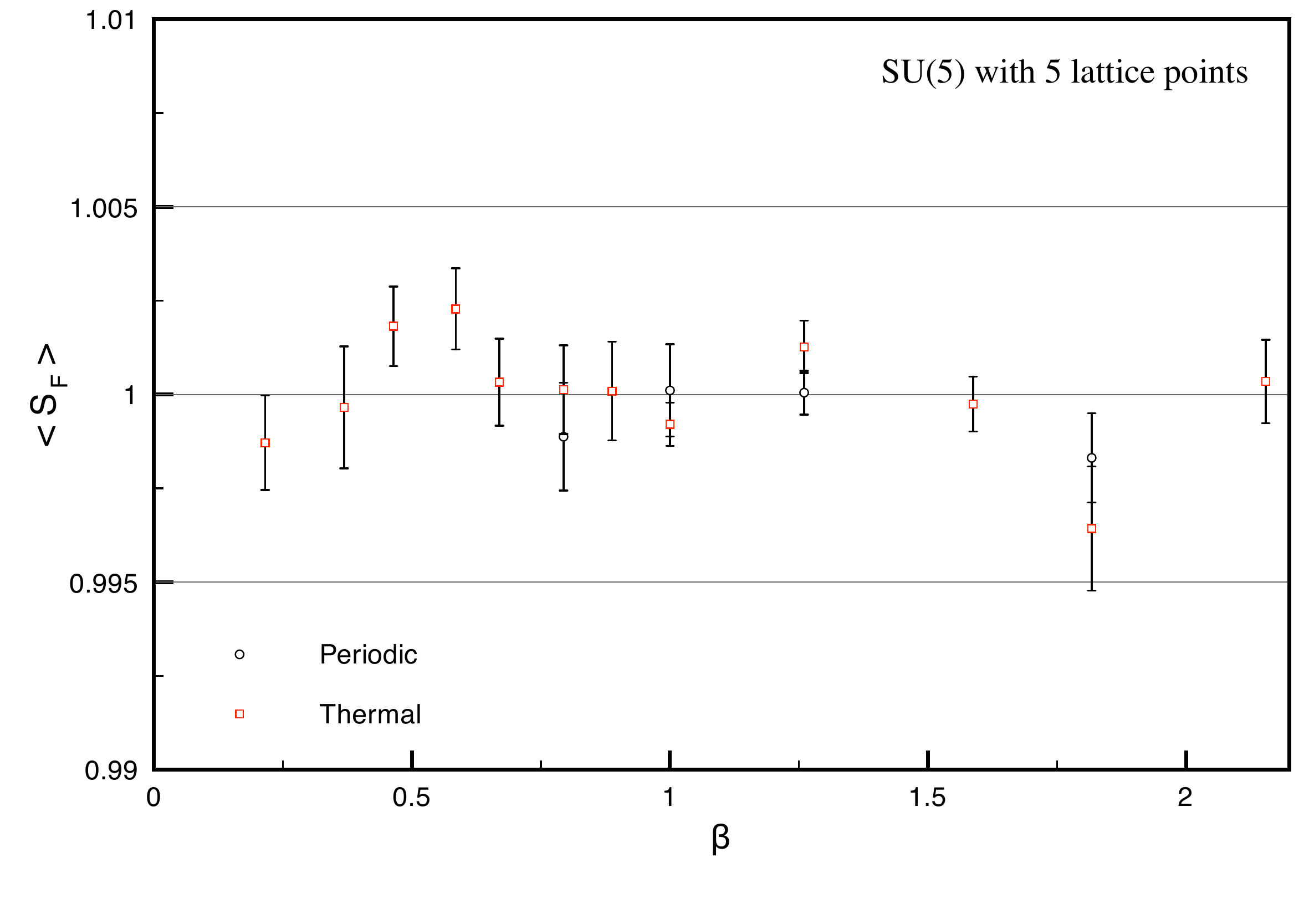}}
        \caption{Plots of the fermion action $S_F$ for periodic theory with $N=3$ (left) and periodic and thermal with $N=5$ (right), both for 5 lattice points.}
   \label{fig:check1}
}

We conclude with figure \ref{fig:check1} which shows data from the naive implementation of the fermion action $S_F$ for the naive theory for representative values of $N$ and 5 lattice points. Scaling arguments imply this quantity should equal unity and we see the data confirms this to within statistical errors, providing a non-trivial check of the RHMC routines. A similar check was performed on the supersymmetric implementation.

%
\bibliographystyle{JHEP}
\bibliography{ref}
%

\end{document}